\documentclass[reprint,aps,prc,amsmath,amssymb,nofootinbib,superscriptaddress]{revtex4-2}
 
\usepackage[T1]{fontenc}
\usepackage[utf8]{inputenc}
\usepackage{graphicx,color,rotating,pifont}
\usepackage{mathtools}
\usepackage{siunitx}
  
\usepackage{bm}
\usepackage{ae}

\usepackage{siunitx}
\usepackage{dcolumn}
\usepackage{txfonts}
\usepackage{tensor}
\usepackage{braket}
\usepackage{booktabs}
\usepackage{xspace}
\usepackage{mathrsfs}
\usepackage{amssymb} 
\usepackage{amsmath}
\usepackage{esvect}
\usepackage{diagbox}

\usepackage{hyperref}

\DeclareSIUnit{\fm}{\femto\meter}

\renewcommand{\vec}[1]{\ensuremath{\bm{#1}}}

\newcommand{\beq}{\begin{equation}}
\newcommand{\eeq}{\end{equation}}
\newcommand{\beqn}{\begin{eqnarray}}
\newcommand{\eeqn}{\end{eqnarray}}
\newcommand{\bsub}{\begin{subequations}}
\newcommand{\esub}{\end{subequations}}
\newcommand{\bpm}{\begin{pmatrix}}
\newcommand{\epm}{\end{pmatrix}}

\begin{document}
 
\title{Multireference covariant density-functional theory  for the low-lying states of odd-mass nuclei}

 \author{E. F. Zhou}
  \affiliation{School of Physics and Astronomy, Sun Yat-sen University, Zhuhai 519082, P.R. China}  

\author{X. Y. Wu}
\address{College of Physics and Communication Electronics, Jiangxi Normal University, Nanchang 330022, P.R. China}
\affiliation{School of Physics and Astronomy, Sun Yat-sen University, Zhuhai 519082, P.R. China}  
 
  \author{J. M. Yao}   
 \email{Corresponding author: yaojm8@sysu.edu.cn}
  \affiliation{School of Physics and Astronomy, Sun Yat-sen University, Zhuhai 519082, P.R. China}

\date{\today}

\begin{abstract}
We extend multireference covariant density-functional theory (MR-CDFT) based on a relativistic point-coupling energy functional to describe the low-lying states of odd-mass nuclei. The nuclear wave function is constructed as a superposition of quadrupole-octupole deformed mean-field configurations, with projection onto angular momentum, particle numbers, and parity within the framework of the generator coordinate method. Using $^{25}$Mg as an example, we calculate the energy spectrum, electric multipole, and magnetic dipole transition strengths based on three different schemes for the mean-field configurations of odd-mass nuclei. We find that the low-energy structure of $^{25}$Mg is reasonably reproduced in all three schemes. In particular, the effect of octupole correlation is illustrated in the application to the low-lying parity doublets of $^{21}$Ne. This work demonstrates the success of the MR-CDFT for the low-lying states of odd-mass nuclei with possible strong quadruple-octupole correlations.
\end{abstract}

\maketitle


  \section{Introduction}\label{introduction}

 The low-lying states of odd-mass nuclei contain rich nuclear structure information, including energy spectra, electromagnetic moments, and transition strengths. These quantities are highly sensitive to the underlying configurations of the states~\cite{Bohr:1998,Ring:1980}. A detailed study of the corresponding observables is crucial for understanding nuclear shell structure and collective excitations. Moreover, precise information regarding the low-lying states of odd-mass nuclei is essential for interpreting measurements related to dark matter candidates~\cite{Klos:2013} and nonzero atomic electric dipole moments~\cite{Engel:2013PPNP}.  The latter provides a valuable constraint on the strengths of time-reversal symmetry-violating interactions.  However, due to the interplay of single-particle and collective motions and the presence of half-integer angular momenta, the low-energy structure of odd-mass nuclei is complex, presenting a considerable challenge for nuclear models. 

The low-lying states of odd-mass nuclei have primarily been investigated utilizing a (quasi)particle-rotor model (PRM)~\cite{Bohr:1998,Ring:1980} or a (quasi)particle-phonon model~\cite{Malov:2023}, where the unpaired valence (quasi)particle couples with nuclear collective rotations or vibrations. The PRM has proven successful in describing the level structures of rotational and even transitional nuclei. However, it introduces some phenomenological parameters that need determination from data of the corresponding nucleus. Notably, it does not account for the effects of the Pauli exclusion principle between valence nucleons and core nucleons, core-polarization excitations, and rotation-induced internal structure changes. These effects can be well incorporated in a cranking model, originally introduced by Inglis~\cite{Inglis:1954,Inglis:1956}, where nucleons inside an atomic nucleus are treated as independent particles moving in an average field rotating with the coordinate frame. The cranking model, an approximate variation after projection onto angular momentum, when combined with modern energy density functionals (EDFs), offers a self-consistent microscopic method for nuclear rotational motions~\cite{Bonche:1987,Koepf:1989,Konig:1993,Egido:1993,Vretenar:2005}. Since the cranking model is formulated in the intrinsic frame, nuclear wave function is characterized by a definite value of cranking frequency but lacks a well-defined angular momentum. Consequently, interpreting nuclear spectroscopic data within this model can be challenging. Furthermore, the cranking model fails to explain the experimentally observed "Coriolis attenuation" effect~\cite{Almberger:1980}, and it cannot capture the interplay of collective rotational motions and shape vibrations in transitional nuclei.

 Over the past few decades, the majority of studies on odd-mass nuclei, starting from self-consistent mean-field approaches that employ various EDFs~\cite{Bonneau:2007,Martin:2008EFA,Schunck:2010EFA,Pan:2022,Giuliani:2023}, has primarily concentrated on the mean-field level. In the Hartree-Fock (HF) approach with the Bardeen-Cooper-Schrieffer (BCS) theory for nuclear pairing correlation, the ground state of an odd-mass nucleus is typically obtained similarly to that of an even-even nucleus, with the unpaired nucleon fixed onto one of the single-particle orbitals. Specifically, the occupation of the blocked orbital is $v^2_k=1$ and $v^2_{\bar k} = 0$, where $\ket{\bar{k}}$ is the time-reversal partner state of $\ket{k}$. In other words, the blocked orbitals $(\ket{k},\ket{\bar{k}})$ do not contribute to pairing correlations~\cite{Ring:1980}.   When the polarization effect of the unpaired nucleon is considered self-consistently, the valence nucleon of the blocked orbital generates nonzero currents, breaking the time-reversal invariance of the mean-field state~\cite{Rutz:1998,Hofmann:1988,Yao:2006}. In this case, the HF+BCS approach is often replaced by the Hartree-Fock-Bogoliubov (HFB) approach. For simplicity, the equal filling approximation (EFA) is frequently adopted to handle the unpaired nucleon~\cite{Bender:2000fqv,Martin:2008EFA,Schunck:2010EFA,Guzman:2010EFA,Li:2012EFA}, i.e., $v^2_k = v^2_{\bar k} =0.5$ in the canonical basis, where time-reversal symmetry is restored, and polarization effect from time-odd fields is omitted. The studies with the schemes of both EFA and exact blocking, where the time-odd mean field is fully taken into account demonstrated that the impact of time-odd fields is of the order of $100-200$~keV~\cite{Schunck:2010EFA,Giuliani:2023}. In the HFB approach, the wave function of an odd-mass nucleus is usually approximated as a one-quasiparticle state, and the wave function can be obtained simply by exchanging one column of the Bogoliubov transformation matrices\cite{Ring:1980}. Alternatively,  the wave function of an odd-mass nucleus can be approximated with one quasiparticle excitation on top of a false quantum vacuum (FQV) state with an odd average particle number~\cite{Duguet:2001}. 
  
  Mean-field wave functions are typically determined using the variational principle,  leading to solutions corresponding to local energy minima within the restricted Hilbert space. However, this approach does not guarantee the full retention of the symmetries of a given nuclear Hamiltonian in the mean-field wave functions, making it challenging to compare with data on nuclear low-lying states. To address these issues,  the core-(quasi)particle coupling model~\cite{Meyer:1979NPA,Libert:1982PRC,Quan:2017CQC,Sun:2019CQC} and interaction boson-fermion models~\cite{Nomura:2016IBFM,Nomura:2017IBFM} have been developed. These hybrid models employ phenomenological parameters determined by the mean-field calculations for the even-even nuclear core and have proven successful in global studies of low-lying states of odd-mass nuclei.  In the pursuit of a full microscopic and self-consistent description of nuclear low-lying states, where mean-field configurations and collective wave functions are derived from the same underlying effective nuclear interactions or EDFs, multireference density-functional theory (MR-DFT) has been developed \cite{Bender:2003SMF}. The MR-DFT incorporates projection techniques to restore broken symmetries in mean-field solutions and utilizes the generator coordinate method (GCM) to explicitly consider large-amplitude fluctuations around the most probable mean-field solution. Review papers~\cite{Bender:2003SMF,Niksic:2011_PPNP,Robledo:2019JPG,Sheikh:2021JPG,Yao:2022_HB} provide comprehensive insights into this field. Compared to even-even nuclei, extending MR-DFT for odd-mass or odd-odd nuclei is more complex, requiring  mixing of many nearly degenerate single-(quasi)particle excitation configurations. 
  
  Early studies of the low-lying states of odd-mass nuclei employed angular-momentum projection (AMP) based on a Hartree-Fock (HF) state defined within one major shell~\cite{Bassichis:1965,Gunye:1967,Rath:1993}. More advanced approaches include both AMP and particle-number projection (PNP) on top of one or more quasiparticle states defined within three shells using a pairing-plus quadrupole interaction~\cite{Hara:1984,Hara:1995}, as well as AMP+PNP on top of one quasiparticle state using a microscopically-derived Brueckner $G$-matrix with additional phenomenological terms~\cite{Hammaren:1985}. Other methods such as antisymmetrized molecular dynamics (AMD) combined with GCM~\cite{Kimura:2007,Kimura:2011}, resonating-group method (RGM) combined with AMD for one-neutron halo structure in \nuclide[31]{Ne}\cite{Minomo:2012}, and three-dimensional angular momentum projection (3DAMP) and PNP combined with GCM based on triaxially-deformed HFB wave functions with time-reversal symmetry using Skyrme-type EDFs for \nuclide[25]{Mg}\cite{Bally:2014prl}, \nuclide[129]{Xe}\cite{Bally:2022_EPJA}, and \nuclide[197]{Au}\cite{Bally:2023_EPJA} have been developed. Similar studies using the Gogny force have been carried out for odd-mass magnesium isotopes~\cite{Borrajo:2016vfz,Borrajo:2017,Borrajo:2018sqj}.

In the past decades, covariant density-functional theory (CDFT) has proven highly successful in various aspects of nuclear physics\cite{Meng:2017}. Within CDFT, Lorentz invariance imposes stringent restrictions on the number of parameters in the EDF. Notably, the relativistic framework of CDFT naturally incorporates the spin-orbit potential, and the time-odd  fields can be included without introducing any additional free parameters. This feature is particularly important for describing odd-mass nuclei and rotating nuclei.  On the mean-field level, the single-reference (SR) CDFT, or the relativistic mean-field (RMF) theory with pairing correlation considered in the BCS theory or with a general Bogoliubov transformation, provides a good description of the static ground-state properties for finite nuclei~\cite{Ring:1996PPNP,Vretenar:2005,Meng:2006,Meng:2017}. Similar to other nuclear DFT, however, SR-CDFT suffers from the drawbacks of symmetry violation. To restore the broken symmetries, SR-CDFT has been extended to a multireference framework with the  AMP~\cite{Niksic:2006PRC,Yao:2009traxi,Zhao:2016,SunXX:2021}, PNP, and parity projection for axially, triaxially~\cite{Yao:2011_Mg,Yao:2014_76Kr,Wang:2019_TPSM,Wang:2021triaxial}, and octupole-deformed~\cite{Yao:2015_Octupole,Zhou:2016_Ne20,Rong:2023} nuclei, respectively. The shape fluctuation effects can be taken into account with GCM. The MR-CDFT has also been applied to the low-lying states of hypernuclei~\cite{Mei:2016PRC,Xia:2023}, and the nuclear matrix element of neutrinoless double-beta decay~\cite{Song:2014,Yao:2015_NLDBD,Song:2017,Ding:2023}. See, for instance, the review papers~\cite{Niksic:2011_PPNP,Yao:2022PPNP,Zhou:2023_IJMPE}. All these studies have been restricted to even-even nuclei. In view of the success of the MR-CDFT, we extend this framework to the low-lying states of odd-mass nuclei.

The paper is arranged as follows. In Sec.\ref{sec:method}, we present the framework of MR-CDFT for odd-mass nuclei, including the construction of mean-field configurations for odd-mass nuclei in three different schemes, the restoration of broken symmetries in the configurations, and the mixing of these configurations with the GCM. Taking \nuclide[25]{Mg} as an example, we present the results on the energy spectra, electric multipole, and magnetic dipole transitions in the low-lying states in Sec.\ref{sec:results}. Besides, we demonstrate the applicability of the method for \nuclide[21]{Ne} with strong quadrupole-octupole correlations in the low-lying parity doublet states.  A conclusion is drawn in Sec.~\ref{sec:summary}.

 \section{Theoretical framework}
 \label{sec:method} 
 \subsection{GCM for nuclear low-lying states}
 
In the MR-CDFT, the wave functions of low-lying states for an odd-mass nucleus are constructed with the symmetry-projected GCM as follows, 
\begin{equation}
\label{eq:gcmwf}
\vert \Psi^{J\pi}_\alpha\rangle
=\sum_{c} f^{J\alpha \pi}_{c}  \ket{NZ J\pi; c},
\end{equation} 
Here, $\alpha$ distinguishes the states with the same angular momentum $J$, and the symbol $c$ is a collective label for the indices $(K,\kappa,\mathbf{q})$. The basis function with correct quantum numbers ($NZJ\pi$) is given by
\begin{equation}
\label{eq:basis}
\ket{NZ J\pi; c} 
=  \hat P^J_{MK} \hat P^N\hat P^Z\hat P^\pi\ket{\Phi^{\rm (OA)}_\kappa(\mathbf{q})},
\end{equation}
where $\hat P^{J}_{MK}$, $\hat{P}^{N, Z}$, and $\hat P^\pi$ are projection operators that select components with the angular momentum $J$, neutron number $N$, proton number $Z$, and parity $\pi=\pm$~\cite{Ring:1980},
\begin{subequations}
\begin{align}
\label{eq:Euler_angles}
\hat P^{J}_{MK} &= \frac{2J+1}{8\pi^2}\int d\Omega D^{J\ast}_{MK}(\Omega) \hat R(\Omega),\\
\label{eq:PNP_gauge}
\hat P^{N_\tau} &= \frac{1}{2\pi}\int^{2\pi}_0 d\varphi_{\tau} e^{i\varphi_{\tau}(\hat N_\tau-N_\tau)},\\
\hat P^\pi &= \frac{1}{2}(1+\pi\hat{\mathscr{P}}).
\end{align}
\end{subequations}
The operator $\hat P^J_{MK}$ extracts the component of angular momentum along the intrinsic axis $z$ defined by $K$. The Wigner $D$-function is defined as $D^{J}_{MK}(\Omega)\equiv\bra{JM}\hat R(\Omega)\ket{JK}=\bra{JM}e^{i\phi\hat J_z}e^{i\theta\hat J_y}e^{i\psi\hat J_z}\ket{JK}$, where $\Omega=(\phi, \theta, \psi)$ represents the three Euler angles. The $\hat N=\sum_k a^\dagger_k a_k$ and $\hat{\mathscr{P}}$ are particle-number operator and space-inversion operator, respectively. The mean-field configurations $\ket{\Phi^{\rm (OA)}_\kappa(\mathbf{q})}$ for odd-mass nuclei are chosen as one quasiparticle state,
\begin{eqnarray}
\label{eq:odd-mass-wfs}
 \ket{\Phi^{\rm (OA)}_\kappa(\mathbf{q})}  =\alpha^\dagger_\kappa \ket{\Phi_{(\kappa)}(\mathbf{q})},
\end{eqnarray} 
where $\ket{\Phi_{(\kappa)}}$ is a quasiparticle vacuum from the SR-CDFT calculation, 
\begin{eqnarray} 
 \alpha_\kappa \ket{\Phi_{(\kappa)}(\mathbf{q})}=0.
\end{eqnarray}  
The quasiparticle operators $(\alpha, \alpha^\dagger)$ are connected to single-particle operators $(a, a^\dagger)$ via the Bogoliubov transformation~\cite{Ring:1980}, 
\begin{eqnarray}
\label{eq:Bogoliubov_transformation}
\left(
\begin{array}{cc}
\alpha\\
\alpha^{\dag}\\
\end{array}
\right)=\left(
\begin{array}{cc}
U^\dag&V^{\dag}\\
V^T&U^T\\
\end{array}
\right)\left(
\begin{array}{cc}
a\\
a^\dag\\
\end{array}
\right).
\end{eqnarray}
More details on the SR-CDFT calculation are introduced in the next subsection and appendix~\ref{app:kernels}.

As usual, we optimize the energy of the state (\ref{eq:gcmwf}) with respect to the weight function $f^{J\alpha \pi}_{c}$ with the variational principle which leads to the following Hill-Wheeler-Griffin (HWG) equation~\cite{Hill:1953,Ring:1980},
\begin{eqnarray}
\label{eq:HWG}
\sum_{c'}
\Bigg[\mathscr{H}^{NZJ\pi}_{cc'}
-E_\alpha^{J\pi }\mathscr{N}^{NZJ\pi }_{cc'} \Bigg]
f^{J\alpha \pi}_{c'}=0,
\end{eqnarray}
where the Hamiltonian kernel  and norm kernel are defined by
\begin{equation}
\label{eq:kernel}
 \mathscr{O}^{NZJ\pi}_{cc'}
 =\bra{NZ J\pi; c}  \hat O \ket{NZ J\pi; c'},
\end{equation}
with the operator $\hat O$ representing $\hat H$ and $1$, respectively.   Similar to that for even-even nuclei, the kernel for odd-mass nuclei is also evaluated with the mixed-density prescription~\cite{Bonche:1990ghh}. The details on the calculaton of the kernels (\ref{eq:kernel}) can be found in the appendix~\ref{app:kernels}.

The HWG equation (\ref{eq:HWG}) for a given set of quantum numbers $(NZJ\pi)$ is solved  in the standard way  as discussed in Refs.\cite{Ring:1980,Yao:2010}. It is accomplished by diagonalizing the norm kernel $\mathscr{N}^{NZJ\pi }_{cc'}$ first. A new set of basis is constructed using the  eigenfunctions of the norm kernel with eigenvalue larger than a pre-chosen cutoff value to remove possible redundancy in the original basis. The Hamiltonian is diagonalized in this new basis. In this way, one is able to obtain the energies $E_\alpha^{J\pi}$ and
the mixing weights $f^{J\alpha\pi}_{c}$ of nuclear states $\vert \Psi^{J\pi}_\alpha\rangle$. Since the basis functions $\ket{NZ J\pi; c}$ are nonorthogonal to each other, one usually introduces the collective wave function $g^{J\pi}_\alpha(K, \mathbf{q})$ as below
\begin{equation}
\label{eq:coll_wf}
g^{J\pi}_\alpha(K, \mathbf{q})=\sum_{c'} (\mathscr{N}^{1/2})^{NZJ\pi}_{c,c'} f^{J\alpha\pi}_{c'},
 \end{equation}
 which fulfills the normalization condition . The distribution of $g^{J\pi}_\alpha(K, \mathbf{q})$ over $K$ and $\mathbf{q}$ reflects the contribution of each basis function to the nuclear state $\vert \Psi^{J\pi}_\alpha\rangle$. 

 \subsection{CDFT for the mean-field configurations of odd-mass nuclei}
  
 The wave functions of the mean-field configurations $\ket{\Phi^{\rm (OA)}_\kappa(\mathbf{q})}$ in (\ref{eq:basis}) for an odd-mass nucleus are generated from the SR-CDFT calculation in which the following EDF is optimized with respect to single-particle wave functions,
 \beqn
 \label{eq:EDF}
    && E'[\rho_i, \nabla^2\rho_i, j^\mu_i, \nabla^2j^\mu_i,\kappa,\cdots] \nonumber\\
    &=& E_{\rm CDFT}[\rho_i, \nabla^2\rho_i, j^\mu_i, \nabla^2j^\mu_i,\kappa,\cdots] \nonumber \\
    && +\sum_{\tau=n,p} \lambda_\tau \Bigg(\braket{\hat N_\tau} - N_\tau\Bigg) + \sum_{\lambda,\mu}  C_{\lambda\mu}\Bigg(\braket{\hat Q_{\lambda\mu}} - q_{\lambda\mu}\Bigg)^2.
\eeqn 
 Here, the second term is introduced to ensure the correct average particle number  $\langle\Phi_{(\kappa)}(\mathbf{q})| \hat N_\tau |\Phi_{(\kappa)}(\mathbf{q})\rangle = N_\tau$ with $N_\tau$ standing for the number of neutrons or protons. The quadratic constraint terms are added with the stiff parameters $C_{\lambda\mu}$ for the multipole moments. The $E_{\rm CDFT}[\rho_i, \nabla^2\rho_i, j^\mu_i, \nabla^2j^\mu_i,\kappa,\cdots]$
 represents the energy of nucleus which is a functional of various types of densities $\rho_i$, currents $j^\mu_i$ and their derivatives\cite{Yao:2009traxi},
\beqn
\label{eq:ECDFT}
  && E_{\rm CDFT}[\rho_i, \nabla^2\rho_i, j^\mu_i, \nabla^2j^\mu_i,\kappa,\cdots] \nonumber\\
 &=& \int d^3\vec{r} \Bigg[ 
   \sum_k v_k^2\psi_k^\dag\left(\vec{r})(\vec{\alpha}\cdot\vec{p}+\beta m_N-m_N)\psi_k(\vec{r}\right) \nonumber\\
   &&+\cfrac{\alpha_S}{2}~\rho_S^2 +\cfrac{\beta_s}{3}~\rho_S^3
     +\cfrac{\gamma_S}{4}~\rho_S^4 +\cfrac{\delta_S}{2}~\rho_S\nabla^2 \rho_S \nonumber\\
     &&+\cfrac{\alpha_V}{2}~(j_V)_\mu j^\mu_V + \cfrac{\gamma_V}{4}~[(j_V)_\mu j^\mu_V]^2 +\cfrac{\delta_V}{2}~(j_V)_\mu \nabla^2j^\mu_V\nonumber\\
     && + \cfrac{\alpha_{TV}}{2}~\vv{j}^\mu_{TV} (\vv{j}_{TV})_\mu
     + \cfrac{\delta_{TV}}{2} \vv{j}^\mu_{TV} \nabla^2(\vv{j}_{TV})_{\mu}\nonumber\\
     &&+ \cfrac{1}{4}~F_{\mu\nu}F^{\mu\nu}-F_{0\mu}\partial_0 A_{\mu} + e A_\mu j_\rho^{~\mu} 
   \Bigg]+  E_{\rm pair}[\kappa,\kappa^*].
\eeqn
Here, the densities and currents are defined by    
    \begin{subequations} 
    \label{currents}
\begin{align}
         \rho_S(\vec{r}) &= \sum_kv_k^2\bar \psi_k(\vec{r})\psi_k(\vec{r}),\\
     j_V^{~\mu}(\vec{r}) &= \sum_kv_k^2 \bar \psi_k(\vec{r}) \gamma^\mu\psi_k(\vec{r}),\\
     \vv{j}_{TV}^\mu(\vec{r}) &= \sum_kv_k^2 \bar \psi_k(\vec{r}) \vv{\tau} \gamma^\mu\psi_k(\vec{r}).
\end{align}
\end{subequations} 
The subscript $S$ stands for scalar-isoscalar, $V$ for vector-isoscalar, and $TV$ for vector-isovector types of vertices, and  $\vv{\tau}$ for the isospin operator. The values of the coupling constants are taken from the PC-F1 parametrization~\cite{Burvenich:2001rh}. The use of the PC-PK1~\cite{Zhao:2010} does not change the conclusion of the work. The $v^2_k$ is the occupation probability of the $k$-th single-particle state, determined by the BCS theory. In this work, the pairing energy $E_{\rm pair}[\kappa,\kappa^*]$ is derived on the basis of a contact pairing interaction, 
\begin{equation}
    E_{\rm pair}[\kappa,\kappa^*]=-\sum_{\tau=n,p} \cfrac{V_\tau}{4} \int d^3\vec{r}~\kappa_\tau^*(\vec{r})\kappa_\tau(\vec{r}),
\end{equation}
where $V_\tau$ is a parameter for the pairing strength. The pairing tensor $\kappa_\tau(\vec{r})$  of neutrons or protons  
 in coordinate space is given by
\begin{eqnarray}
    \kappa_\tau(\vec{r}) = -2\sum^\tau_{k>0} f_ku_kv_k|\psi_k(\vec{r})|^2,
\end{eqnarray}
where an energy-dependent cutoff factor $f_k$ is introduced to regulate the contribution from high-energy single-particle states~\cite{Bender:2000fqv},
\begin{equation}
\label{eq:cutoff}
    f_k= \cfrac{1}{1+\exp{[(\varepsilon_k-\lambda_\tau-\Delta E_\tau)/\mu_\tau]}}.
\end{equation}
Here, $\varepsilon_k$ is the energy of the $k$-th single-particle state. The symbol $\lambda_\tau$ in (\ref{eq:EDF}) and (\ref{eq:cutoff}) is called Fermi energy. Following Ref.\cite{Bender:2000fqv}, we choose the energy-window parameters $\Delta E_\tau$ and $\mu_{\tau} = \Delta E_\tau/10$ so that the condition $2\sum_{k>0} f_k=N_\tau +1.65 N_\tau^{2/3}$ is fulfilled.  

The variation of the EDF in (\ref{eq:EDF}) with respect to single-particle wave function $\psi_k(\vec{r})$ leads to a Dirac equation. {Since the Dirac equation contains scalar and vector potentials depending on nucleon densities and currents, it is solved iteratively.} See for instance Refs.~\cite{Yao:2009traxi,Zhao:2010} for more details. After the solution, the total energy of an atomic nucleus is given by,
\beq 
E_{\rm Tot} = 
E_{\rm CDFT}[\rho_i, \nabla^2\rho_i, j^\mu_i, \nabla^2j^\mu_i,\kappa,\cdots]
 + E_{\rm cm},
\eeq 
where the energy correction from the center of motion is determined by~\cite{Bender:2000com,Yao:2009traxi,Yao:2010}
\begin{equation}\label{cm collection}
    E_{\rm cm} = -\cfrac{1}{2m_NA}\langle \Phi_{(\kappa)}|\hat{\vec{P}}_{cm}^2|\Phi_{(\kappa)}\rangle,
\end{equation}
with $m_N$ being the mass of neutron or proton, and $A$ mass number, and $\vec{\hat P}_{cm}$ the operator of total momentum in the center of mass frame.

The mean-field wave functions $\ket{\Phi^{\rm (OA)}_\kappa}$ for the odd-mass nuclei have been given in (\ref{eq:odd-mass-wfs}), where the $\ket{\Phi_{(\kappa)}}$ can be constructed with three schemes in the BCS theory.
\begin{itemize}
\item Scheme I: the quasiparticle vacuum $\ket{\Phi_{(\kappa)}}$ in (\ref{eq:odd-mass-wfs}) is chosen as a BCS wave function for the even-mass nuclear core,
 \begin{eqnarray}
 \label{eq:EAC}
 \ket{\Phi_{(\kappa)}}_{\rm EAC} = \prod^{\rm EA}_{j>0}(u_j+v_ja^\dagger_{j}a^\dagger_{\bar j})\ket{0},
\end{eqnarray}  
where ${}_{\rm EAC}\bra{\Phi_{(\kappa)}} \hat N \ket{\Phi_{(\kappa)}}_{\rm EAC}$ is even for both neutron and proton.  This scheme is referred to even-mass-core (EAC) hereafter.
\item Scheme II:  the quasiparticle vacuum $\ket{\Phi_{(\kappa)}}$ is chosen as a FQV state~\cite{Duguet:2001},
 \begin{eqnarray}
 \label{eq:FQV}
  \ket{\Phi_{(\kappa)}}_{\rm FQV} = 
 \prod^{\rm OA}_{j>0}(u_j+v_ja^\dagger_{j}a^\dagger_{\bar j})\ket{0},
\end{eqnarray}
 where ${}_{\rm FQV}\bra{\Phi_{(\kappa)}} \hat N \ket{\Phi_{(\kappa)}}_{\rm FQV}$ is odd either for neutron or proton, depending on the nucleus of interest. 
 
\item Scheme III:  the quasiparticle vacuum $\ket{\Phi_{(\kappa)}}$ is chosen according to the EFA~\cite{Martin:2008EFA}, 
 \begin{eqnarray}
 \label{eq:EFA}
  \ket{\Phi_{(\kappa)}}_{\rm EFA} =  
(1+a^\dagger_{k}a^\dagger_{\bar k}) \prod^{\rm EA}_{j\neq k}(u_j+v_ja^\dagger_{j}a^\dagger_{\bar j})\ket{0}.
\end{eqnarray}
In this scheme, the wave function $\ket{\Phi_{(\kappa)}}_{\rm EFA}$ depends on the index $k$ of the blocked single-particle orbital.  
\end{itemize}
It is noted that the number parity of the wave function $\ket{\Phi_{(\kappa)}}$ is even in the above three schemes. The introduction of the one quasiparticle creation operator $\alpha^\dagger_\kappa$ in (\ref{eq:odd-mass-wfs}) changes the number parity from even to odd. Therefore, the mean-field wave function $ \ket{\Phi^{\rm (OA)}_\kappa}$ has the correct number parity of odd-mass nuclei. Besides, time-reversal symmetry is maintained in the three schemes. Therefore, in the SR-CDFT calculation, the space-like components of the currents $j^\mu_i$ vanish.

The single-particle wave function $\psi_k(\vec{r})$ for nucleons is a Dirac spinor which is composed of large  and small components. Both are expanded using the eigenfunctions $|n\rangle=|n_z,n_r,m_l,m_s\rangle$ of a harmonic oscillator in a cylindrical coordinate system,
\begin{eqnarray}
\label{eq:HO_basis_expansion}
    |\psi_k\rangle = \left(
\begin{array}{c}
\sum_n f_{nk}|n\rangle\\
i\sum_{\bar{n}}g_{\bar n k}|\bar n\rangle \\
\end{array}
\right) \chi_{t_k}(t),
\end{eqnarray}
where $\chi_{t_k}(t)$ represents the isospin part.   The wave function for the harmonic oscillator basis states is as follows
\begin{eqnarray}\label{hobs}
   \bra{\vec{r}} n\rangle =\phi_{n_z}(z)\phi_{n_r}^{m_l}(r_\bot)\cfrac{1}{\sqrt{2\pi}}e^{im_l\varphi}\chi_{m_s}(s)
\end{eqnarray}
where $\chi_{m_s}(s)$  is the spin component and 
\begin{eqnarray}
    \phi_{n_z}(z)&=&\cfrac{N_{n_z}}{\sqrt{b_z}}H_{n_z}(\xi)e^{-\xi^2/2},\quad \xi=\cfrac{z}{b_z}\\
    \phi_{n_r}^{m_l}(r_\bot)&=&\cfrac{N_{n_r}^{m_l}}{b_\bot}\sqrt{2}\eta^{m_l/2}L_{n_r}^{m_l}(\eta)e^{-\eta/2},\quad \eta=\cfrac{r_\bot^2}{b_\bot^2}.
\end{eqnarray}
In the equation, $H_{n_z}(\xi)$ and $L_{n_r}^{m_l}(\eta)$ are the Hermite polynomials and the Laguerre polynomials, respectively. The expression for the normalization coefficient is as follows
\begin{eqnarray}
    N_{n_z}=\cfrac{1}{\sqrt{\sqrt{\pi}2^{n_z}n_z!}},\quad 
    L_{n_r}^{m_l}=\sqrt{\cfrac{n_r!}{(n_r+m_l)}}.
\end{eqnarray}
After the solution of Dirac equation for $\ket{\psi_k}$, we carry out the basis transformation from cylindrical coordinate system to  the  spherical harmonic oscillator basis $\ket{m}=\ket{nljm_j}$,
\begin{eqnarray}
    |\psi_k\rangle=\left(
\begin{array}{c}
\sum_m F_{mk}|m\rangle\\
i\sum_{\bar{m}}G_{\bar m k}|\bar m\rangle \\
\end{array}
\right).
\end{eqnarray}
The expansion coefficients $F_{km}$ and $G_{k\bar{m}}$ are determined by
\begin{eqnarray}
    F_{mk}=\sum_n f_{nk}\langle m|n\rangle,\quad 
    G_{\bar m k}=\sum_{\bar n}g_{\bar nk}\langle \bar m|\bar n\rangle.
\end{eqnarray}
where the  transformation coefficient $\langle m|n\rangle$ is calculated in Ref.~\cite{Chasman:1967}.
 The representation of the rotation operator $\hat R(\Omega)$ in this basis is simply given by the Wigner $D$-function~\cite{Yao:2009traxi}.


\subsection{Electromagnetic transition strengths}

The strength of the electromagnetic multipole transition from the initial state $\ket{\Psi^{J_i\pi_i}_{\alpha_i}}$ to the final state $\ket{\Psi^{J_f\pi_f}_{\alpha_f}}$ is determined by 
\begin{eqnarray}
&& B(T\lambda,J_i\alpha_i\pi_i\rightarrow J_f\alpha_f\pi_f) = \cfrac{1}{2J_i+1} \nonumber\\
&&\times\left| \sum_{c_f, c_i} f^{J_i\alpha_i \pi_i}_{c_i}f^{J_f\alpha_f \pi_f}_{c_f} 
    \langle NZJ_f\pi_f,c_f||\hat T_\lambda ||NZJ_i\pi_i,c_i\rangle
    \right|^2,\nonumber\\
\end{eqnarray} 
where the configuration-dependent reduced matrix element is simplified as follows, 
\begin{eqnarray}
\label{eq:reduced_matrix_element}
    &&\langle NZ J_f\pi_f; c_f ||\hat  T_\lambda|| NZ J_i\pi_i; c_i\rangle\nonumber\\
    =&& \delta_{\pi_f\pi_i,(-1)^\lambda}(-1)^{J_f-K_f}
    \cfrac{\hat{J}_i^2\hat{J}_f^2}{8\pi^2} \sum_{\nu M} \left(
 \begin{array}{ccc}
J_f&\lambda&J_i\\
-K_f &\nu&M\\
\end{array}
    \right) \nonumber\\
   &&\times  \int d\Omega D_{MK_i}^{J_i\ast}(\Omega)
\bra{\Phi^{\rm (OA)}_{\kappa_f}(\mathbf{q}_f)}
    \hat T_{\lambda\nu}
  \hat R(\Omega) \hat P^Z\hat P^N\hat P^{\pi_i}
   \ket{\Phi^{\rm (OA)}_{\kappa_i}(\mathbf{q}_i)},\nonumber \\
\end{eqnarray}
where $\hat J=\sqrt{2J+1}$.

For the electric dipole ($E1$) and quadrupole ($E2$) transitions, the tensor operator is simply given by
\beq 
\hat T_{\lambda\nu} = er^\lambda Y_{\lambda\nu},
\eeq 
with $Y_{\lambda \nu}$  standing for the spherical harmonic function of rank $\lambda$.  For  the magnetic dipole ($M1$) transitions, the tensor operator is defined as 
 \bsub\begin{eqnarray}
    \hat T_{1\nu} &= &\sqrt{\cfrac{3}{4\pi}}~\hat{\mu}_{1\nu},\\
    \hat{\mu}_{1\nu}&=& \left(
    \cfrac{1}{\sqrt{2}}(\hat \mu_x - i\hat \mu_y), \hat{\mu}_z,
    -\cfrac{1}{\sqrt{2}}(\hat \mu_x + i\hat \mu_y)
    \right).
\end{eqnarray}
\esub
 Here, the symbol $\hat\mu_{k=x,y,z}$ represents the $k$-component of the magnetic dipole moment operator determined by the effective  current operator, which is composed of Dirac current and anomalous current~\cite{Serot:1981,Yao:2006},
 \begin{eqnarray}
        \hat{\boldsymbol{\mu}}
        &=& \int d^3 r\left[\frac{m_Nc^2}{\hbar c}e\psi^{\dagger}(\mathbf{r}) \mathbf{r} \times \boldsymbol{\alpha} \psi(\mathbf{r})
        +\kappa \psi^{\dagger}(\mathbf{r}) \beta \boldsymbol{\Sigma} \psi(\mathbf{r})\right],\nonumber\\
 \end{eqnarray}
with $e$ being the electric charges of nucleons,  $\kappa$ the free anomalous gyromagnetic ratio (called $g$ factor) of the nucleon, i.e., $\kappa_p=1.793$ for protons, $\kappa_n=-1.913$ for neutrons, and $m_N$ the mass of nucleon. The Dirac gamma matrices read
\beqn 
\boldsymbol{\alpha}=\begin{pmatrix}
0 & \boldsymbol{\sigma} \\
 \boldsymbol{\sigma} & 0
\end{pmatrix},\quad \beta \boldsymbol{\Sigma} =\begin{pmatrix}
\boldsymbol{\sigma} & 0\\
0 & -\boldsymbol{\sigma}
\end{pmatrix}.
\eeqn 
Here, $\boldsymbol{\sigma}$ is the Pauli spinor. The units of the magnetic dipole moment $\hat{\boldsymbol{\mu}}$ is nuclear magneton ($\mu_N=e\hbar/m_N$).

The magnetic dipole moment $\mu(J_\alpha^\pi)$ of the state  $\ket{\Psi^{J\pi}_{\alpha}}$   can be calculated by 
\begin{eqnarray}
\label{eq:magnetic_mu}
    \mu(J_\alpha^\pi) 
    &=& \bra{\Psi^{J\pi}_\alpha}\hat \mu_{10}\ket{\Psi^{J\pi}_\alpha}\nonumber\\
   &=& \left(
     \begin{array}{ccc}
J&1&J\\
-J&0&J\\
\end{array}
   \right)
  \sum_{c_f, c_i} f^{J_i\alpha_i \pi_i}_{c_i} f^{J_f\alpha_f \pi_f}_{c_f} \nonumber\\
   &&\times 
   \langle NZ J_f\pi_f; c_f ||\hat  \mu_1|| NZ J_i\pi_i; c_i\rangle.
\end{eqnarray}
 The spectroscopic quadrupole moment $Q^{\rm s}(J^\pi_\alpha)$ of the state $\ket{\Psi^{J\pi}_{\alpha}}$ is given by
\begin{eqnarray}
\label{eq:spectroscopic_Q}
    Q^{\rm s}(J^\pi_\alpha) 
    &=&  \sqrt{\cfrac{16\pi}{5}}\bra{\Psi^{J\pi}_\alpha}\hat Q_{20}\ket{\Psi^{J\pi}_\alpha}\nonumber\\
    &=&  \sqrt{\cfrac{16\pi}{5}}
    \left(
     \begin{array}{ccc}
J&2&J\\
-J&0&J\\
\end{array}
   \right)
  \sum_{c_f, c_i} f^{J_i\alpha_i \pi_i}_{c_i} f^{J_f\alpha_f \pi_f}_{c_f} \nonumber\\
   &&\times 
   \langle NZ J_f\pi_f; c_f ||er^2Y_{2}|| NZ J_i\pi_i; c_i\rangle.
\end{eqnarray}
The reduced $E2$ transition matrix elements in (\ref{eq:magnetic_mu}) and (\ref{eq:spectroscopic_Q}) can be computed with Eq.(\ref{eq:reduced_matrix_element}). Since the transition operator is defined in the full single-particle space in the calculation, the bare electric charges for protons and  neutrons, i.e., $e_p=1, e_n=0$, are employed.

\section{Results and discussion}
\label{sec:results}

\subsection{Numerical details}

\begin{figure*}[bt]
\centering
\includegraphics[width=0.8\paperwidth]{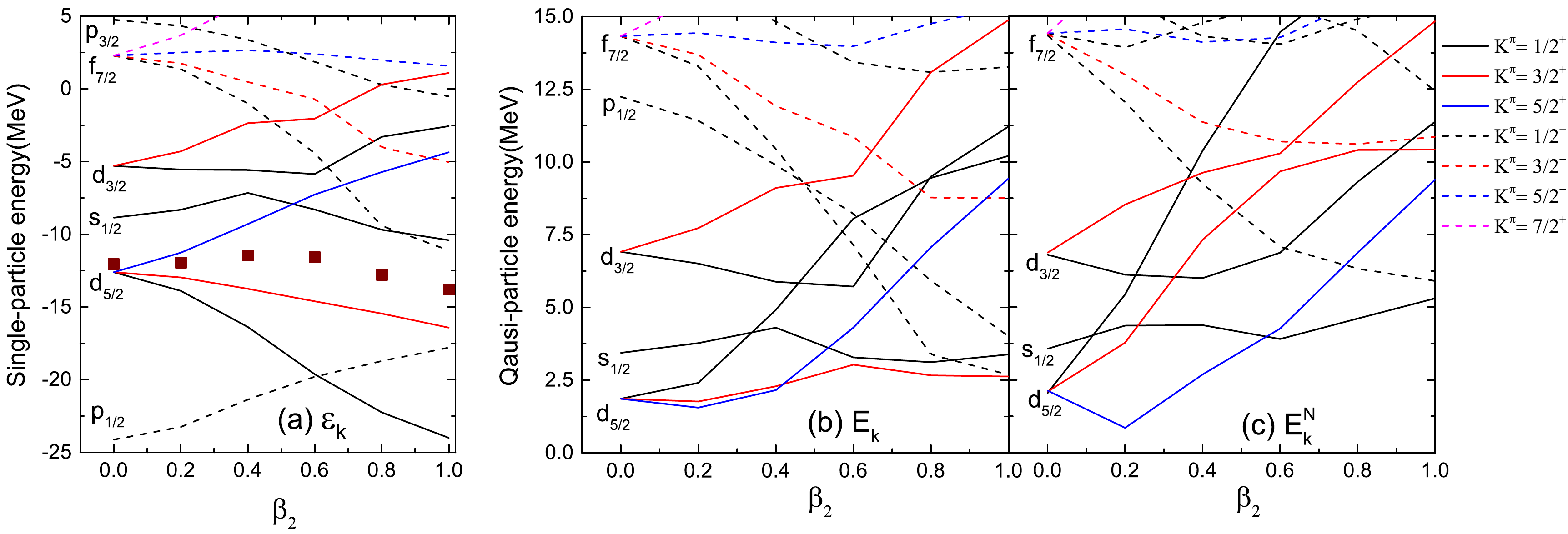}
\caption{\label{fig:single-particle-energy} (Color online) (a) The single-particle energies  $\varepsilon_k$ of neutrons in \nuclide[24]{Mg} as a function of quadrupole deformation parameter $\beta_2$. (b) The energies $E_k$ of one-quasiparticle neutron states as a function of $\beta_2$.  (c) The energies  $E^N_k$ of one-quasiparticle neutron states projected onto the correct nucleon numbers as a function of $\beta_2$. See main text for details.   }
\end{figure*}

The Dirac spinor $\ket{\psi_k}$ in Eq. (\ref{eq:HO_basis_expansion}) for the nucleons is expanded in a set of harmonic oscillator basis functions comprising 8 major shells. This choice, as demonstrated in Ref.~\cite{Yao:2009traxi}, has proven sufficient to yield a reasonably convergent mean-field binding energy curve for $^{24}$Mg. The oscillator frequency is defined as $\hbar\omega=41A^{-1/3}$. Research conducted in Ref.~\cite{Borrajo:2018sqj} indicates that the level of $K$ mixing caused by triaxiality in the low-lying states of \nuclide[25]{Mg} is minimal. Therefore, for the sake of simplicity, mean-field configurations are constrained to possess axial symmetry, conserving the $K$ value in each state. Consequently, only one of the three Euler angles in the AMP requires numerical integration within the interval $\theta \in [0,\pi]$, where the number of mesh points in this interval is set as $N_\theta=12$.

Table \ref{table:convergence_check_PNP} demonstrates the convergence of results with the number of mesh points for the gauge angles $\varphi_\tau$ within the interval $[0, \pi]$. It is observed that choosing $N_\varphi=5$ allows for the reproduction of particle numbers and yields a convergent solution for the total energy of \nuclide[25]{Mg}. Further details regarding the problems of singularity and finite steps are presented in appendix~\ref{app:singularity_PNP}. 

\begin{table}[bt]
    \centering
    \tabcolsep=3pt 
    \caption{\label{table:convergence_check_PNP}  The convergence of the neutron numbers ($N$), proton numbers ($Z$) and the total energy $E_{\rm Tot}$  (MeV) of \nuclide[25]{Mg} in the PNP calculation with respect to the number $N_{\varphi}$ of mesh points in the gauge angles for both neutrons and protons, where the mean-field  configuration is constructed with the EAC scheme (\ref{eq:EAC}), and the quasiparticle vacuum state is constrained to have quadrupole deformation $\beta_2=0.3$.}
 \begin{tabular}{cccccc}
   \hline
  $N_{\varphi}$ & 1 & 3& 5 & 7&9\\
  \hline
  $N$ & 12.901586 & 12.999993 & 12.999999 & 13.000000&13.000000\\ 
  $Z$ & 12.000000 & 11.999991 & 12.000000 & 12.000000 &12.000000\\ 
  $E_{\rm Tot}$ &  $-198.453$ & $-200.042$ & $-200.041$ & $-200.041$ & $-200.041$ \\
  \hline
\end{tabular} 
\end{table}

\subsection{Application to \nuclide[25]{Mg} with quadrupole correlations}


\subsubsection{Selection of lowest-energy quasiparticle configurations}

 The mean-field configurations for  odd-mass  nuclei are chosen as the lowest-energy quasiparticle configurations with different deformation parameters $\mathbf{q}$. In the canonical basis, the energy $E_k$ of the quasiparticle state $\ket{\Phi^{\rm (OA)}_\kappa(\mathbf{q})}$ defined in Eq.~(\ref{eq:odd-mass-wfs}) can be computed as follows,
\beq 
\label{eq:1qp_E}
E_\kappa=\sqrt{(\varepsilon_k-\lambda_n)^2+f^2_k\Delta_k^2}.
\eeq 
Alternatively, one can define the energy $E^{N}_\kappa$ of the state with the restoration of particle numbers, which for \nuclide[25]{Mg} is given by,
\beqn
\label{eq:quasiparticle_energy_PNP}
E^{N}_\kappa (^{25}{\rm Mg})
&=&\frac{\bra{\Phi^{\rm (OA)}_\kappa(\mathbf{q})}   \hat H \hat P^{Z=12} \hat P^{N=13} \ket{\Phi^{\rm (OA)}_\kappa(\mathbf{q})}}{\bra{\Phi^{\rm (OA)}_\kappa(\mathbf{q})}    \hat P^{Z=12} \hat P^{N=13} \ket{\Phi^{\rm (OA)}_\kappa(\mathbf{q})}}\nonumber\\
&&-\frac{\bra{\Phi(\mathbf{q})}  \hat H \hat P^{Z=12}\hat P^{N=12}\ket{\Phi(\mathbf{q})}}{\bra{\Phi(\mathbf{q})}   \hat P^{Z=12}\hat P^{N=12}\ket{\Phi(\mathbf{q})}}-\lambda_n,
\eeqn 
where $\lambda_n$ is the Fermi energy of neutrons introduced in (\ref{eq:EDF}).

 Figure~\ref{fig:single-particle-energy}(a) displays the change of the single-particle energy levels of neutrons in $^{24}$Mg from the SR-CDFT calculations with the quadrupole deformation parameter $\beta_2$,
 where the dimensionless deformation parameter $\beta_{\lambda }$ for the multipole moment is defined as 
 \begin{equation}
\label{eq:deformation}
\beta_{\lambda} = \dfrac{4\pi}{3AR^\lambda}\bra{\Phi(\mathbf{q})} r^\lambda Y_{\lambda 0}\ket{\Phi(\mathbf{q})}.
\end{equation}
 The quasiparticle energies $E_\kappa$ and $E^N_\kappa$ defined  in (\ref{eq:1qp_E}) and (\ref{eq:quasiparticle_energy_PNP})  are shown in Fig.\ref{fig:single-particle-energy}(b) and (c), respectively.
 It is seen that the two lowest-energy configurations in the weak deformation region have $K^\pi=3/2^+$ and $K^\pi=5/2^+$, both of which are originated from the splitting of the spherical $d_{5/2}$ orbital located near the neutron Fermi energy $\lambda_n$. In the largely deformed region, the configuration with $K^{\pi}=1/2^+$ splitting from the spherical $d_{3/2}$ orbital falls down in energy. In particular, the orderings of the configurations by the $E_\kappa$ and $E^N_\kappa$ could be very different, indicating the importance of using the quasiparticle energy $E^N_\kappa$ with PNP  as a guideline in the selection of the lowest-energy configuration. However, if the quantum numbers $K^\pi$ of the configuration are fixed, one can still use the $E_\kappa$ as a convenient guideline to generate the lowest-energy one quasiparticle state $\ket{\Phi^{\rm (OA)}_\kappa(\mathbf{q})}$.

\begin{figure}[tb]
\centering
\includegraphics[width=0.8\columnwidth]{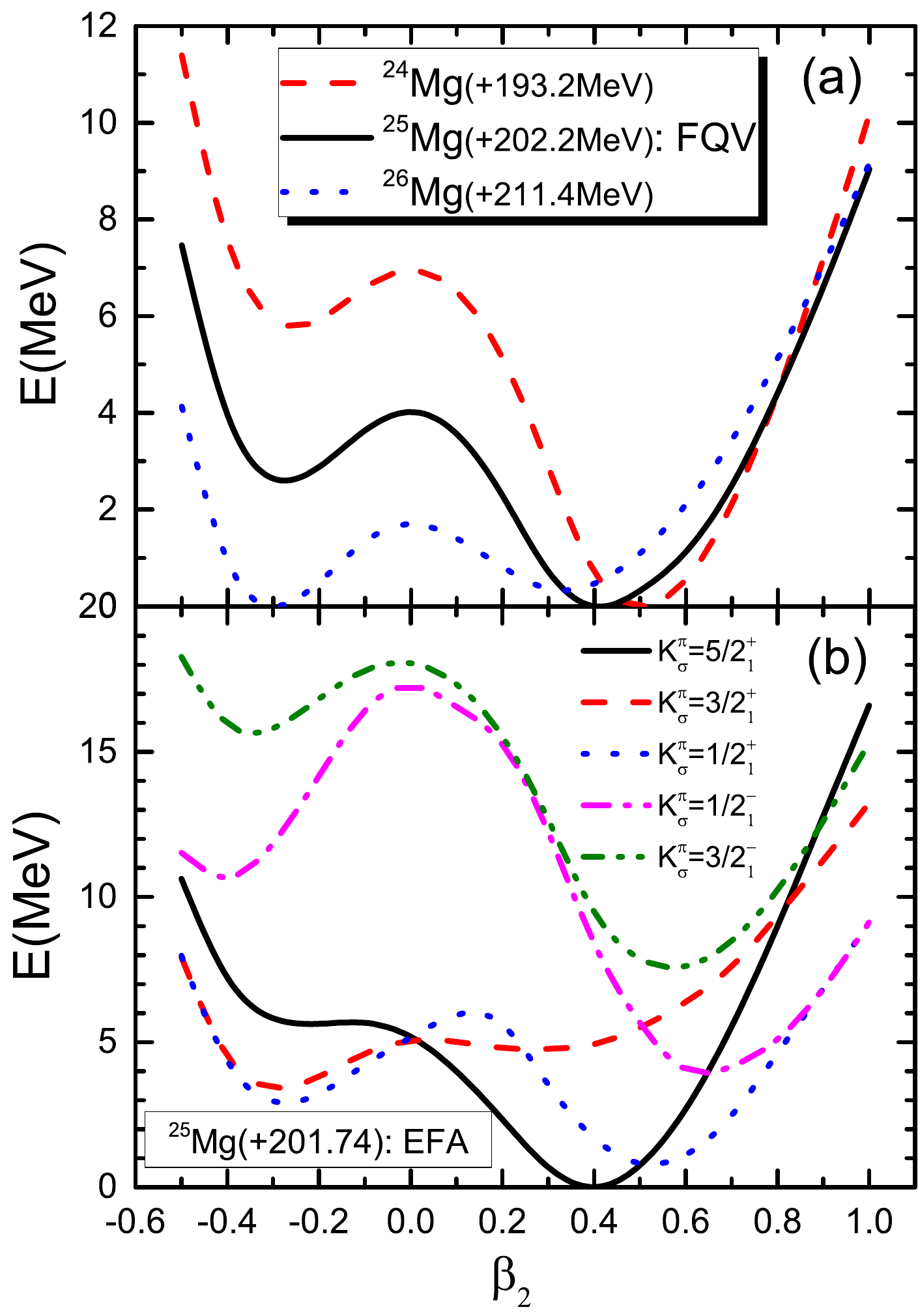}
\caption{\label{fig:MF4PES} (Color online) (a)  The energies of mean-field states of $^{24,25,26}$Mg as a function of the quadrupole deformation parameter $\beta_2$. Energies are normalized to the corresponding energy minima whose values are 193.2 MeV, 202.2 MeV, and 211.4 MeV, respectively.  The mean-field states $\ket{\Phi(\beta_2)}$ for $^{25}$Mg are quasiparticle vacua from the calculation based on the FQV scheme.  (b) The energies of mean-field states   (normalized to 201.74 MeV) from the calculation based on the EFA scheme for \nuclide[25]{Mg}, where the block orbitals are selected as the lowest quasiparticle states ($\sigma=1$) with $K^\pi_\sigma=5/2^+_1, 3/2^+_1, 1/2^+_1, 1/2^-_1$, and $3/2^-_1$, respectively.}
\end{figure}

\begin{figure}[tb]
\centering
\includegraphics[width=\columnwidth]{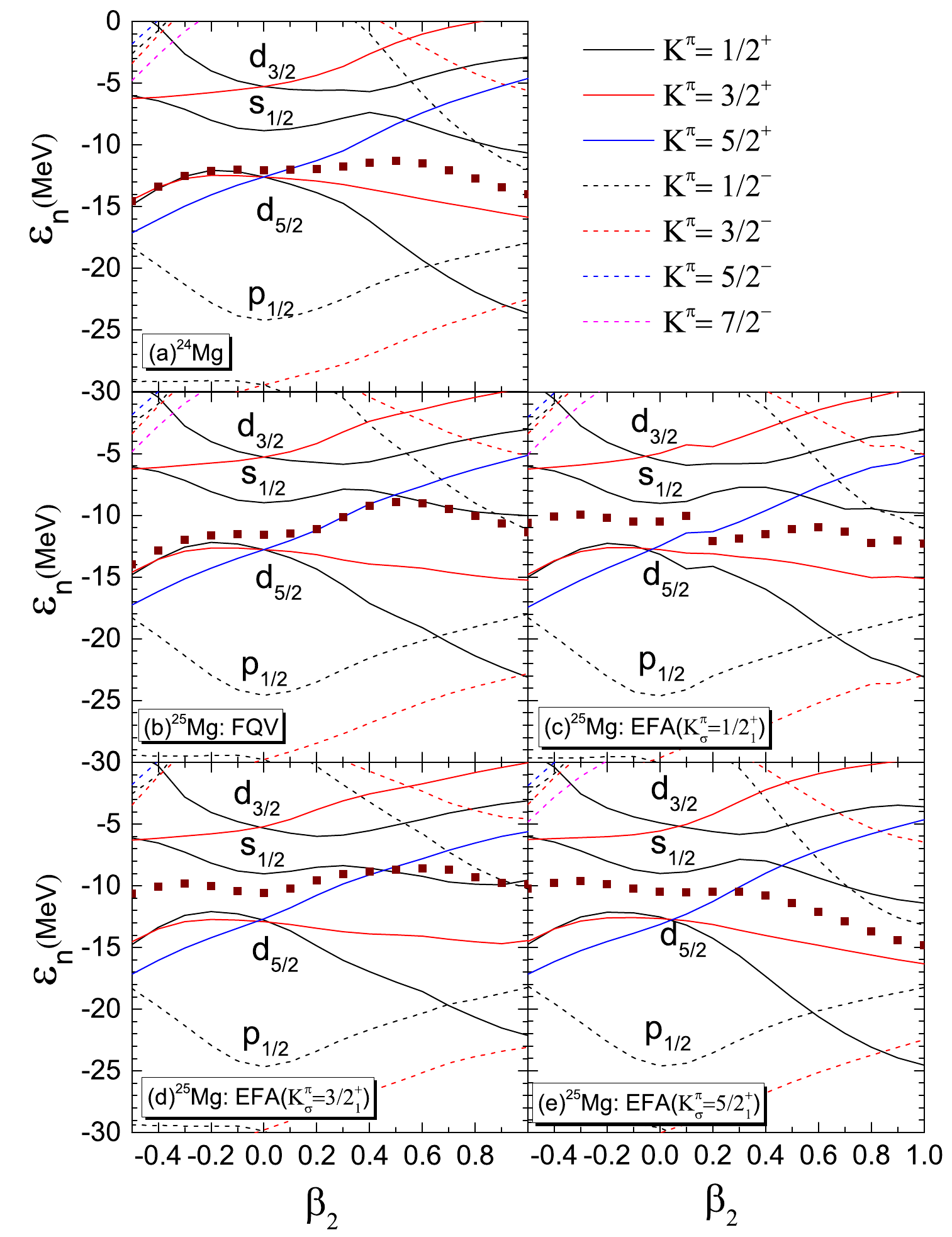}
\caption{\label{fig:spe4Mg24} (Color online) Comparison of Nilsson diagrams for the single-particle energies of neutrons in (a) \nuclide[24]{Mg} (EAC), (b) \nuclide[25]{Mg} (FQV), and (c-e) \nuclide[25]{Mg} (EFA), where the block orbitals in the EFA scheme are chosen as the lowest quasiparticle states of \nuclide[25]{Mg}(FQV) with $K^\pi_\sigma=1/2^+_1, 3/2^+_1$, and $5/2^+_1$, respectively. The Fermi energies of neutrons in each case are indicated with square symbols. }
\end{figure}

\begin{figure}[tb]
\centering
\includegraphics[width=7cm]{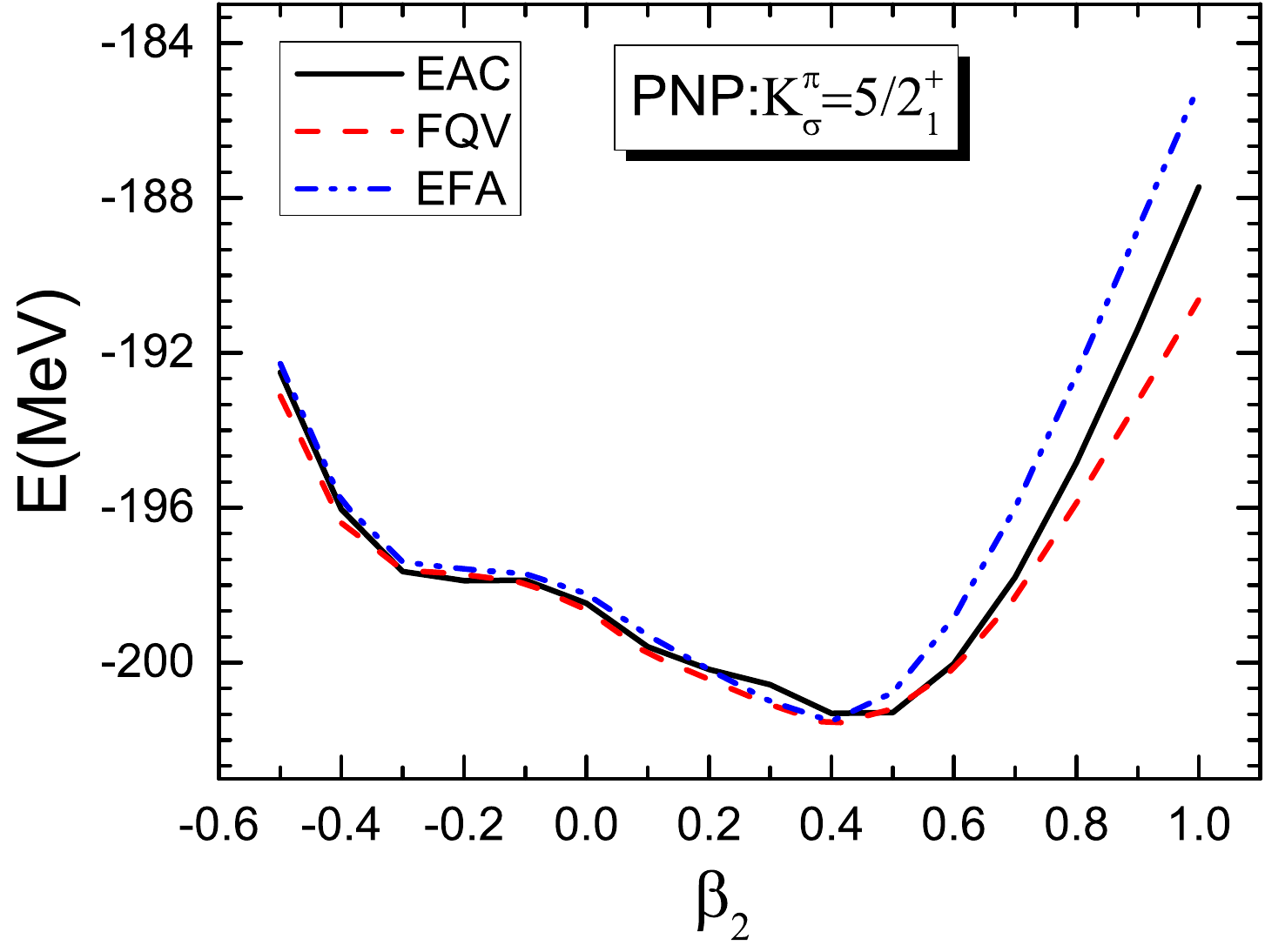}
\caption{\label{fig:PES_PNP} (Color online) The energies of particle-number projected states for $^{25}$Mg as a function of quadrupole deformation $\beta_2$ from the three schemes, where the intrinsic quasiparticle configurations have the same quantum numbers $K^\pi_\sigma=5/2^+_1$.}
\end{figure}

 \begin{figure}[tb]
 \centering
\includegraphics[width=8cm]{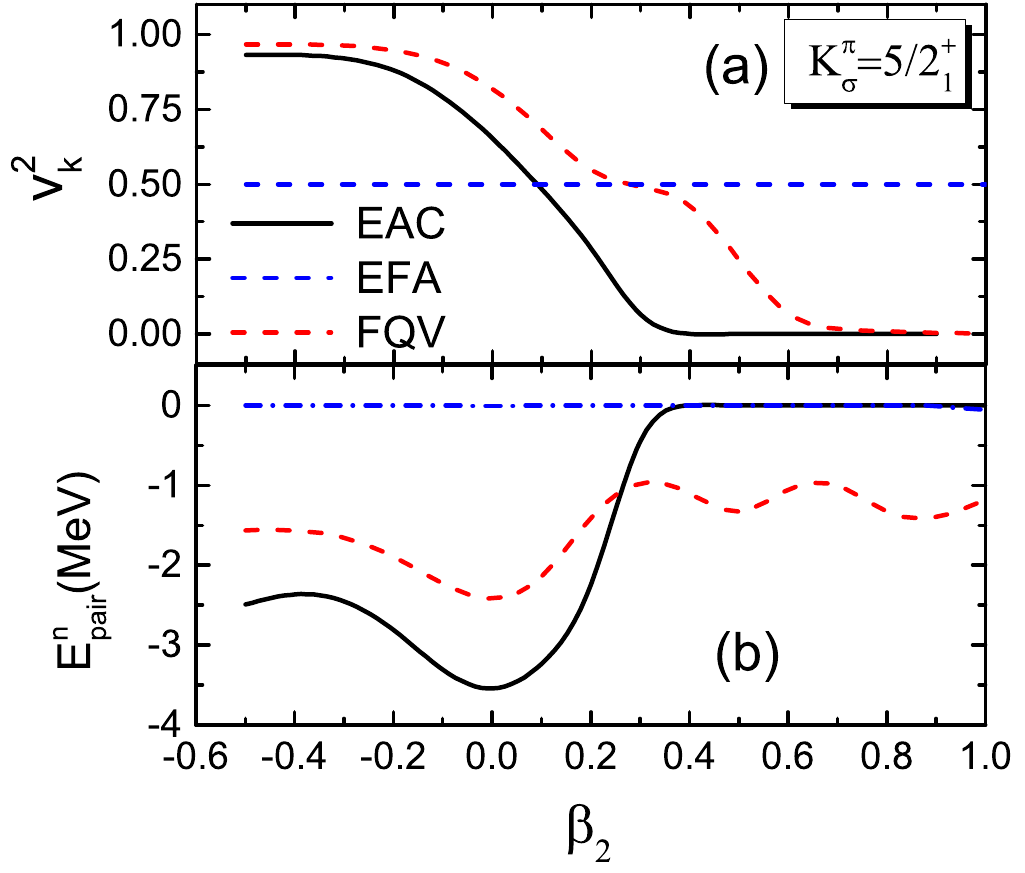}
\caption{\label{fig:pairing_energy} (Color online) 
(a) The occupation probability of the lowest-energy quasiparticle state with $K^\pi_\sigma=5/2^+_1$  in \nuclide[25]{Mg} from the calculations with the three schemes. (b) Pairing energy of neutrons in the mean-field states for \nuclide[24]{Mg},  \nuclide[25]{Mg}(FQV), and \nuclide[25]{Mg}(EFA), respectively, as a function of quadrupole deformation $\beta_2$.
 }
 \end{figure}

Figure~\ref{fig:MF4PES}(a) displays the mean-field energies of quasiparticle vacuum states for $^{24,25,26}$Mg, where the FQV scheme (\ref{eq:FQV}) is employed for $^{25}$Mg. The results from the EFA scheme (\ref{eq:EFA}) with the valence neutron blocked in the lowest-energy orbital with different quantum numbers $K^\pi$ are plotted in  Fig.~\ref{fig:MF4PES}(b).  It is seen that the energy difference between the oblate and prolate energy minima is decreasing from $^{24}$Mg to $^{26}$Mg. It has been shown in Ref.~\cite{Yao:2011_Mg} that $^{26}$Mg is actually a triaxial $\gamma$-soft nucleus. For $^{25}$Mg, the energy curve by the FQV scheme (\ref{eq:FQV}) is generally different from those by the EFA scheme. In the latter case, the energy curve depends on the blocked orbital labeled with $K^\pi$, representing the energy of different quasiparticle configurations $\ket{\Phi^{\rm (OA)}_\kappa(\mathbf{q})}$.

Figure~\ref{fig:spe4Mg24} displays the single-particle energies of neutrons in \nuclide[24]{Mg}, and the energies of one-quasiparticle states in $^{25}$Mg by the FQV and EFA schemes, respectively. In the EFA case, the blocked orbital is chosen as $K^\pi=1/2^+$, $3/2^+$, and $5/2^+$, respectively. The single-particle energies are almost the same in all the cases. However, the location of the neutron Fermi energy is much different, which may bring changes into the energy ordering of quasiparticle configurations, cf. Fig.~\ref{fig:single-particle-energy}.  The energies of the configurations with $K^\pi=5/2^+$ by the three schemes and with projection onto correct particle numbers for \nuclide[25]{Mg}  are compared in Fig.~\ref{fig:PES_PNP}. It is shown that the energy curves by the three schemes are similar, but with evident differences in the deformation region ($\beta_2>0.3$) where pairing correlation collapses, as shown in Fig.~\ref{fig:pairing_energy}. One can see that pairing correlation among neutrons in the EFA scheme collapses in the entire deformation space, but it does not happen in the FQV case.  From this point of view, the FQV provide a better choice to produce quasiparticle configurations for odd-mass nuclei.

\subsubsection{Angular-momentum projection}

\begin{figure}[h!]
\centering
\includegraphics[width=7cm]{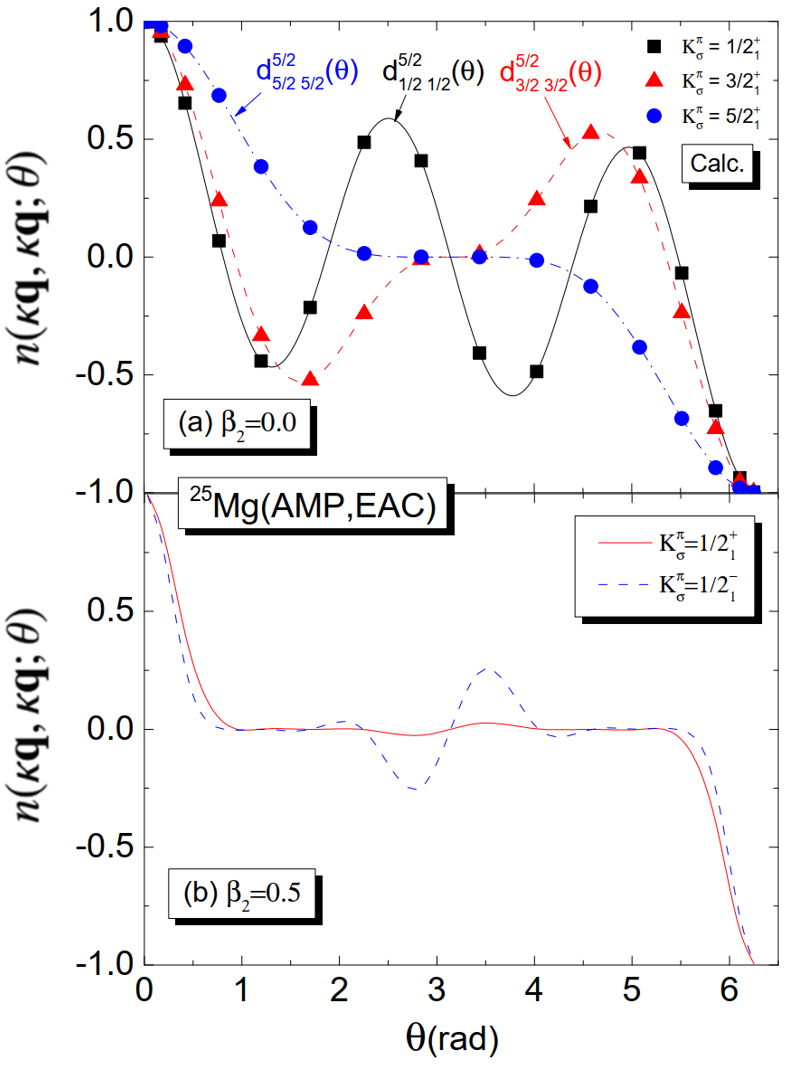} 
\caption{\label{fig:norm_odd_mass} (Color online)  The norm overlaps $n(\kappa\mathbf{q},\kappa\mathbf{q}; \theta)$ of (a) spherical \nuclide[25]{Mg} (EAC, $\beta_2=0$)  and (b) deformed \nuclide[25]{Mg} (EAC, $\beta_2=0.5$) mean-field states characterized with different $K^\pi$, as a function of the Euler angle $\theta$. In the panel (a), the three mean-field states of \nuclide[25]{Mg} (EAC) correspond to the configurations with the valence neutron occupying the multiplets $K^\pi=1/2^+, 3/2^+$, and $5/2^+$ of the spherical $d_{5/2}$ orbital. The results are compared with the Wigner $d$-function $d^{5/2}_{KK}(\theta)$, with $K=1/2,3/2$, and $5/2$, respectively.}
\end{figure}

\begin{figure}[h!]
\centering
\includegraphics[width=8.5cm]{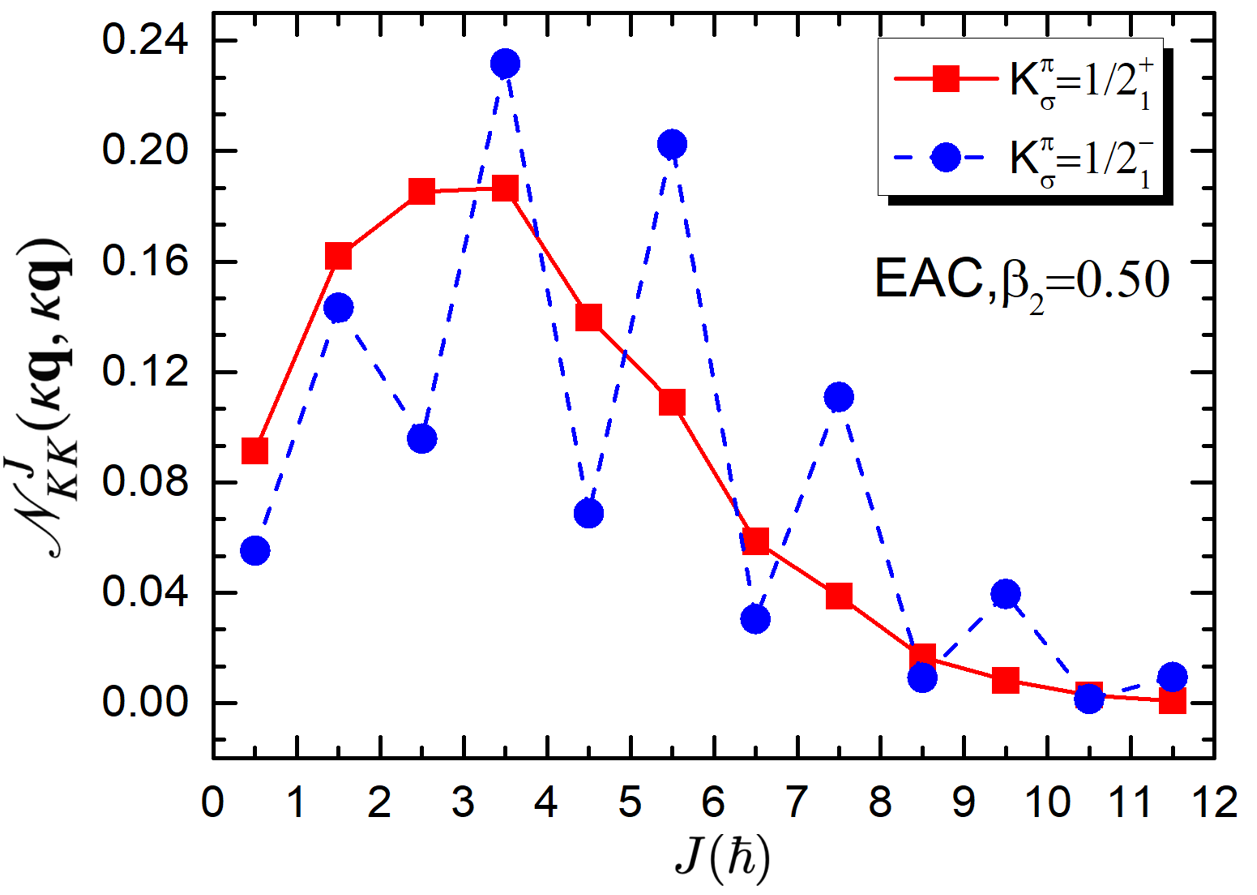} 
\caption{\label{fig:norm_overlap_odd_mass} (Color online) Probability of components with angular momentum $J$ in the axially-deformed  wave function $\ket{\Phi^{\rm (OA)}_\kappa(\mathbf{q})}$ for \nuclide[25]{Mg} (EAC, $\beta_2=0.5$) characterized with  $K^\pi=1/2^{\pm}$.  }
\end{figure}

The axially-deformed mean-field wave function $\ket{\Phi^{\rm (OA)}_\kappa(\mathbf{q})}$ labeled with $K^\pi$ can be decomposed into the eigenstates $\ket{JMK\pi}$ of the angular momentum operator $\hat J^2$, and its projection $\hat J_z$~\cite{Ring:1980}
\beqn 
\ket{\Phi^{\rm (OA)}_\kappa(\mathbf{q})}
=\sum_{J\ge K} C_{JK\pi}(\mathbf{q})) \ket{JMK\pi},
\eeqn 
where  the probability of finding the component $\ket{JMK\pi}$ is
\beqn 
|C_{JK\pi}(\mathbf{q}))|^2  
&=&  \bra{\Phi^{\rm (OA)}_\kappa(\mathbf{q})} \hat P^J_{KK} \hat P^\pi\ket{\Phi^{\rm (OA)}_\kappa(\mathbf{q})}\nonumber\\
&=&  {\mathscr N}^{J}_{KK}(\kappa\mathbf{q},\kappa\mathbf{q}).
\eeqn 
Here, only quadrupole-deformed configurations are employed for \nuclide[25]{Mg} and thus parity is already conserved in the configurations. With the above decomposition, one can rewrite the norm overlap in terms of the weight of each component,
\beqn
\label{eq:norm_overlap_dfunction}
n(\kappa\mathbf{q},\kappa\mathbf{q}; \theta)
&\equiv&\bra{\Phi^{\rm (OA)}_\kappa(\mathbf{q})} \hat R_y(\theta) \ket{\Phi^{\rm (OA)}_\kappa(\mathbf{q})} \nonumber\\
&=& \sum_{J\ge K} |C_{JK\pi}(\mathbf{q})|^2 d^J_{KK}(\theta).
\eeqn 
For an even-even nucleus, the predominant component of the deformed mean-field wave function has the angular momentum $J=0$. Because of $d^{J=0}_{00}=1$,  the norm overlap  $n(\kappa\mathbf{q},\kappa\mathbf{q}; \theta)$ is always positive and can be well approximated with a Gaussian function centered at $\theta=0$~\cite{Ring:1980}. In contrast, the angular momentum  for an odd-mass nucleus is half integer with $J\ge 1/2$. In this case, the norm overlap can change sign during the oscillation with the rotation angle $\theta$. This phenomenon is illustrated in Fig.~\ref{fig:norm_odd_mass}. For the spherical quasiparticle states of \nuclide[25]{Mg} with $J=5/2$ and $K^\pi=1/2^+, 3/2^+$, and $5/2^+$, respectively, the norm overlap $n(\kappa\beta_2=0,\kappa\beta_2=0; \theta)$ is in line with the Wigner $d$-function $d^{5/2}_{KK}(\theta)$. For the deformed quasiparticle states of \nuclide[25]{Mg}, the norm overlap is given by a linear combination of $d^J_{KK}(\theta)$ with the mixing weight given by $|C_{JK\pi}(\mathbf{q}))|^2$, cf.(\ref{eq:norm_overlap_dfunction}). In particular, it is shown that the magnitude of oscillation around the rotation angle $\theta=\pi$ is much larger for the  quasiparticle configuration with $K^\pi=1/2^-$ than that with $K^\pi=1/2^+$.  One can read from Fig.~\ref{fig:single-particle-energy} that the lowest $1/2^+$ quasiparticle state at $\beta_2=0.5$ is dominated by the components split from the spherical $d_{3/2}$ orbital and $s_{1/2}$ orbital, while the $1/2^-$ quasiparticle state at $\beta_2=0.5$ is dominated by the component split from the spherical $f_{7/2}$ orbital. Thus, these two quasiparticle states have different distributions of the weights $|C_{JK\pi}(\mathbf{q}))|^2$ over the angular momentum $J$, as illustrated in Fig.~\ref{fig:norm_overlap_odd_mass}.  Moreover, the norm overlaps in all cases have the symmetry that $n(\kappa\mathbf{q},\kappa\mathbf{q}; 2\pi-\theta)=-n(\kappa\mathbf{q},\kappa\mathbf{q}; \theta)$, which is attributed to the symmetry property of the $d^J_{KK}(2\pi-\theta)=-d^J_{KK}(\theta)$ for all the half-integer $J$, see Eq.(\ref{eq:norm_overlap_dfunction}). 

\begin{figure}[]
\centering
\includegraphics[width=0.75\columnwidth]{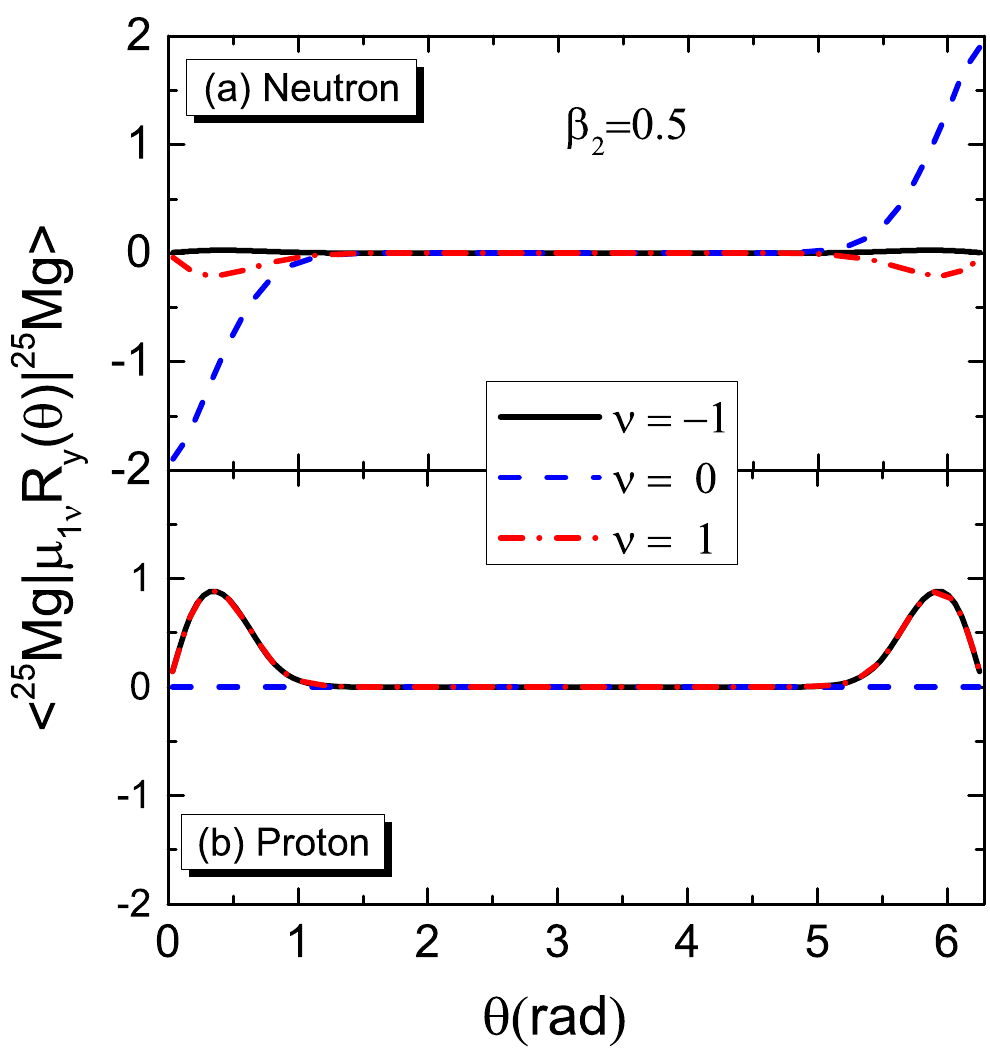}
\caption{\label{fig:overlap_magnetic_moment} (Color online) The overlap  of the magnetic dipole moment operators $\mu_{1\nu}(\kappa\mathbf{q}, \kappa\mathbf{q}; \theta)$ defined in (\ref{eq:magnetic_moment_overlap}) for neutrons (a) and protons (b) as a function of the Euler angle $\theta$, where the mean-field configurations for $^{25}$Mg is constructed with the EAC scheme and the quadrupole deformation of
nuclear core is $\beta_2=0.5$.}
\end{figure}

Figure~\ref{fig:overlap_magnetic_moment} displays the overlap of the magnetic dipole moment operators 
\beq 
\label{eq:magnetic_moment_overlap}
\mu_{1\nu}(\kappa\mathbf{q}, \kappa\mathbf{q}; \theta)
\equiv \bra{\Phi^{\rm (OA)}_\kappa\mathbf{q}} \hat \mu_{1\nu} e^{i\theta\hat J_y} \ket{\Phi^{\rm (OA)}_\kappa(\mathbf{q})}, 
\eeq 
 for both neutrons and protons in \nuclide[25]{Mg}. It is shown that  
 \beq 
 \mu_{1\nu}(\kappa\mathbf{q}, \kappa\mathbf{q}; 2\pi-\theta) = (-1)^{\nu}\mu_{1\nu}(\kappa\mathbf{q}, \kappa\mathbf{q}; \theta),
 \eeq 
 which has to be convoluted with the Wigner $d$-function $d^J_{KK}(\theta)$. It is worth noting that different from the case of even-even nuclei, the norm overlap has the symmetry that $n(\mathbf{q},\mathbf{q};\pi-\theta)=n(\mathbf{q},\mathbf{q};\theta)$~\cite{Yao:2009traxi}, the norm overlap for the odd-mass nuclei does not have this property any more. Therefore, one has to compute the norm overlap explicitly for the Euler angle $\theta$ in the interval $[0, \pi]$. This argument is also applied to the overlaps of other quantities.

\begin{figure}[]
\centering
\includegraphics[width=0.8\columnwidth]{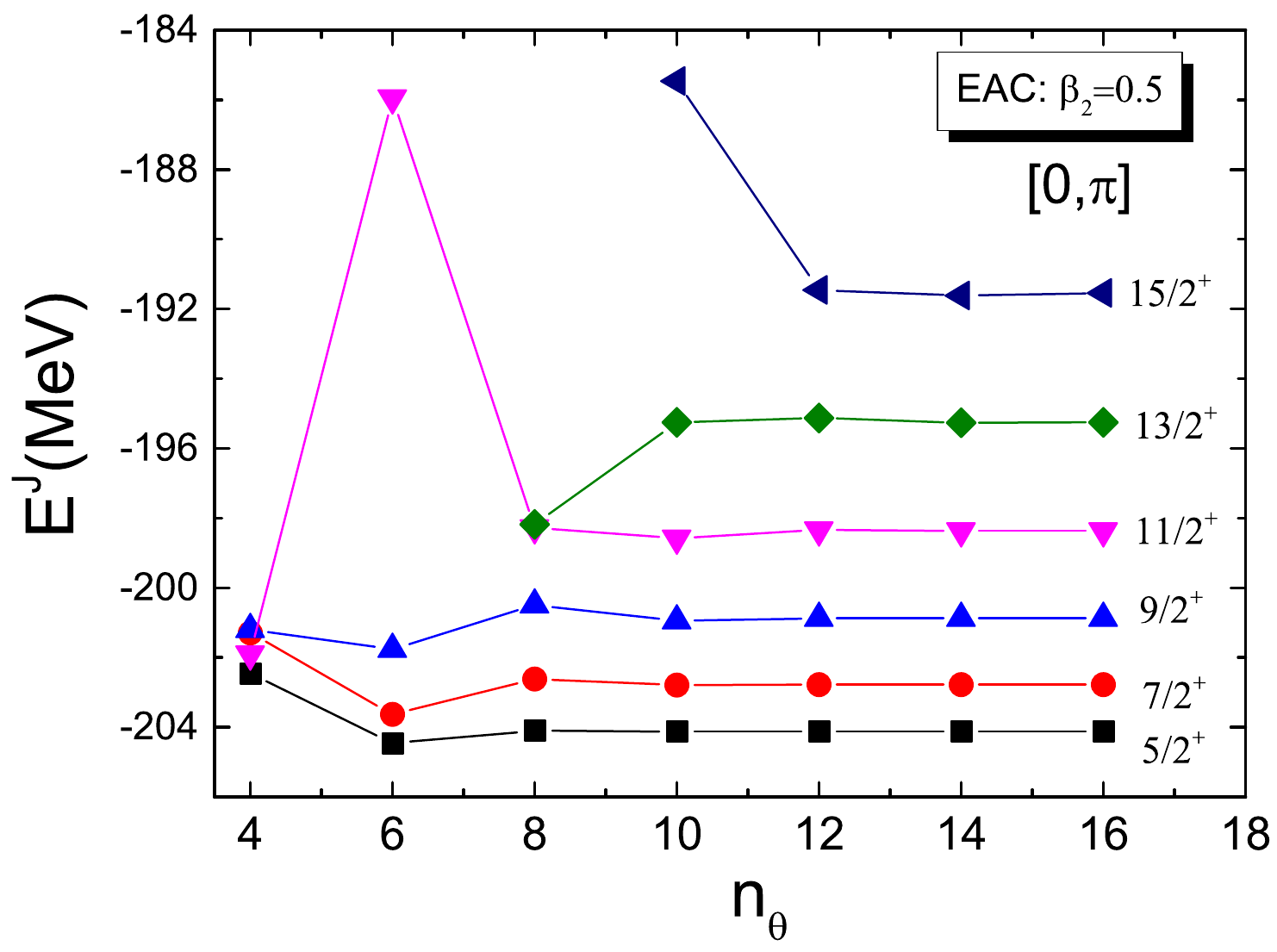}
\caption{\label{fig:AMP_meshpoint_number} (Color online) The convergence  of the energies of angular-momentum projected states for  \nuclide[25]{Mg} (EAC) with $J^\pi=5/2^+, 7/2^+,\cdots, 15/2^+$ as a function of the number $N_\theta$ of mesh points for the Euler angle $\theta\in[0,\pi]$, where  the quadrupole deformation of the intrinsic state is $\beta_2=0.5$. See main text for details. }
\end{figure}

\begin{figure}[]
\centering
\includegraphics[width=0.8\columnwidth]{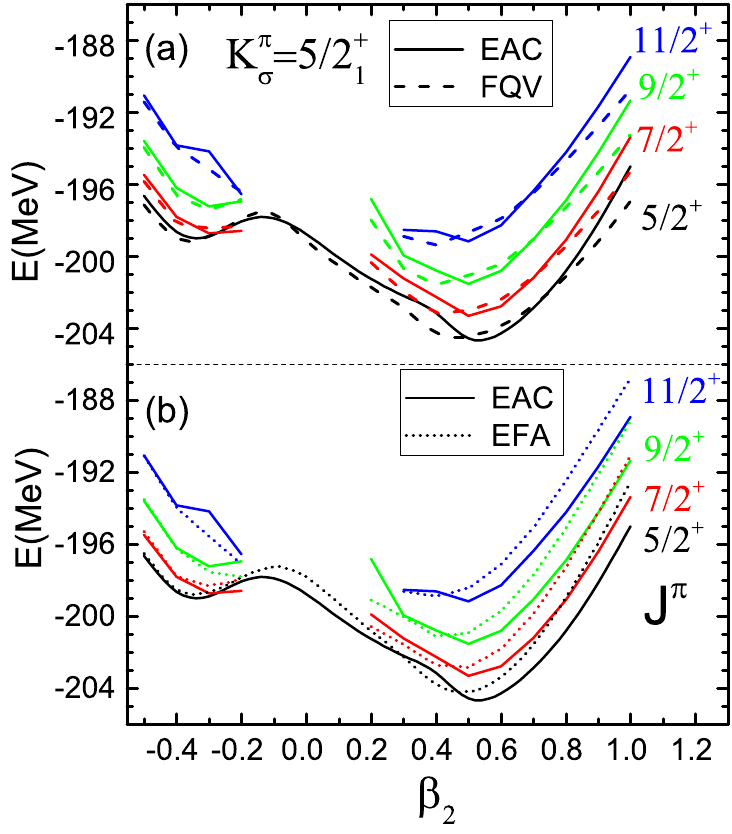}
\caption{\label{fig:PEC_AMP} (Color online) The energies of states with projection onto correct particle numbers and different angular momenta $J$ for $^{25}$Mg from the three  schemes  as a function of the quadrupole deformation parameter $\beta_2$. The quasiparticle states are characterized with quantum numbers $K^\pi_\sigma=5/2^+_1$.
}
\end{figure}

\begin{figure*}[]
\centering
\includegraphics[width=0.7\paperwidth]{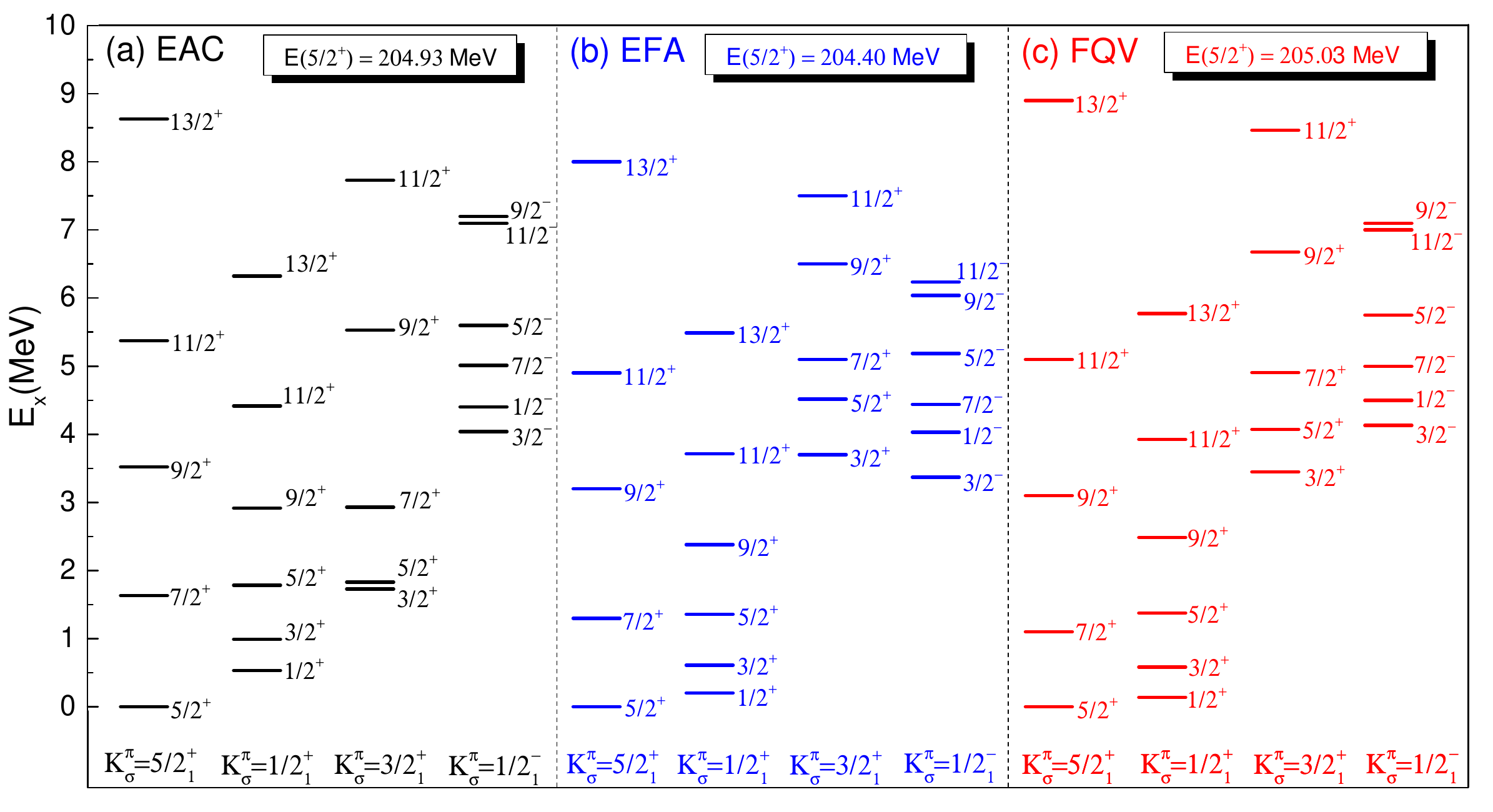}
\caption{\label{fig:spectra_three_method} (Color online) 
Comparison of low-energy spectra for \nuclide[25]{Mg} from the calcualtions based on the three different schemes.  See main text for details.
}
\end{figure*}

Figure~\ref{fig:AMP_meshpoint_number} displays the convergence of the energies of angular-momentum projected states for \nuclide[25]{Mg} as a function of the number $N_\theta$ of mesh points in  the Euler angle $\theta$, where the mean-field configuration is generated with the EAC scheme and the quadrupole deformation of the nuclear core $\nuclide[24]{Mg}$ is fixed at $\beta_2=0.5$.  It is shown that the choice of $N_\theta=12$ is sufficient to provide a convergent solution to the state with $J\leq 15/2$. In Fig.~\ref{fig:PEC_AMP}, we compare the energies of states for \nuclide[25]{Mg} with projection onto the correct particle numbers and different angular momenta based on different axially deformed states with $K^\pi=5/2^+$. The projected energy curves of the same spin-parity by the three schemes are similar to each other, except for the weakly-deformed regions where the norm kernels for the states of interest are so small that the energies can be strongly affected by numerical noises. Similar phenomena are found in the projected states based on the quasiparticle configurations with $K^\pi=1/2^+, 3/2^+$, and $1/2^-$.   We note that the observed discontinuity around $\beta_2=0.3$ in the projected energy curves by the EAC is due to the fact that pairing correlation collapses in the mean-field states of \nuclide[24]{Mg} with $\beta_2\ge0.3$, cf.Fig.~\ref{fig:pairing_energy}.

\subsubsection{Low-energy spectra from configuration-mixing calculations}

\begin{figure}[]
\centering
\includegraphics[width=8.5cm]{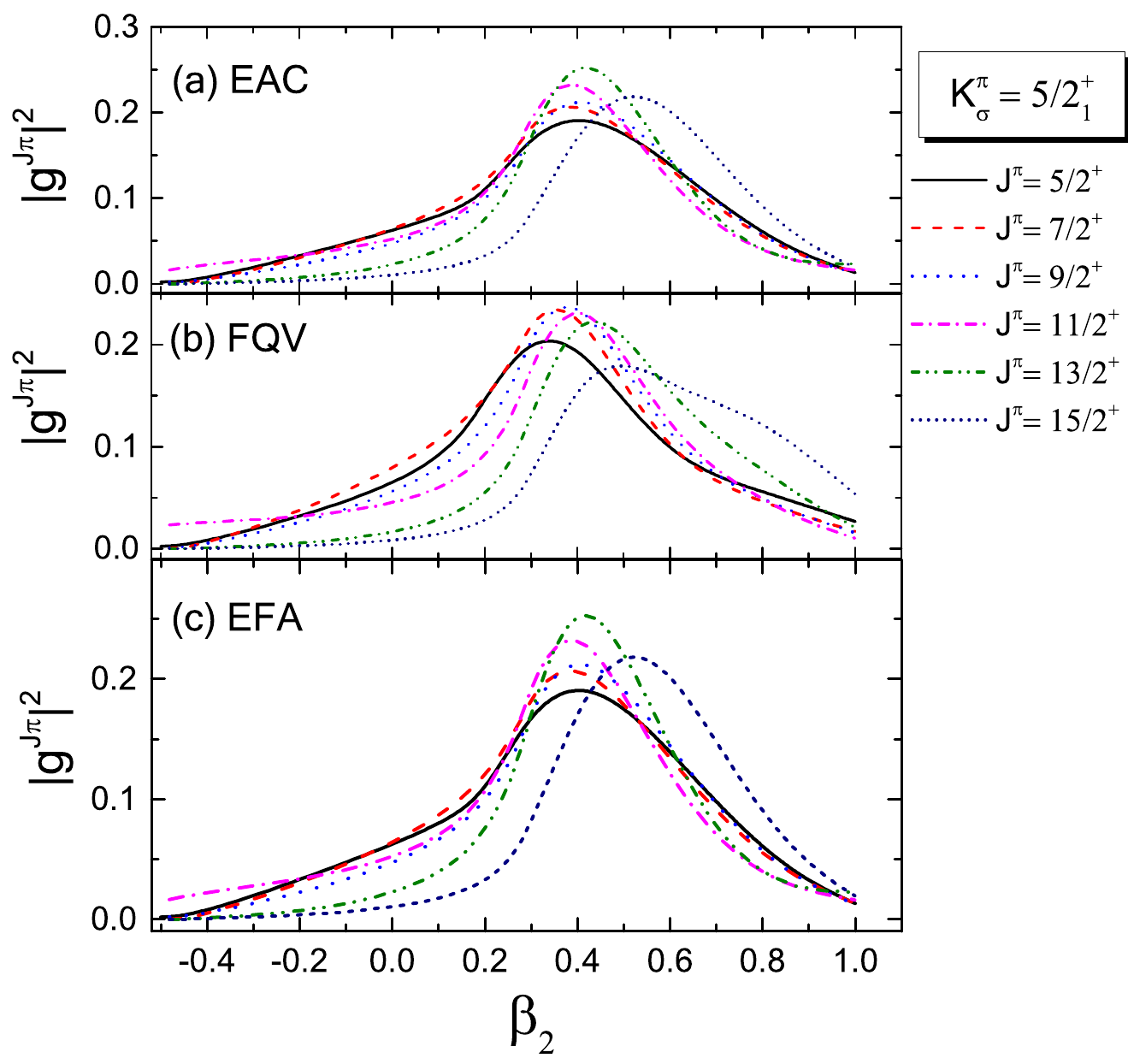} 
\caption{\label{fig:wfs_Kpi5} (Color online)  The distributions of collective wave functions $|g^{J\pi}|^2$ defined in Eq.(\ref{eq:coll_wf}) as a function of quadrupole deformation $\beta_2$ for the low-lying states of $^{25}$Mg with different spin-parity quantum numbers $J^\pi$ from three different types of calculations.}
\end{figure}

\begin{figure}[]
\centering
\includegraphics[width=8.5cm]{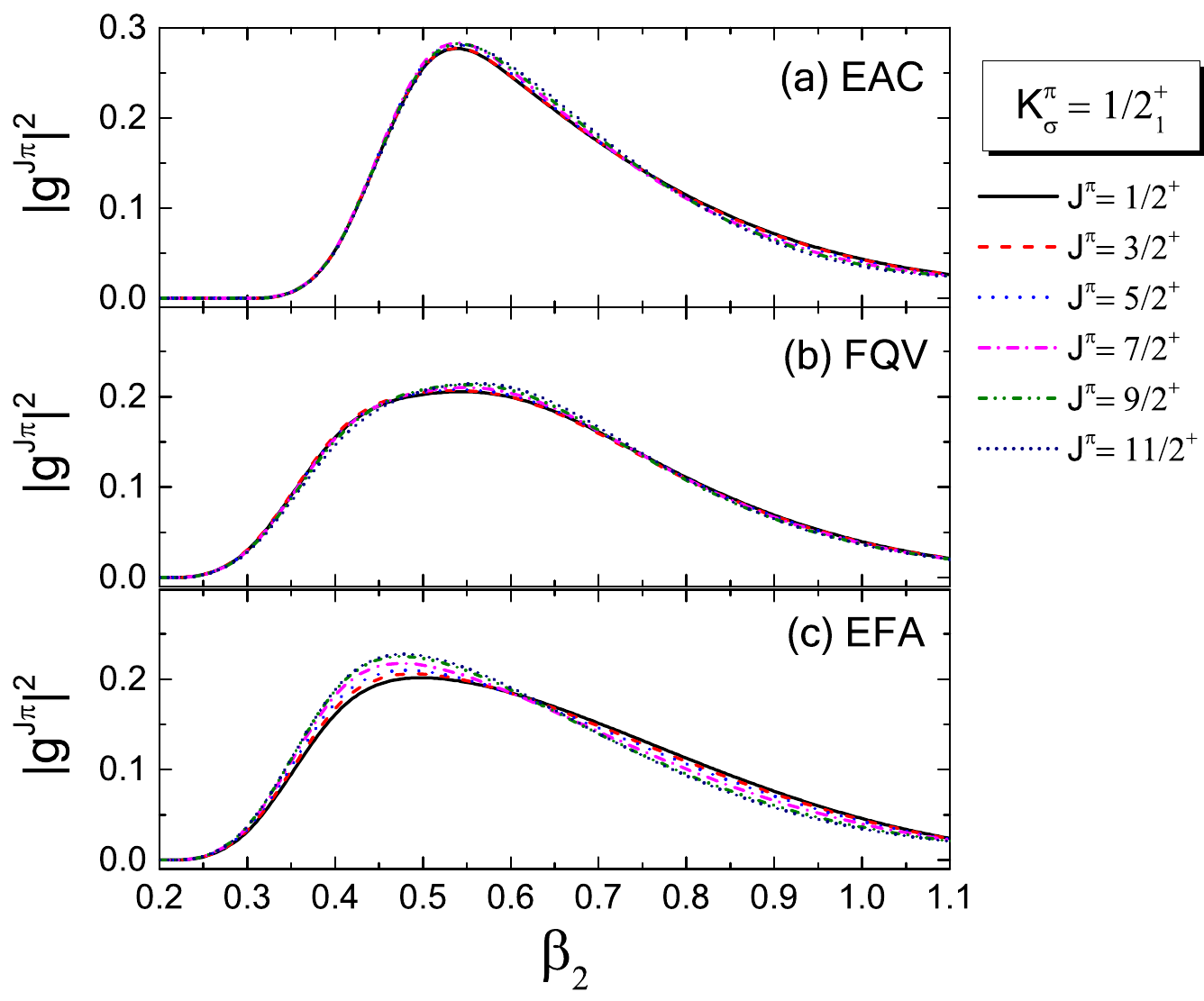}
\caption{\label{fig:wfs_Kpi1} (Color online) 
Same as Fig.~\ref{fig:wfs_Kpi5}, but for the states of $K^\pi=1/2^+$ band.
}
\end{figure}

\begin{figure}[]
\centering
\includegraphics[width=8.5cm]{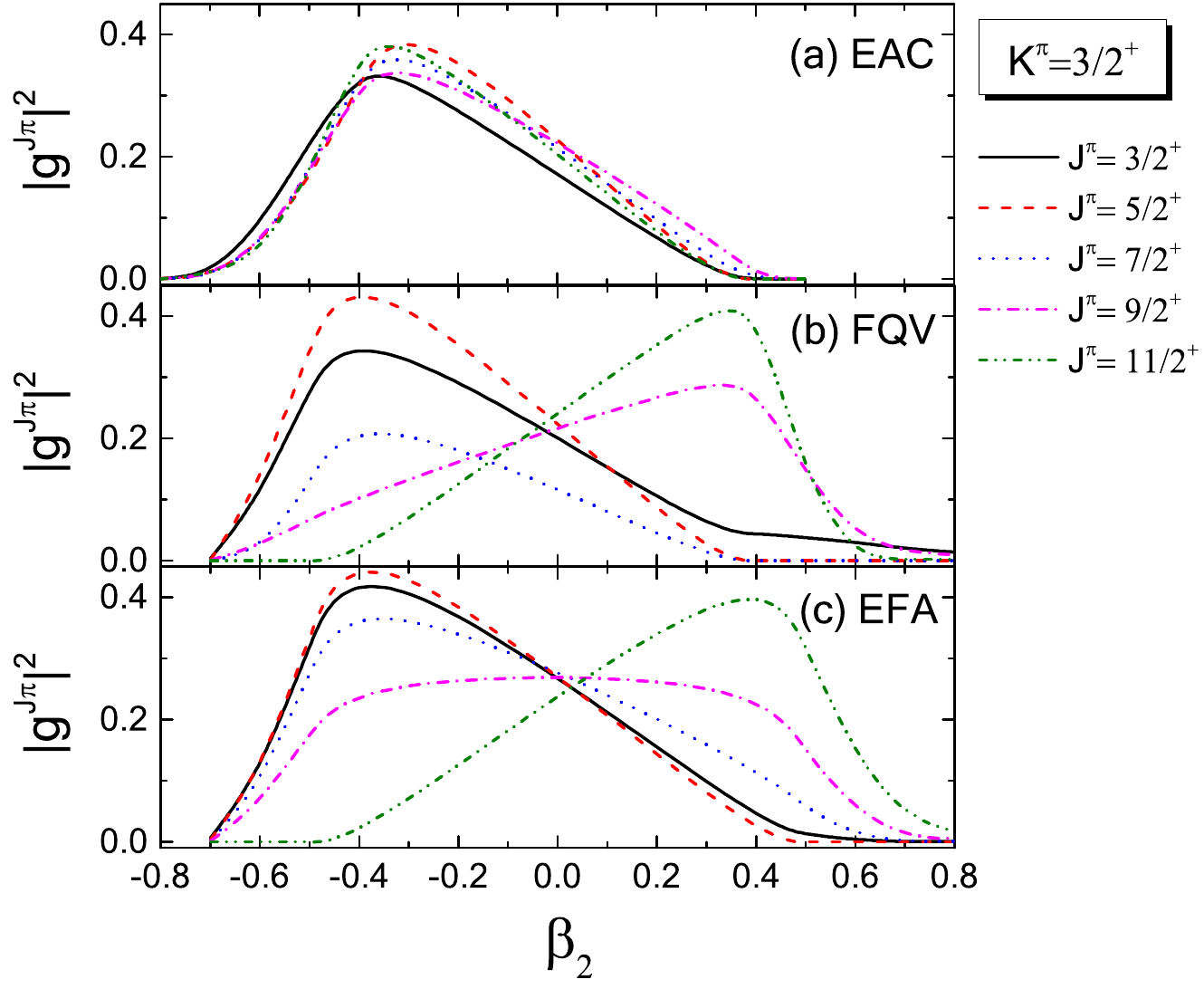}
\caption{\label{fig:wfs_Kpi3} (Color online) 
Same as Fig.~\ref{fig:wfs_Kpi5}, but for the states of $K^\pi=3/2^+$ band.
}
\end{figure}

\begin{figure}[]
\centering
\includegraphics[width=8.5cm]{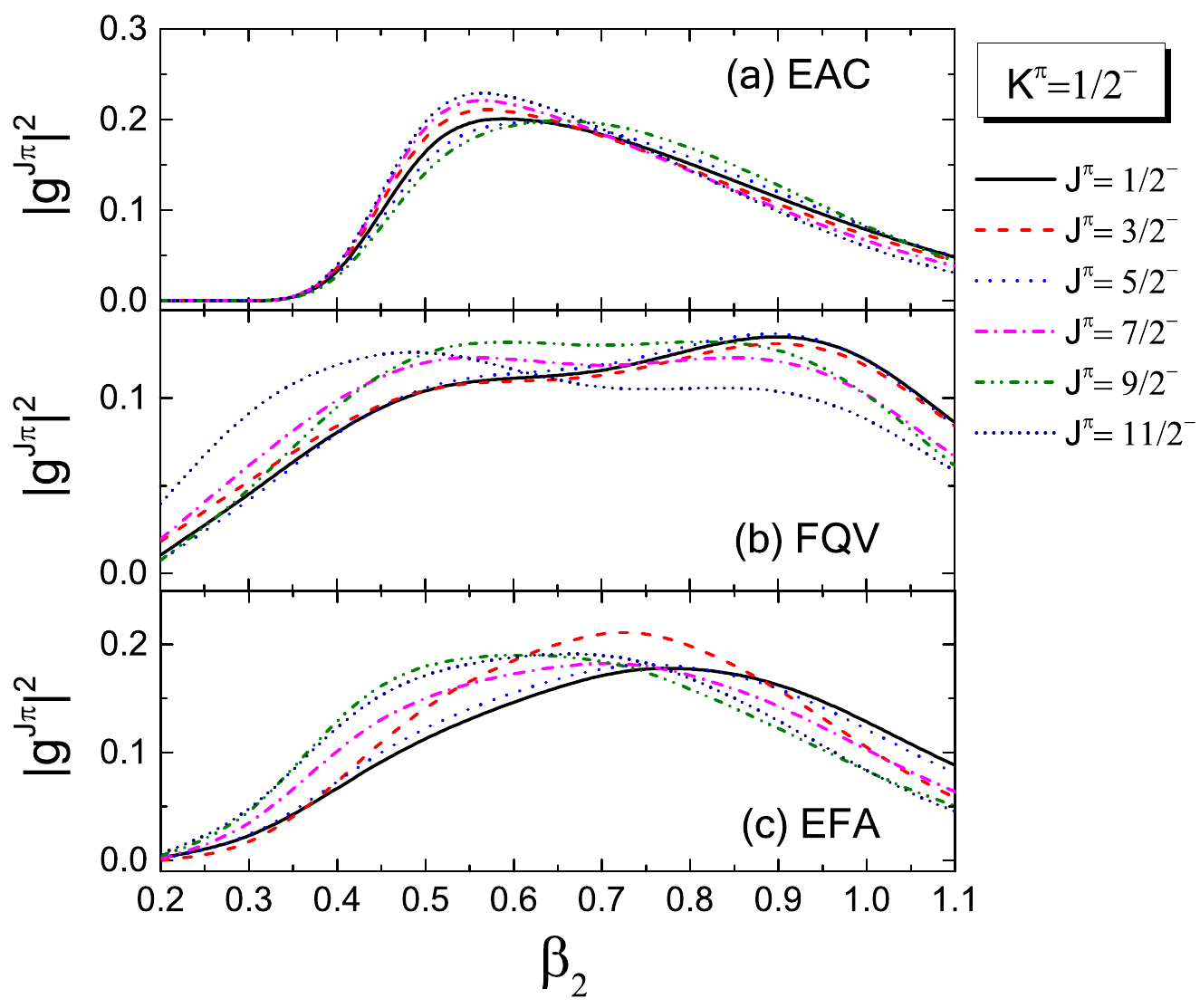}
\caption{\label{fig:wfs_Kpi1minus} (Color online) 
Same as Fig.~\ref{fig:wfs_Kpi5}, but for the states of $K^\pi=1/2^-$ band.
}
\end{figure}

Figure~\ref{fig:spectra_three_method} shows the comparison of the energy spectra for \nuclide[25]{Mg} by the three  schemes. The low-lying states are obtained from the mixing of quantum-number projected configurations with different axial deformation parameters. The states with the same $K^\pi$ are put into the same column. One can see that the entire energy spectra by the three calculations are qualitatively the same. Quantitatively, 
the predicted binding energy of the ground state with $J^\pi=5/2^+$ is $204.93$ MeV, $204.40$ MeV, and $205.03$ MeV by the EAC, EFA, and FQV schemes, respectively, slightly under-bound compared with the data $205.59$ MeV~\cite{NNDC}. The states of the $K^\pi=1/2^+$ band by the EAC is slightly higher than the other two cases. However, the states of the $K^\pi=3/2^+$ band by the EAC is significantly lower than other two schemes.   Otherwise, it is difficult to distinguish the low-energy spectra by the three schemes. Figure~\ref{fig:wfs_Kpi5} shows the comparison of the collective wave functions for the states of the $K^\pi=5/2^+$ band. It is seen that in all the three calculations, the predominant component of the state is slightly moving to larger deformed configurations with the increase of angular momentum. In contrast, the distributions of the collective wave functions for the states of  $K^\pi=1/2^+$ band are not evidently changed with the angular momentum, as shown in Fig.~\ref{fig:wfs_Kpi1}. The collective wave functions of the states in the $K^\pi=3/2^+$ band are displayed in Fig.~\ref{fig:wfs_Kpi3}, which shows clearly that these states by the EAC scheme are significantly different from those by the FQV and EFA schemes. For the $K^\pi=3/2^+$ band by the EAC, the predominant configuration of the state remains in the oblate deformed state with $\beta_2=-0.4$ with the increase of angular momentum. In contrast, the predominant configuration by the FQV and EFA is shifted gradually from the oblate state to a prolate state. It explains the observed difference in the energy spectra of the $K^\pi=3/2^+$ band, shown in Fig.~\ref{fig:spectra_three_method}. For all the states in the $K^\pi=1/2^-$ band, their collective wave functions are broadly distributed in the prolate deformed region and only slightly change with the increase of angular momentum, {as shown in Fig.~\ref{fig:wfs_Kpi1minus}.} In short, in most cases, the results by the FQV and EFA are closer to each other, in comparison with those by the EAC.

\begin{figure}[tb]
\centering
\includegraphics[width=8.5cm]{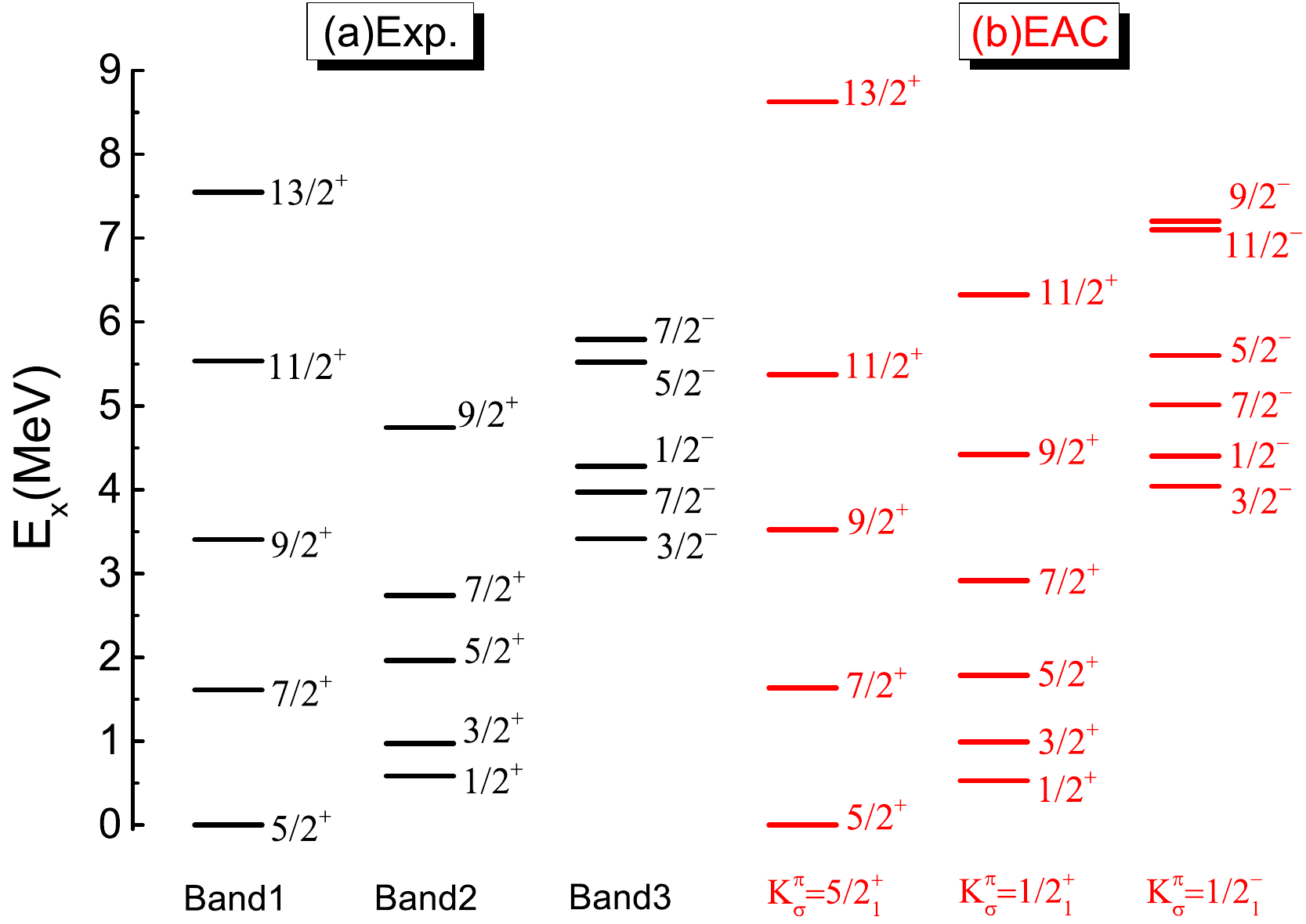}
\caption{\label{fig:excitation_pec_data} (Color online)  
The energy spectra of low-lying states in the $K^\pi=5/2^+, 1/2^+$ and $1/2^-$ bands of \nuclide[25]{Mg} from the calculation based on the EAC scheme, in comparison with available data.   }
\end{figure}

\begin{figure}[]
\centering
\includegraphics[width=\columnwidth]{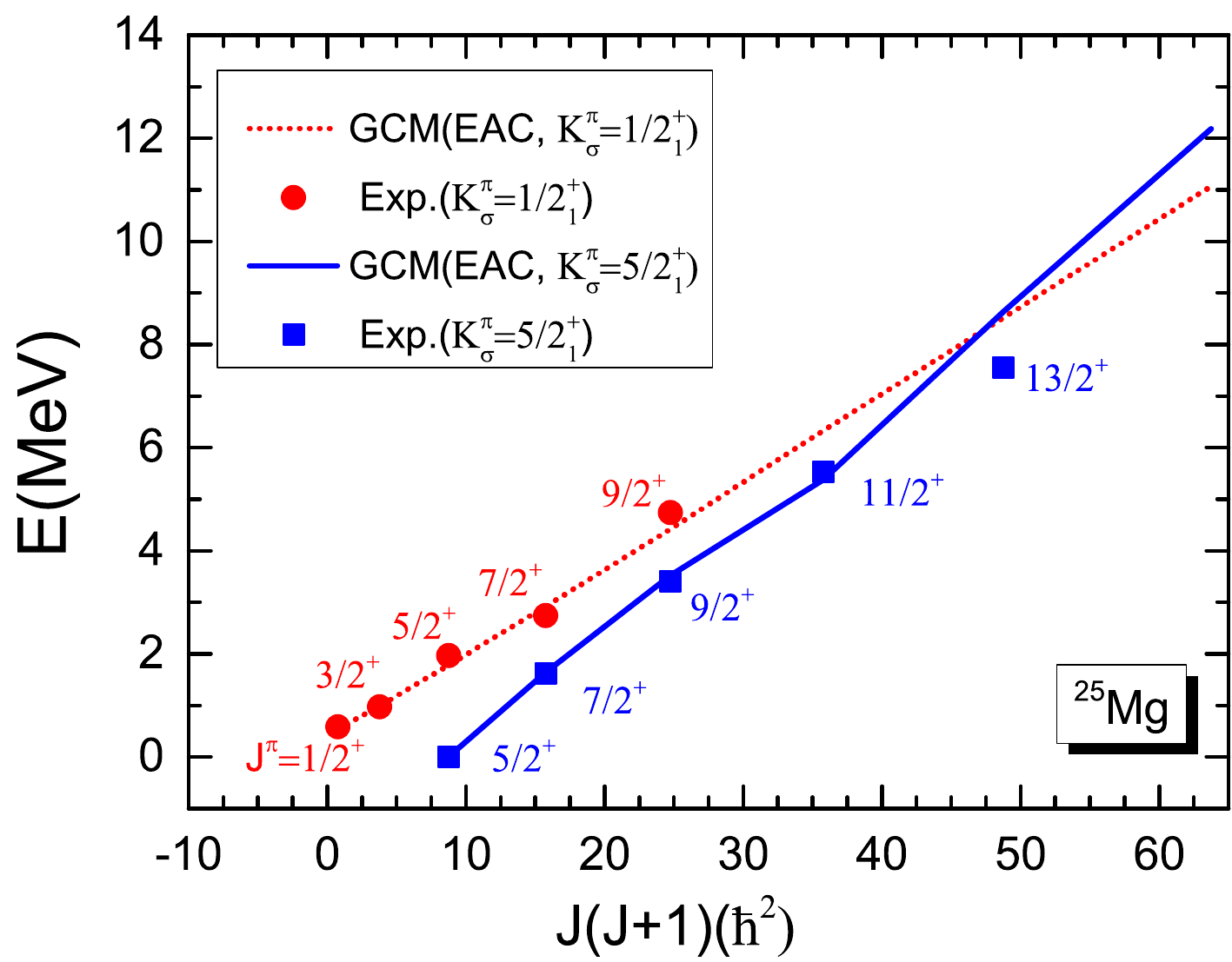}
\caption{\label{fig:excitation_energy_vs_J} (Color online) The excitation energies of the states in the $K^\pi=5/2^+$ and $1/2^+$ bands of \nuclide[25]{Mg} from the calculation with the EAC scheme as a function of $J(J+1)$, in comparison with corresponding data. 
}
\end{figure}

The energy spectra by the EAC are compared to available data in Fig.~\ref{fig:excitation_pec_data}. One can see that the observed three bands are in excellent agreement with the predicted $K^\pi=5/2^+, 1/2^+$, and $1/2^-$ bands, except for the wrong energy orderings of some states in the high-lying $1/2^-$ band which may need to consider other quansiparticle configurations.  Figure~\ref{fig:excitation_energy_vs_J} shows the excitation energies of the states in the $K^\pi=5/2^+, 1/2^+$ bands of \nuclide[25]{Mg}. If these states are pure rotational excitations, their energies  as a function of the $J(J+1)$ should fall in a straight line. This phenomenon is shown in the $K^\pi=1/2^+$ band, consistent with the distribution of the collective wave functions in Fig.~\ref{fig:wfs_Kpi1}. For the $K^\pi=5/2^+$ band, the excitation energies of the states in the low-spin region are in excellent agreement with the data. However, a slight deviation shows up at relatively high-spin regions. It can be understood from  Fig.~\ref{fig:wfs_Kpi5} that the distributions of the collective wave functions change with the increase of the angular momentum. 

\subsubsection{Electromagnetic properties of low-lying states}

The magnetic dipole moment of an axially deformed quasiparticle state with projection onto the correct particle numbers and  angular momentum $J$ is defined as
  \beqn 
  \label{eq:mu_JK_q}
 \mu(JK^\pi, \mathbf{q})  
   &=& \left(
     \begin{array}{ccc}
J&1&J\\
-J&0&J\\
\end{array}
   \right) 
   \frac{\langle NZ J\pi; c  ||\hat  \mu_1|| NZ J\pi; c\rangle}{\langle NZ J\pi; c  | NZ J\pi; c\rangle},
\eeqn
where the norm kernel and reduced matrix element have been defined in Eq.(\ref{eq:kernel}) and (\ref{eq:reduced_matrix_element}), respectively.

\begin{figure}[]
\centering
\includegraphics[width=0.8\columnwidth]{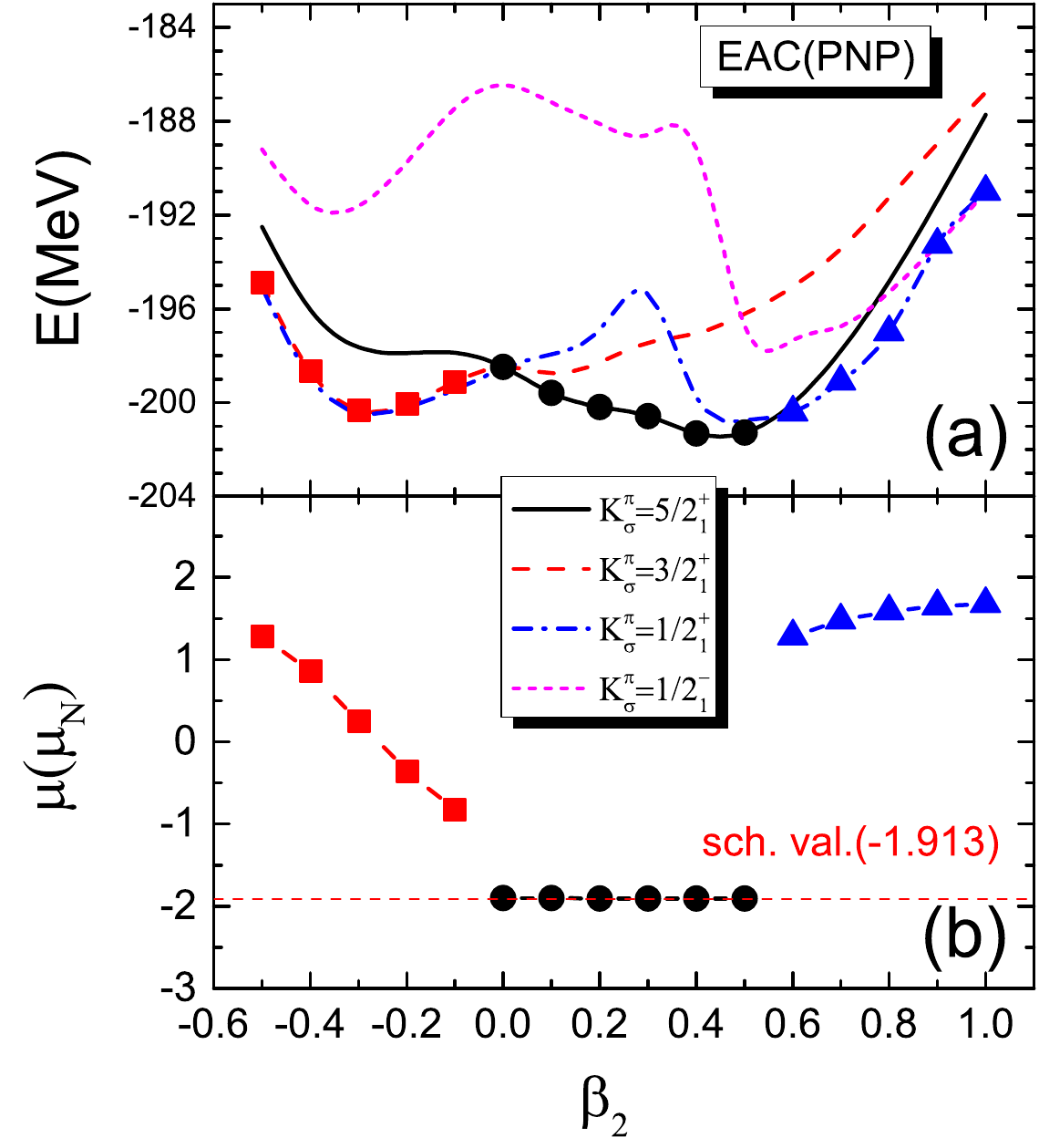}
\caption{\label{fig:dipole_moment_deformation_25Mg} (Color online) 
 (a) The energies of states projected onto the correct particle numbers for \nuclide[25]{Mg} (EAC) 
  with the $K_\sigma^\pi=5/2_1^+,3/2_1^+,1/2_1^+$, and 
 $1/2_1^{-}$, respectively, as a function of intrinsic  quadrupole deformation $\beta_2$.  (b) The magnetic moments of the lowest-energy configurations indicated in the panel (a). 
}
\end{figure}

\begin{figure}[]
\centering
\includegraphics[width=0.8\columnwidth]{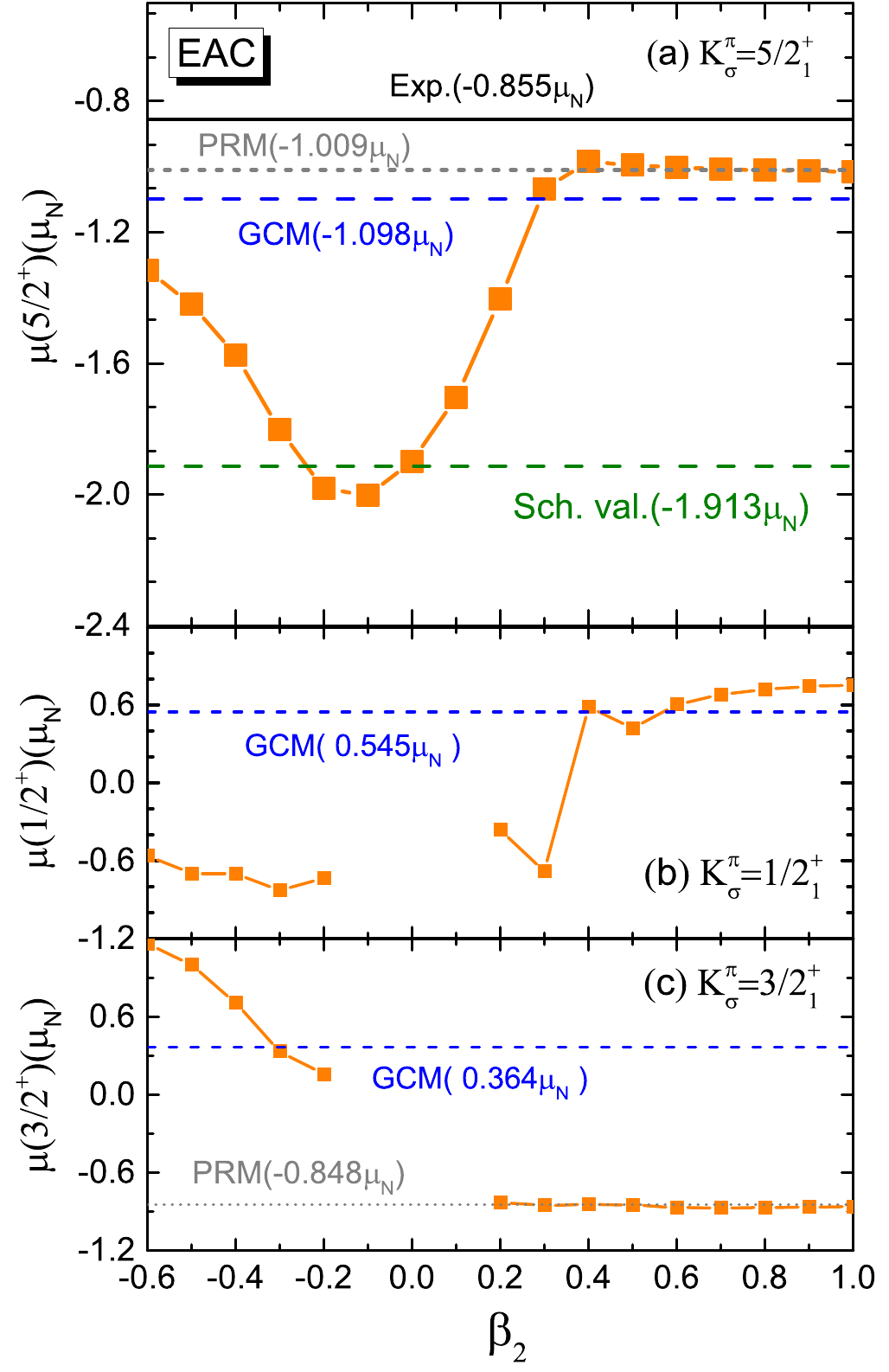}
\caption{\label{fig:magnetic_moment_AMP_25Mg} (Color online) 
The magnetic dipole moments of the band head states of the three bands with 
$K_\sigma^\pi=5/2_1^+,1/2_1^+$, and $3/2_1^+$ in $^{25}$Mg from the pure PNP+AMP calculation as a function of intrinsic quadrupole deformation $\beta_2$, where the mean-field wave function is constructed with the EAC scheme. The results by the configuration-mixing (GCM) calculation, PRM, and the Schmidt formula are also given for comparison.
}
\end{figure}

To illustrate the importance of restoration of rotational symmetry on the magnetic moment, we carry out the calculations for the magnetic moments with and without AMP. The magnetic dipole moments $\mu(K^\pi, \mathbf{q})$ of the lowest-energy quasiparticle state as a function of the quadrupole deformation $\beta_2$ without AMP are shown in Fig.~\ref{fig:dipole_moment_deformation_25Mg}, where $\mu(K^\pi, \mathbf{q})$ corresponds to $\mu(JK^\pi, \mathbf{q})$ in Eq.(\ref{eq:mu_JK_q}) without AMP. The energy of the corresponding state is also shown for comparison. It is seen that the lowest-energy quasiparticle state corresponds to $K^\pi=3/2^+$, $5/2^+$, and $1/2^+$ in the oblate, weakly and largely prolate deformed regions, respectively. The magnetic dipole moments are very different for the three quasiparticle configurations. In particular, the quasiparticle configuration with $K^\pi=5/2^+$ is in line with the Schmidt value by the independent particle model~\cite{Ring:1980}. After restoration of rotational symmetry with projection onto $J=5/2, 1/2$, and $3/2$ respectively for the quasiparticle configurations of $K^\pi=5/2^+, 1/2^+$, and $3/2^+$, the corresponding magnetic dipole moments are shown in Fig.~\ref{fig:magnetic_moment_AMP_25Mg}. The results with the  mixing
of different deformed configurations with GCM are also presented in comparison with available data. It is interesting to note that the magnetic dipole moment of the particle-number projected quasiparticle configuration with $K^\pi=5/2^+$ varies significantly with the quadrupole deformation from oblate to weakly prolate deformed shapes after projected onto $J=5/2$.  In contrast,  the magnetic moments of the projected states with the quadrupole deformation $\beta_2>0.5$ are almost constant. To understand this behavior, we compare the results with those by the PRM with an axially deformed rotor. In the strong coupling limit, the magnetic moment for the state $\ket{JK^\pi}$ with $K\neq 1/2$ is simply given by~\cite{Ring:1980}
\beq 
\label{eq:PRM_magnetic_moment}
\mu(JK^\pi) = g_R\frac{J(J+1)-K^2}{J+1} + g_K \frac{K^2}{J+1},
\eeq 
where the value $g_R$ for \nuclide[25]{Mg} is $g_R=(Z/A) \mu_N$, and  the $g_K$ in the PRM is only determined by the spin $g$-factor of the valence neutron ($g_l=0$ for the orbital $g$-factor)~\cite{Ring:1980}
\beq 
g_KK=-1.913 \mu_N.
\eeq 
According to (\ref{eq:PRM_magnetic_moment}), the magnetic moment of the state with $J^\pi=5/2^+$ and $K=5/2$ is $\mu(5/2^+)=-1.009 \mu_N$, and that of $3/2^+$ state with $K=3/2$ is $\mu(3/2^+)=-0.848 \mu_N$. The magnetic moments of the states with $K=1/2$ in the PRM contain an additional contribution which depends on an unknown decoupling factor~\cite{Ring:1980}.

It is seen from Fig.~\ref{fig:magnetic_moment_AMP_25Mg} that the PRM can nicely reproduce the  magnetic moments of the projected states with the quadrupole deformation $\beta_2>0.5$. It indicates that the picture of strong coupling limit of the PRM is realized in those largely prolate deformed states. After configuration mixing with GCM, the magnetic dipole moment of the $J^\pi=5/2^+$ state $\mu(5/2^+)=-1.098 \mu_N$, slightly different from the value by the PRM. Both values are in reasonable agreement with the data $\mu(5/2^+)=-0.855\mu_N$. Moreover, it is shown in Fig.~\ref{fig:magnetic_moment_AMP_25Mg} that for the largely deformed states projected onto $J^\pi=3/2^+$, the magnetic dipole moments also agree with the value $\mu(3/2^+)=-0.848\mu_N$ by the PRM. However, after mixing of both oblate and prolate deformed states with the GCM, the magnetic dipole moment of this state becomes $\mu(3/2^+)=0.364\mu_N$, which is very different from the value of the PRM. This configuration mixing effect  
is to be confirmed by the future measurement.

\begin{figure}[]
\centering
\includegraphics[width=8.5cm]{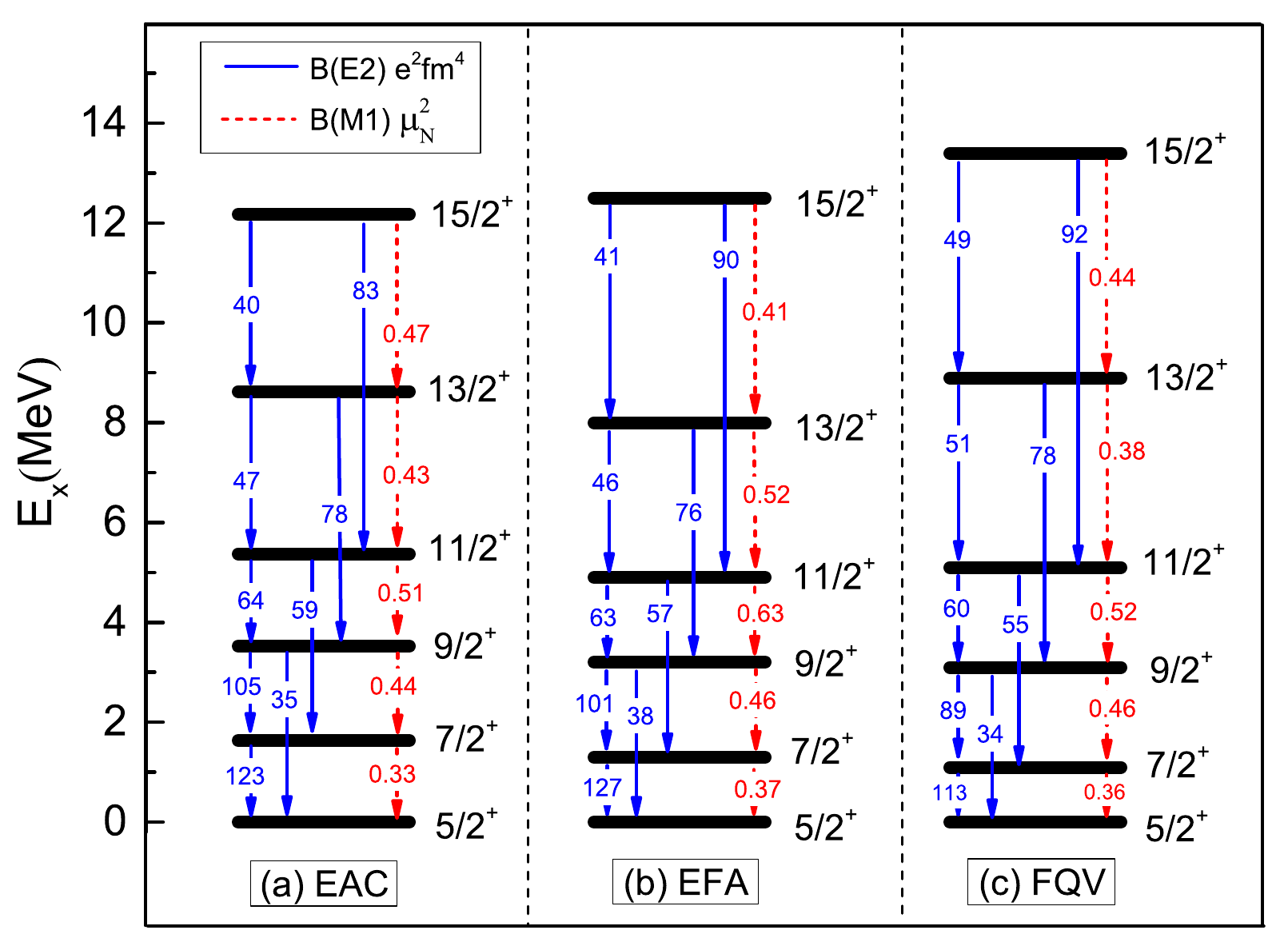}
\caption{\label{fig:comparison_EM_schemes} (Color online) 
Comparison of the $E2$ (blue) and $M1$ (red) transition strengths in the low-lying states of the ground-state band of $^{25}$Mg from the MR-CDFT calculations based on the three schemes. See main text for details.
}
\end{figure}

\begin{figure}[]
\centering
\includegraphics[width=8.5cm]{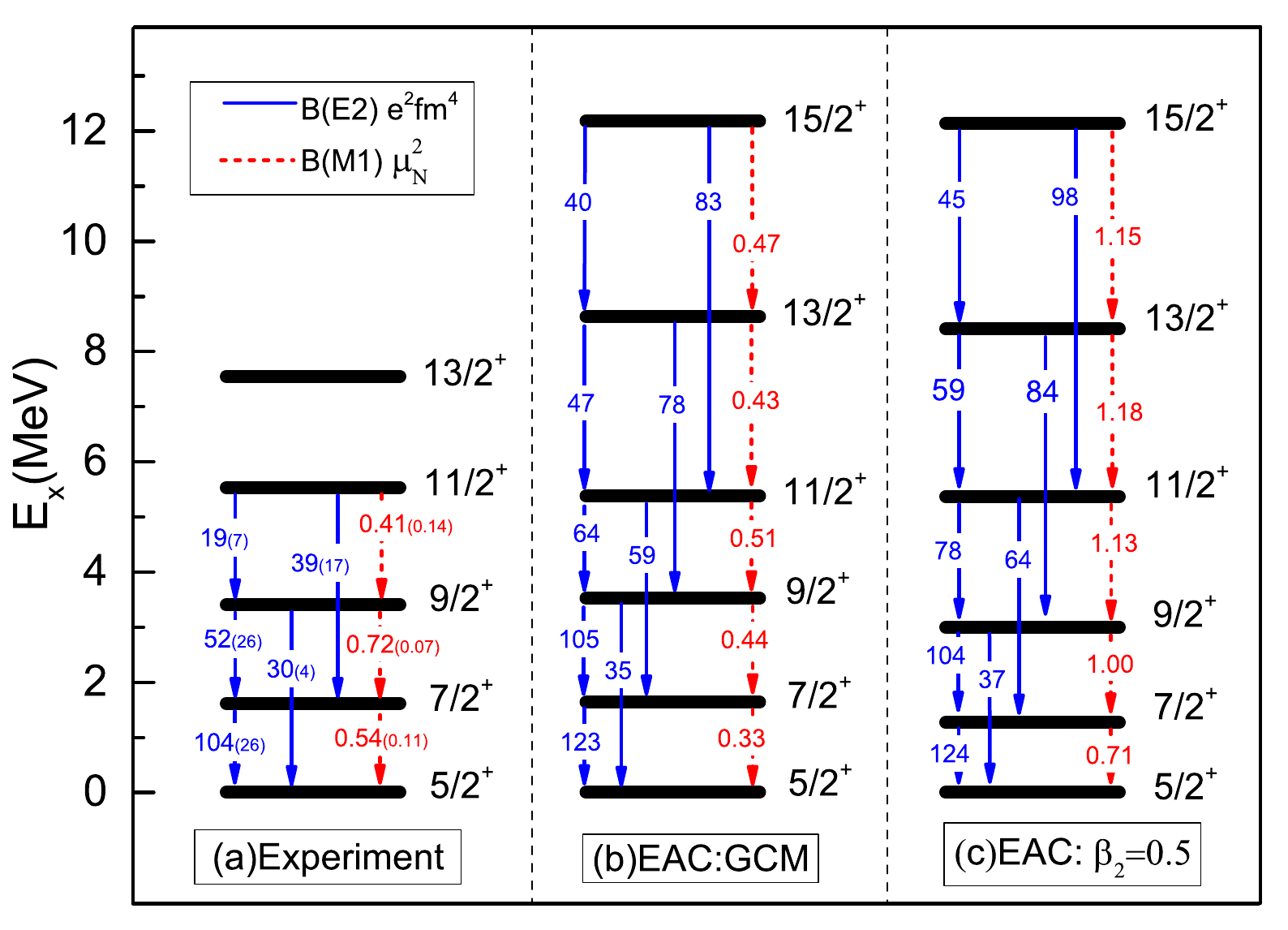} 
\caption{\label{fig:spectrum_PEC_GCM_beta05} (Color online) 
Comparison of the $E2$ (blue) and $M1$ (red) transition strengths of the states in the ground-state band ($K^\pi=5/2^+$) of $^{25}$Mg from the MR-CDFT (EAC) calculation, in comparison with corresponding data~\cite{Firestone:2009}. In panel (c), the configuration with $K^\pi=5/2^+$ and quadrupole deformation parameter $\beta_2=0.5$ is employed. 
}
\end{figure}

\begin{figure}[]
\centering
\includegraphics[width=\columnwidth]{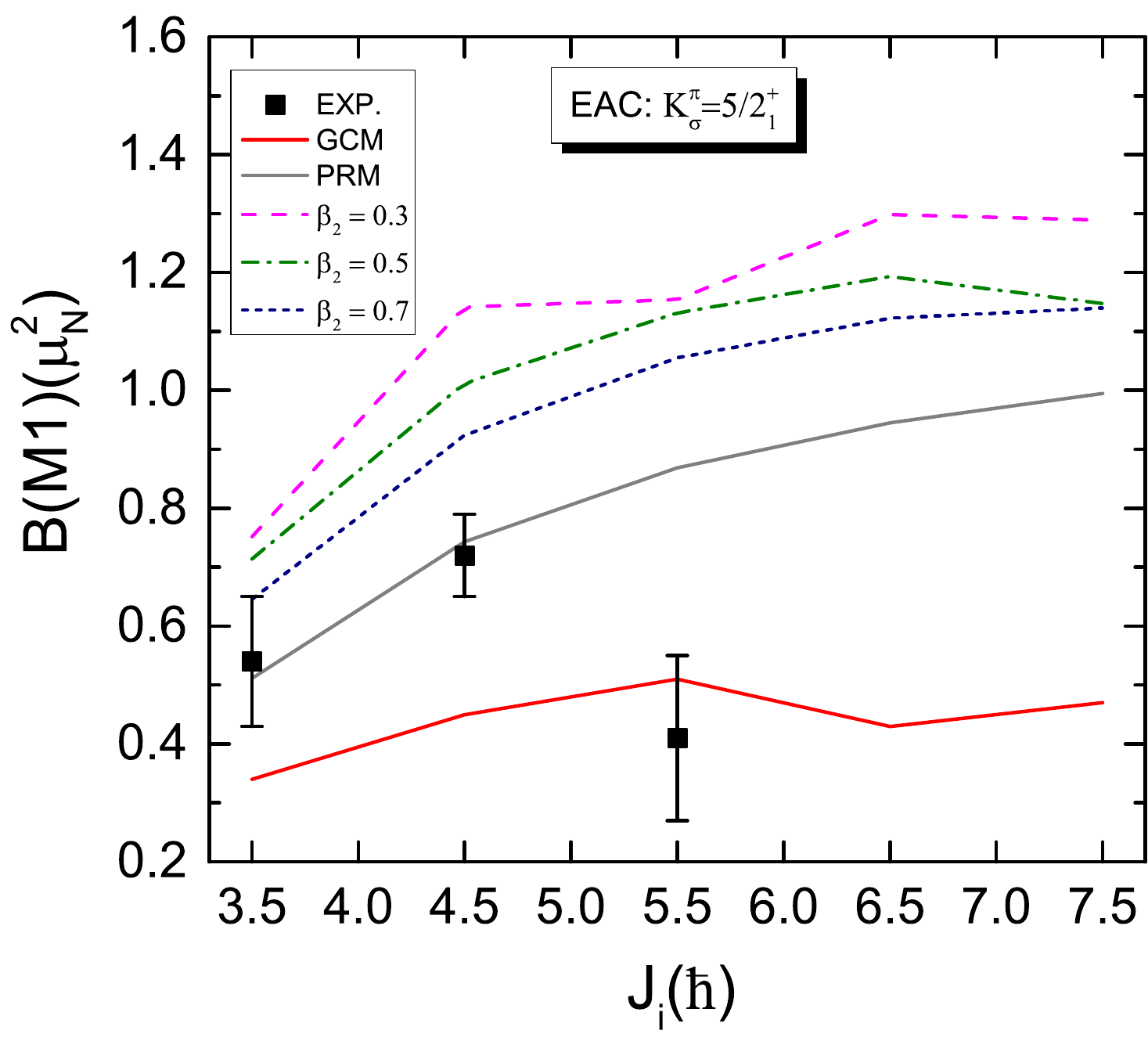}
\caption{\label{fig:BM1_vs_J} (Color online) The $M1$ transition strengths $B(M1, J^\pi_i\to J^\pi_f)$ ($\mu^2_N$) between the low-lying states in the $K^{\pi}=5/2^+$ band of \nuclide[25]{Mg} from different calculations as a function of the angular momentum $J_i$ of the initial state, in comparison with available data~\cite{Firestone:2009}.  See main text for details.
}
\end{figure}

Figure~\ref{fig:comparison_EM_schemes} shows the comparison of the $E2$  and $M1$ transition strengths in the low-lying states of $K^\pi=5/2^+$ band in $^{25}$Mg from the calculations based on the three schemes. The results by the MR-CDFT (EAC) and those based on the quadrupole deformed state with $\beta_2=0.5$ are compared with available data in Fig.~\ref{fig:spectrum_PEC_GCM_beta05}. The energy spectra and transition strengths are reasonably reproduced in all the three cases. It is seen from Fig.~\ref{fig:spectrum_PEC_GCM_beta05} that the $B(M1)$ values by the single-configuration with  $\beta_2=0.5$ are reduced by more than a factor of two after considering the configuration mixing effect.  This is shown more clearly in Fig.~\ref{fig:BM1_vs_J}, where  the  $M1$ transition strengths $B(M1, J^\pi_i\to J^\pi_f)$ from different model calculations are compared with the data.

\begin{figure}[]
\centering
\includegraphics[width=\columnwidth]{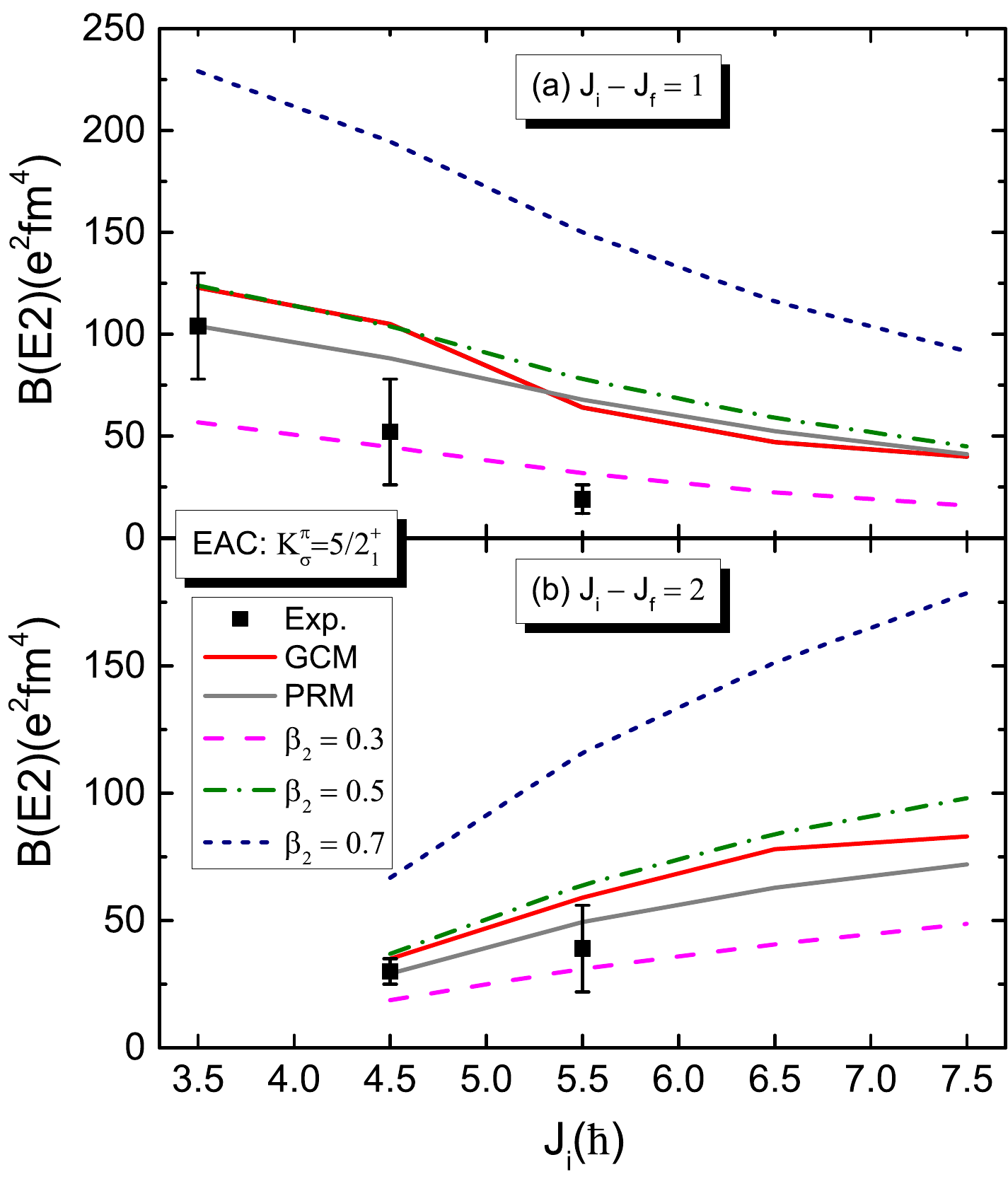}
\caption{\label{fig:BE2_PRM_comparison} (Color online) The $E2$ transition strengths $B(E2, J_i\to J_f)$ (e$^2$fm$^4$) between the low-lying states in the $K^{\pi}=5/2^+$ band of \nuclide[25]{Mg} from different calculations as a function of the angular momentum $J_i$ of the initial state, in comparison with available data~\cite{Firestone:2009}. See main text for details.
}
\end{figure} 

 In the PRM, the $M1$ transition strengths between the states within the same band with $K\neq 1/2$ are given by~\cite{Ring:1980}
\beq 
\label{eq:PRM_BM1}
B(M1, J_i K^\pi \to J_f K^\pi) = \frac{3}{4\pi}(g_K-g_R)^2K^2|\bra{J_iK10} J_fK\rangle|^2.
\eeq 
The change of the $B(M1)$ value with the increase of $J_i$ by the PRM is mainly attributed to the Clebsch–Gordan coefficient $\bra{J_iK10}J_fK\rangle$, and this behavior is also seen in the results of single-configuration calculations with $\beta_2>0.5$. In contrast, the data follow this behavior in low-spin states, but start to deviate at $J_i=11/2$, which is qualitatively reproduced by the GCM, even though the $B(M1)$ values are about two thirds of the data for $J_i=7/2$ and $J_i=9/2$. It is shown in Fig.~\ref{fig:BM1_vs_J} that the use of configurations with different quadrupole deformation $\beta_2$ has an impact on the $B(M1)$ values of the low-lying states of \nuclide[25]{Mg}. This is different from the observation in $^9_\Lambda$Be that the $B(M1)$ values are barely dependent on the quadrupole deformations of the configurations~\cite{Yao:2023_NPA}. This difference is attributed to the fact that the orbital of the valence neutron changes significantly with $\beta_2$, while the $\Lambda$ hyperon remains in the lowest orbital coming from the spherical $s_{1/2}$ orbital in $^9_\Lambda$Be.

 The theoretical results (except for those by the single-configuration calculation with  $\beta_2=0.5$) exhibit the feature of transition from strong-coupling limit to decoupling limit, where a strong $E2$ transition connects the states with $\Delta J=1$ and $\Delta J=2$ states, respectively. 
In particular, one notices that the $E2$ transition between states with $\Delta J=2$ increases with $J$, while that between states with $\Delta J=1$ decreases with $J$.   This feature is also exhibited in the available data and can be understood from the prediction by the PRM~\cite{Ring:1980}
\beq 
\label{eq:BE2_PRM}
B(E2, J_i K^\pi \to J_fK^\pi) = Q^2_{20} \frac{5}{16\pi} |\langle J_i K 20|J_fK\rangle|^2,
\eeq 
where the Clebsch–Gordan coefficient $\langle J_i K 20|J_fK\rangle$ between states with $\Delta J=2$ ($\Delta J=1$) is increasing (decreasing) with $J_i$. Figure~\ref{fig:BE2_PRM_comparison} displays clearly the change of the $B(E2, J_i K^\pi \to J_fK^\pi)$ for both $\Delta J=1$ and $\Delta J=2$ cases  as a function of the angular momentum $J_i$ of the initial state determined from the single-configuration calculation based on different quadrupole deformed states and from the configuration-mixing calculation. The results by the PRM formula and available data are also given, where the parameter $Q^2_{20}(=2927 e^2{\rm fm^4})$ in Eq.(\ref{eq:BE2_PRM}) is determined by the data on the $B(E2, J_i K^\pi \to J_fK^\pi)$ at $J_i=7/2$ for the transition with $\Delta J=1$. 

The magnetic dipole moments and spectroscopic quadrupole moments of the low-lying states in the $K^\pi=5/2^+$ band of $^{25}$Mg from different calculations are presented in Table~\ref{tab:magnetic_moment_Qs_comparison}, in comparison with available data. One can see that there is no essential difference among the calculations of the three schemes, and those of a similar calculation based on the Skyrme~\cite{Bally:2014prl} and Gogny force~\cite{Borrajo:2018sqj}, where triaxiality effect is included. Both the magnetic dipole moment and spectroscopic quadrupole moment of the $J^\pi=5/2^+$ are reasonably reproduced in all the cases.

\begin{table}[]
    \centering
    \tabcolsep=1pt
    \caption{The magnetic dipole moments (in unites of $\mu_N$) and spectroscopic quadrupole moments (in units of $e$fm$^2$) of the low-lying states in the $K^\pi=5/2^+$ band of $^{25}$Mg from the GCM calculations on the basis of different EDFs, in comparison with the prediction of PRM.}
 \begin{tabular}{cccccccc}
\hline
              & Exp.     &   CDFT     & CDFT & CDFT  &  Gogny  & SLyMR0  & PRM  \\ 
              &      &  EAC  & FQV  &  EFA &  \cite{Borrajo:2018sqj} &  \cite{Bally:2014prl} &    \\
\hline
 $\mu(5/2^+)$ & $-0.855$ &   $-1.098 $ & $-0.982$ & $-0.973$ & - & $-1.054$ & $-1.009$\\
 $\mu(7/2^+)$  & -     &    $0.229 $    & $ 0.297 $ &$0.270 $ & -& - & $-0.007$\\
 $\mu(9/2^+)$   & -    &   $1.350 $    & $1.367 $ & $1.310 $ & -& - & 0.812\\
 $\mu(11/2^+)$   & -    &  $2.178$    & $2.195$ & $2.161$ & - & -& 1.533\\
 \hline
  $Q^{\rm s}(5/2^+)$   & $20.1(3)$    &  $19.5$    & $19.9$ 
  & $21.2$ & $22.2$ & 23.25 & -\\
  $Q^{\rm s}(7/2^+)$   & -    &  $3.1$    & $2.8$ & $3.1$ & $3.8$ & -& -\\
   $Q^{\rm s}(9/2^+)$   & -    &  $-6.3$    & $-6.2$ & $-6.3$ & $-6.8$ & -&- \\
   $Q^{\rm s}(11/2^+)$   & -    &  $-10.9$    & $-10.9$ & $-10.8$ & $-11.6$ & -&- \\
   \hline
\end{tabular}
    \label{tab:magnetic_moment_Qs_comparison}
\end{table}

\subsection{Application to $^{21}$Ne with quadrupole-octupole correlations}

Previous studies with the MR-CDFT have demonstrated the importance of octupole correlations in the low-lying states of \nuclide[20]{Ne}~\cite{Zhou:2016_Ne20} and $^{21}_\Lambda$Ne\cite{Xia:2023}. In this work, we extend these studies to the low-lying states of \nuclide[21]{Ne} by considering fluctuations in both quadrupole and octupole deformed shapes. The calculation for \nuclide[21]{Ne} is more complicated than that for \nuclide[25]{Mg} as the octupole deformed states do not have good parity. The parity $\pi$ of the state can be recovered with the parity projection operator $\hat P^\pi$.

\begin{figure}[]
\centering
\includegraphics[width=\columnwidth]{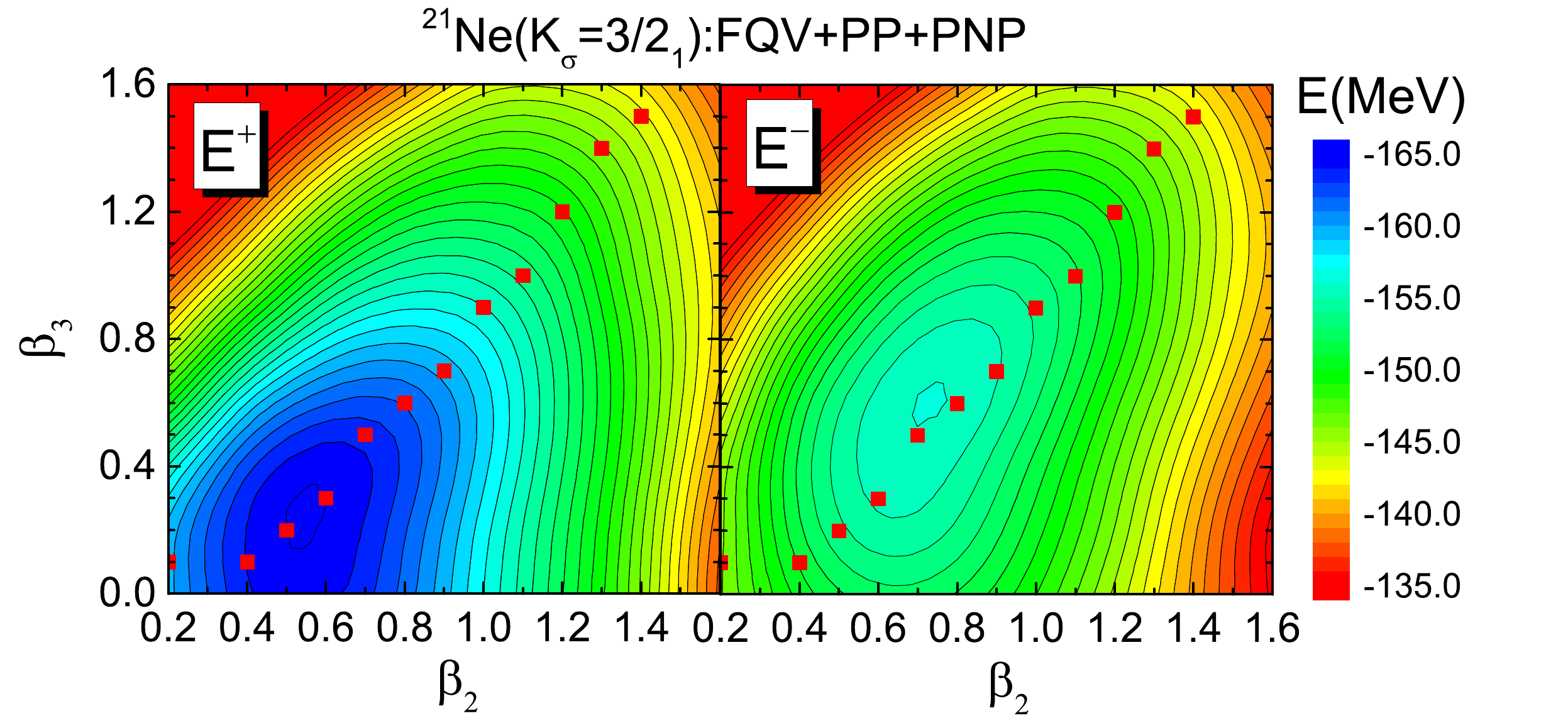}
\caption{\label{fig:Ne21_energy_surface_PNP} (Color online) 
The energies of states with $K=3/2$ (obtained with the FQV scheme) in the quadrupole-octupole deformation plane for $^{21}$Ne with projection onto the correct particle numbers and different parities, where the red squares indicate the optimal configurations. 
}
\end{figure}

\begin{figure}[]
\centering
\includegraphics[width=8.2cm]{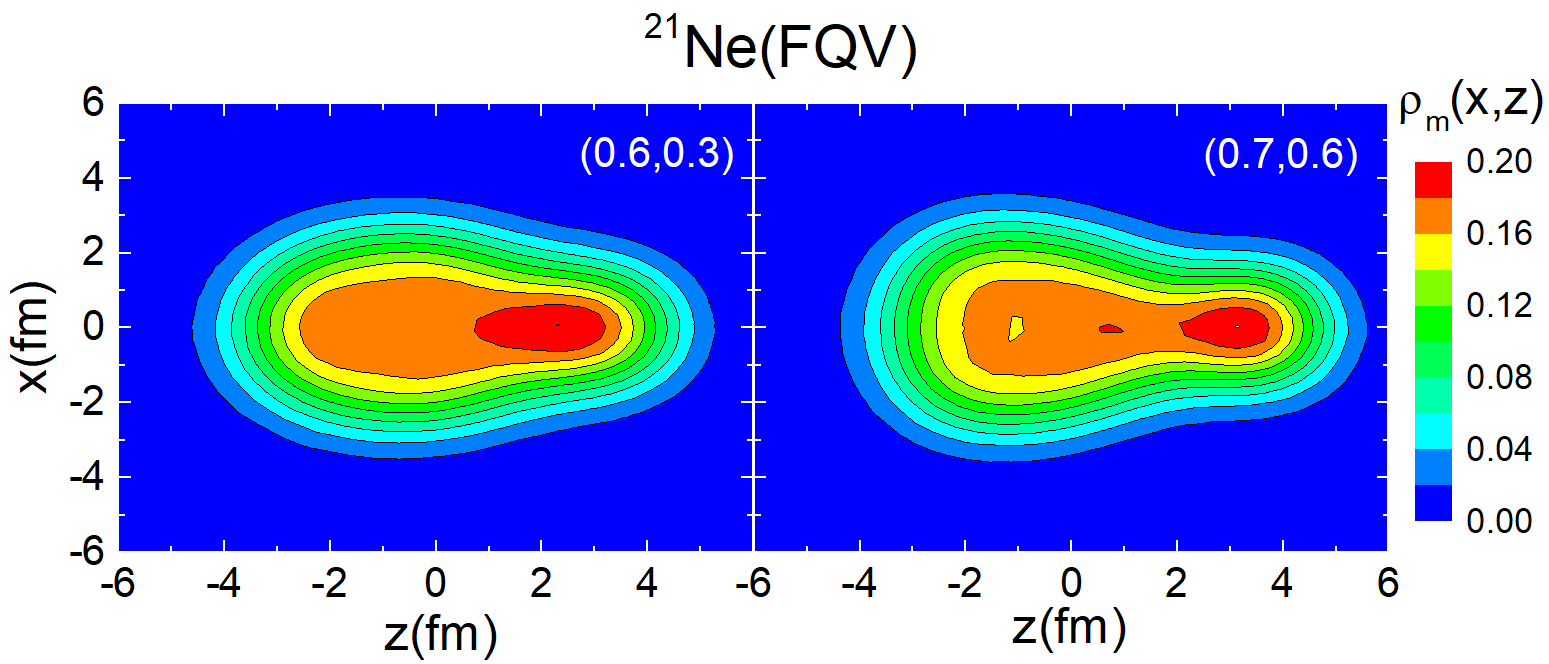}
\caption{\label{fig:Ne21_density} (Color online) 
 Contour plots of total density $\rho_m$  of neutrons and protons in the $x$-$z$ plane ($y=0$), corresponding to the energy-minimal states with positive and negative parities, and the quadrupole-octupole deformation parameters $(\beta_2, \beta_3)=(0.6, 0.3)$ and $(0.7, 0.6)$, respectively. 
}
\end{figure}

\begin{figure}[]
\centering
\includegraphics[width=8.5cm]{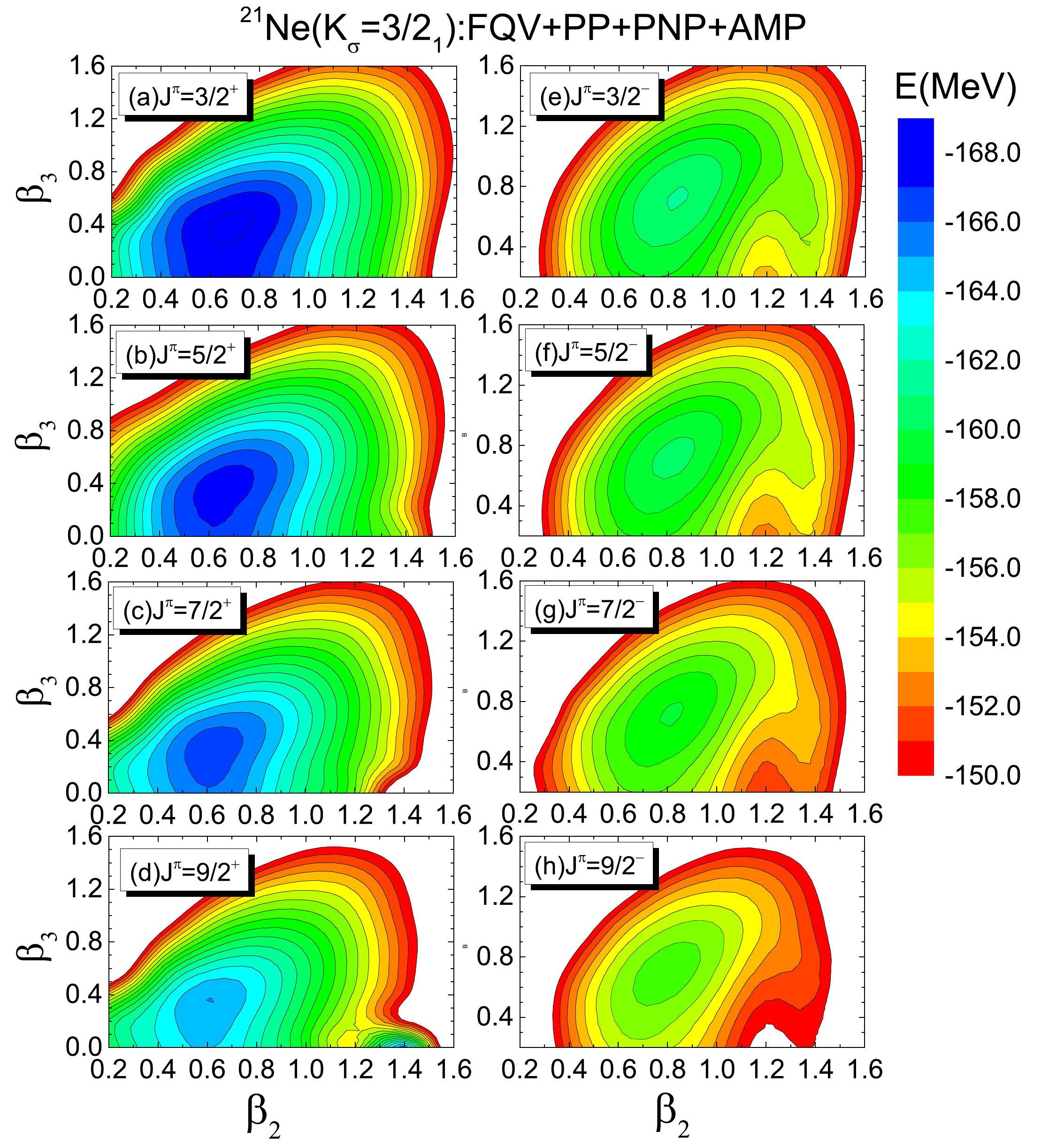}
\caption{\label{fig:Ne21_energy_surface_projection} (Color online) 
Same as Fig.~\ref{fig:Ne21_energy_surface_PNP}, but with additional projection onto different angular momenta and parities.
}
\end{figure}

Figure~\ref{fig:Ne21_energy_surface_PNP} shows the  energy surfaces of $^{21}$Ne in the quadrupole-octupole deformation plane with projection onto correct particle numbers, and positive or negative parities. A quadrupole-octupole deformed energy minimum appears at around $\beta_2=0.6, \beta_3=0.3$ on the energy surface with positive parity, while the energy minimum of negative-parity states is located around $\beta_2=0.7, \beta_3=0.6$.
The distributions of the total densities for these two states are plotted in Fig.~\ref{fig:Ne21_density}, where a pear-like structure similar to that in \nuclide[20]{Ne}~\cite{Zhou:2016_Ne20}  is observed. Figure~\ref{fig:Ne21_energy_surface_projection} displays the energy surfaces of $^{21}$Ne with projection onto $J^\pi=3/2^\pm, 5/2^\pm, 7/2^\pm$, and $9/2^\pm$, respectively. It is seen that the octupole deformation of the positive-parity states is decreasing with the increase of the angular momentum, which is similar to, even though not as much as, that in \nuclide[20]{Ne}. In contrast, the distributions of the collective wave functions for the negative-parity states are rather stable against the rotation, which is different from that in \nuclide[20]{Ne}.

\begin{figure}[]
\centering
\includegraphics[width=8.5cm]{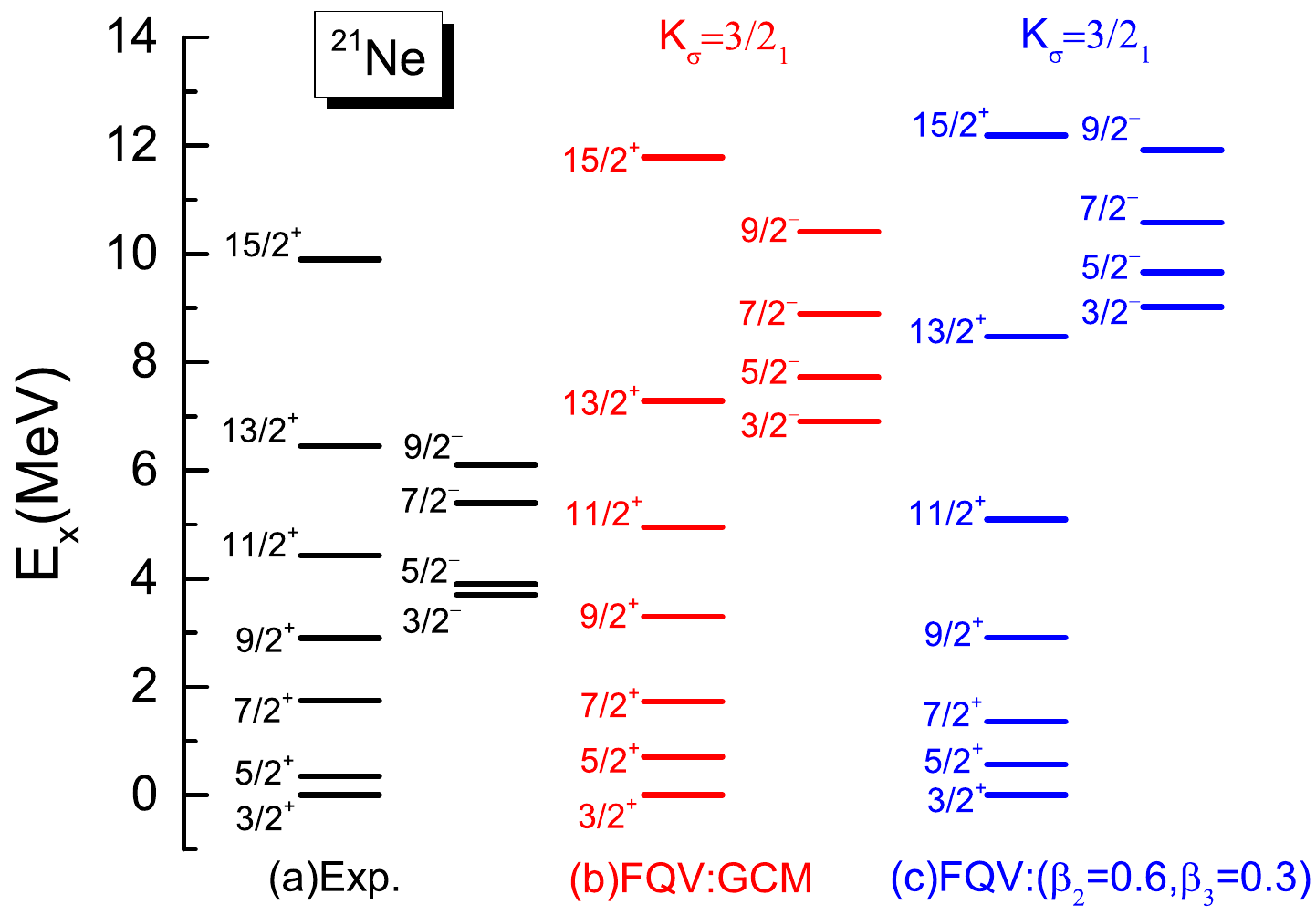}
\caption{\label{fig:Ne21_spectra} (Color online) 
The energy spectra of low-lying parity-doublet states in $^{21}$Ne from two different calculations, in comparison with available data. See main text for details.
}
\end{figure}

\begin{figure}[]
\centering
\includegraphics[width=8cm]{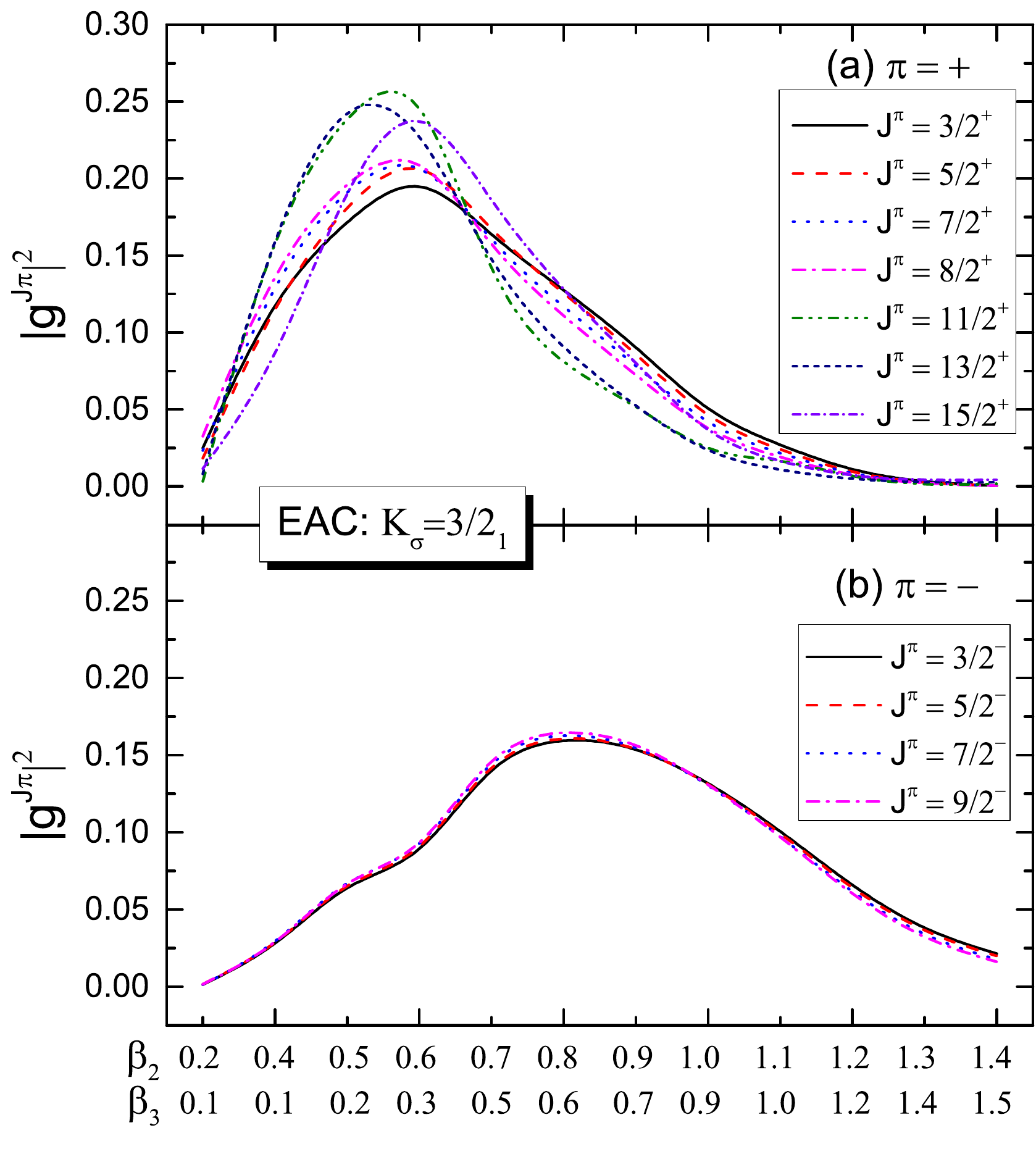}
\caption{\label{fig:Ne21_collective_wfs} (Color online) 
The distributions of the collective wave functions $|g^{J\pi}|^2$ for the low-lying parity-doublet states in  $^{21}$Ne as a function of the deformation parameters $(\beta_2, \beta_3)$ of mean-field configurations by the FQV scheme.
}
\end{figure}

Figure~\ref{fig:Ne21_spectra} shows the comparison of the energy spectra of low-lying parity-doublet states with $K=3/2$ in $^{21}$Ne from both configuration-mixing GCM and single-configuration calculations with the data.  Since the GCM calculation with the mixing of all the deformed  states in the quadrupole-octupole plane is time consuming, only the optimal configurations indicated with red squares in  Fig.~\ref{fig:Ne21_energy_surface_PNP} are included, {similar to Ref.\cite{Xia:2023}.}  It is seen from Fig.~\ref{fig:Ne21_spectra} that the positive-parity states are reasonably reproduced in both calculations. In contrast, the excitation energies of the negative-parity states are significantly overestimated by the single-configuration calculation. With the configuration mixing, the excitation energies of negative-parity states decrease by about 2.0 MeV. Moreover, it is interesting to note that the low-lying energy spectra of \nuclide[21]{Ne} differ from those of $^{21}_\Lambda$Ne~\cite{Xia:2023}. In $^{21}_\Lambda$Ne, the positive parity states follow the pattern $1/2^+, (3/2^+, 5/2^+), (7/2^+, 9^+/2), \cdots$, while the negative parity states follow $(1/2^-, 3/2^-), (5/2^-, 7/2^-), \cdots$. Notably, the two states with $\Delta J=1$ in the parentheses are near degenerate. In contrast, such degeneracy is not observed in \nuclide[21]{Ne}. This distinction arises from the fact that the $\Lambda$ hyperon in $^{21}_\Lambda$Ne is exempt from the Pauli exclusion principle, thus favoring occupation of the lowest-energy $s_{1/2}$ orbital with $K=1/2$. The coupling of the $\Lambda$ with nuclear core states results in nearly degenerate doublets. Conversely, for \nuclide[21]{Ne}, the lowest-energy orbital for the valence neutron is the multiplet of the $d_{5/2}$ orbital with $K=3/2$. We note that the excitation energies of negative-parity states are overestimated. From the recent study on $^{21}_\Lambda$Ne~\cite{Xia:2023}, one anticipates that the mixing of quasiparticle excitation configurations will lower down the negative-parity states significantly.

\begin{figure}[]
\centering
\includegraphics[width=0.8\columnwidth]{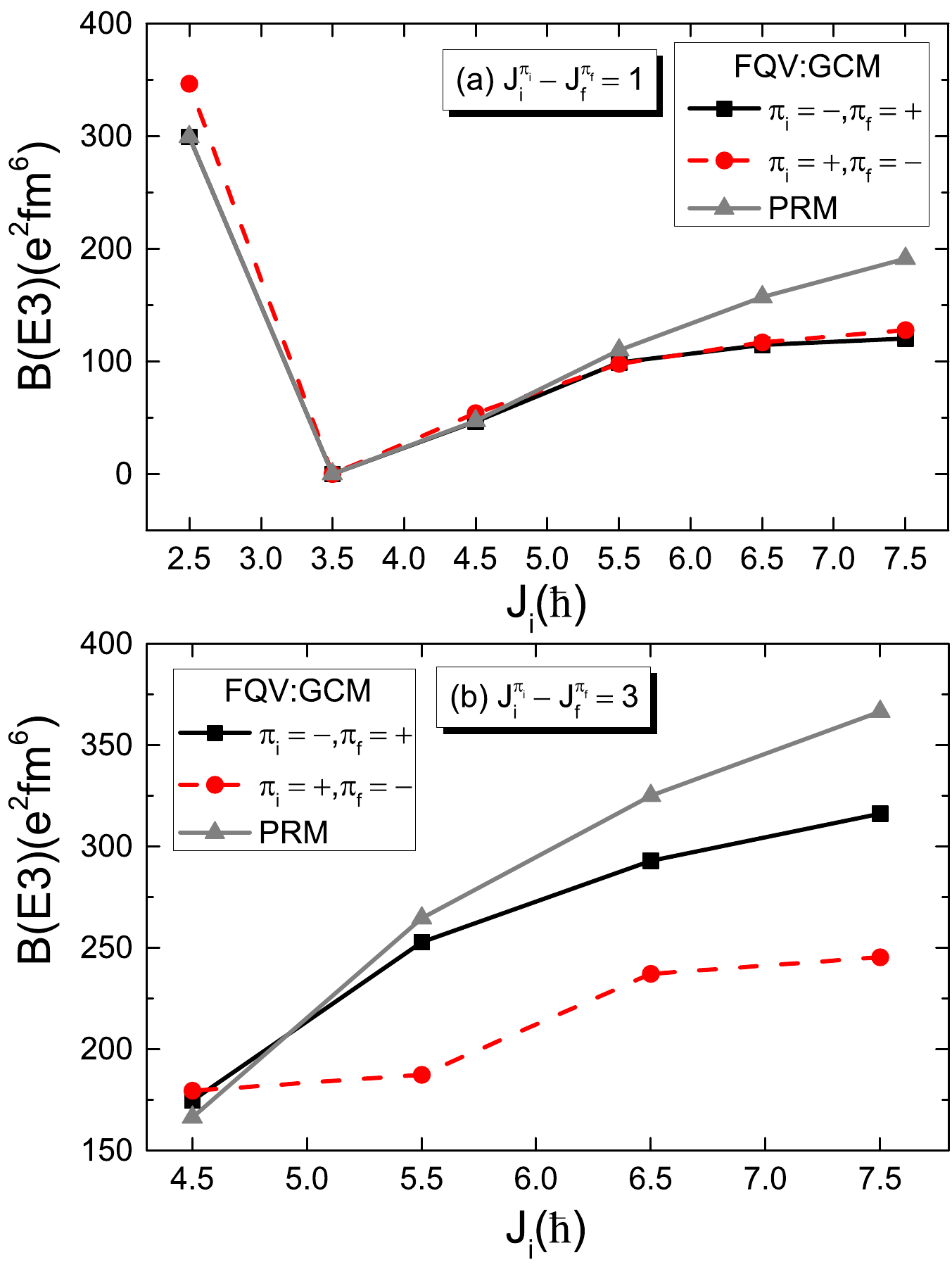}
\caption{\label{fig:BE3_PRM_comparison} (Color online) The $E3$ transition strengths $B(E3, J_i^{\pi_i}\to J_f^{\pi_f})$ (e$^2$fm$^4$) between the low-lying states in the $K_\sigma=3/2_1$ band of \nuclide[21]{Ne} from different calculations as a function of the angular momentum $J_i$ of the initial state, in comparison with the  PRM values (\ref{eq:BE3_PRM}). See main text for details.
}
\end{figure} 

 The collective wave functions of the parity-doublet states  in $\nuclide[21]{Ne}$ are shown in Fig.~\ref{fig:Ne21_collective_wfs}. The positive-parity states are dominated by the configurations around $\beta_2\simeq0.6, \beta_3=0.3$, while the negative-parity states by the configurations around $\beta_2\simeq0.8, \beta_3=0.6$, consistent with the location of the energy minima in Fig.~\ref{fig:Ne21_energy_surface_projection}. With the increase of the angular momentum, the distributions of the collective wave functions of positive-parity states are changing slightly, while those of negative-parity states are barely changed. 

\begin{table}[h!]
\centering
    \tabcolsep=3pt
    \caption{The spectroscopic quadrupole moments $Q^{\rm s}(J^\pi)$, electric dipole ($\lambda=1$) and quadrupole  ($\lambda=2$) transition strengths $B(E\lambda, J^{\pi}_i \to J^{\pi}_f)$, and magnetic dipole moments $\mu(J^\pi)$ and magnetic dipole transition strengths $B(M1, J^\pi_i \to J^\pi_f)$ between the low-lying states in $^{21}$Ne from the MR-CDFT calculation, where the mean-field configurations are constructed in the FQV scheme. The data are taken from Refs.~\cite{Firestone:2015,NNDC}. }
    \begin{tabular}{ccccccc}
      \hline \hline\\
        \multicolumn{3}{c}{$Q^{\rm s}(J^\pi)$ ($e$fm$^2$)}  &  &   \multicolumn{3}{c}{$B(E2, J^\pi_i\to J^\pi_f)$ ($e^2$fm$^4$)} \\
         \cline{1-3}   \cline{5-7} \\
        $J^\pi$ & Exp. & CDFT  &  & $J^\pi_i\to J^\pi_f$  & Exp.  & CDFT \\  
        \hline
        $3/2^+$ & $10.3(8)$ & $10.9$ & & $5/2^+\to 3/2^+$ & $83.6(61)$ & $102.4$\\
        $5/2^+$ & - & $-4.0$ & & $7/2^+\to 3/2^+$ & 32.0(27) & $42.1$\\ 
        $7/2^+$ & - & $-10.7$ & & $7/2^+\to 5/2^+$ & 37.8(137) & $60.2$\\ 
        $9/2^+$ & - & $-14.3$ & & $9/2^+\to 5/2^+$ & $54.0(76)$ & $59.6$\\
        $11/2^+$ & - & $-15.9$ & & $9/2^+\to 7/2^+$ & 31.0(172) & $36.8$\\ 
         $13/2^+$ & - & $-17.3$ & & $11/2^+\to 7/2^+$ & - & $66.9$\\
         $15/2^+$ & - & $-23.0$ & & $11/2^+\to 9/2^+$ & 20.6(130) & $23.2$\\
          &  &  & & $13/2^+\to 9/2^+$ & - & $70.4$\\
          \hline
         $3/2^-$ & - & $14.4$ & & $5/2^-\to 3/2^-$ & - & $177.9$\\ 
         $5/2^-$ & - & $-5.2$ & & $7/2^-\to 3/2^-$ & - & $73.4$\\ 
         $7/2^-$ & - & $-14.4$ & & $7/2^-\to 5/2^-$ & - & $109.1$\\
        $9/2^-$ & - & $-19.6$ & & $9/2^-\to 5/2^-$ & - & $109.3$\\ 
         &   &   & & $9/2^-\to 7/2^-$ & - & $69.8$\\ 
              \hline \hline\\
        \multicolumn{3}{c}{$\mu(J^\pi)(\mu_N)$}  &  &   \multicolumn{3}{c}{$B(M1, J^\pi_i\to J^\pi_f)$ ($\mu_N^2$)} \\
         \cline{1-3}   \cline{5-7} \\
        $J^\pi$ & Exp. & CDFT  &  & $J^\pi_i\to J^\pi_f$  & Exp.  & CDFT \\  
        \hline
        $3/2^+$ & $-0.66$ & $-0.73$ & & $5/2^+\to 3/2^+$ & $0.1275(25)$ & $0.19$\\
        $5/2^+$ & 0.49(4) & $0.24$ & & $7/2^+\to 5/2^+$ & 0.2615(21) & $0.24$\\ 
        $7/2^+$ & - & $0.99$ & & $9/2^+\to 7/2^+$ & 0.43(5)& $0.25$\\ 
        $9/2^+$ & - & $1.68$ & & $11/2^+\to 9/2^+$ & $0.36(7)$ & $0.27$\\
        $11/2^+$ & - & $2.46$ & & $13/2^+\to 11/2^+$ & - & $0.33$\\ 
         $13/2^+$ & - & $3.02$ & & $15/2^+\to 13/2^+$ & - & $0.28$\\
         $15/2^+$ & - & $3.27$ & &  & & \\
          \hline
         $3/2^-$ & - & $-0.73$ & & $5/2^-\to 3/2^-$ & - & $0.36$\\ 
         $5/2^-$ & - & $0.19$ & & $7/2^-\to 5/2^-$ & - & $0.48$\\ 
         $7/2^-$ & - & $0.94$ & & $9/2^-\to 7/2^-$ & - & $0.53$\\
        $9/2^-$ & - & $1.60$ & & $11/2^-\to 9/2^-$ & - & $0.57$\\ 
         \hline\hline\\
          \multicolumn{7}{c}{$B(E1, J_i^\pi\to J_f^{\pi}$) ($10^{-4} e^2$fm$^2$)}\\
          \hline
 \multicolumn{7}{c}{
 \begin{tabular}{cccc}
 \quad \quad $J_i^{\pi_i}\to J_f^{\pi_f}$ \quad\quad& $3/2^-\to 5/2^+$ ~~~& $5/2^-\to 3/2^+$ ~~~& $9/2^-\to 7/2^+$ \\ 
 \hline
 EXP. & $1.11(12)$ & $0.68(11)$ & $1.16(2)$ \\
 CDFT & $4.4$ & $3.4$ & $5.4$ \\ 
 \end{tabular}
 }\\
 \hline\hline
    \end{tabular}
    \label{tab:Qs_BE2_21Ne}
\end{table}

 The spectroscopic quadrupole moment $Q^{\rm s}(J^\pi)$ and $E2$ transition strengths of the low-lying states for the parity doublet bands with $K=3/2$ are given in Table~\ref{tab:Qs_BE2_21Ne}. The results from the MR-CDFT calculation are compared with available data. It is shown that the $Q^{\rm s}(3/2^+)$ of the band head state and the transition strengths are reasonably reproduced.  
 Notably, the MR-CDFT captures the observed trend where $E2$ transition strengths with $\Delta J=1$ ($\Delta J=2$) decrease (increase) with the angular momentum $J$. Additionally, $E1$ transitions between positive and negative-parity states are presented, and it is evident that these transitions are reproduced in the same order, albeit with an enhancement factor of about $4-5$.

 The $E3$ transition strengths $B(E3, J^{\pi_i}_i\to J^{\pi_f}_f)$ for 
the transitions $J^+ \to (J-1)^-$ and $J^-\to (J-1)^+$ from the MR-CDFT calculation with the FQV scheme are shown in Fig.~\ref{fig:BE3_PRM_comparison}(a). Those for the transitions  $J^+ \to (J-3)^-$ and $J^-\to (J-3)^+$ are shown in Fig.~\ref{fig:BE3_PRM_comparison}(b).  These states belong to $K^\pi=3/2^\pm$ bands. It is observed that the values for the transitions with $\Delta J=1$ first decrease to zero at $J_i=7/2$ and then increase steady with the increase of $J_i$. In contrast, the values for the transitions with $\Delta J=3$ increases rapidly with $J_i$. These features are also predicted by the PRM, even though the increasing slopes are different quantitatively. In the PRM, the $B(E3)$ is given by
\beq 
\label{eq:BE3_PRM}
B(E3, J_i K^\pi_i \to J_fK^\pi_f) = Q^2_{30} \frac{7}{16\pi} |\langle J_i K  3 0|J_fK\rangle|^2,
\eeq 
where the parameter $Q^2_{30}(=12543 ~e^2{\rm fm^6})$ for \nuclide[21]{Ne} is determined by the  theoretical value 
$B(E3, 5/2^-\to 3/2^+)$ with $K=3/2$. According to the PRM, the systematic of the $E3$ transition strengths with respect to $J_i$ is again governed by the the Clebsch–Gordan coefficient $\langle J_i K 30|J_fK\rangle$, which is zero for $J_i=7/2, J_f=5/2$ and $K=3/2$.

  \section{Summary}
  \label{sec:summary}

The dynamics of odd-mass nuclei are governed by a complex interplay between single-particle and collective motions, leading to intricate low-energy structures characterized by the mixing of numerous nearly degenerate states. Accurately describing these low-energy structures poses a significant challenge for beyond mean-field approaches. In this study, we extended the MR-CDFT  for investigating the low-lying states of odd-mass nuclei, utilizing a relativistic point coupling energy functional. 

In the MR-CDFT, three different schemes were employed to generate quasiparticle configurations with odd-number parity, where a quasiparticle is created on top of three different quasiparticle vacuum states with either an even or odd average nucleon number. For simplicity, we considered only the lowest-energy one-quasiparticle configurations constrained to have various quadrupole or quadrupole-octupole deformations. These quasiparticle configurations are projected onto correct particle numbers, angular momentum, and parity using projection techniques. The wave function of nuclear state is finally constructed as a linear combination of these projected one-quasiparticle configurations within the framework of the generator coordinate method (GCM).

The MR-CDFT has been tested through its application to the low-lying states of \nuclide[25]{Mg} which are organized into the bands labeled with $K^\pi=5/2^+, 1/2^+, 3/2^+$, and $1/2^-$,  respectively. The energy spectra, electric quadrupole, and magnetic dipole transitions have been discussed in comparison with available data. Our results illustrate that the low-lying states of \nuclide[25]{Mg} are reasonably reproduced by all the three schemes, even though the results by the FQV and EFA schemes are closer to each other compared with the EAC scheme. Furthermore, the MR-CDFT was applied to the parity-doublet states in \nuclide[21]{Ne} with $K^\pi=3/2^\pm$. The main findings of the present study are as follows:
 \begin{itemize}
     \item  The determination of the lowest-energy one-quasiparticle configuration for a given $K$ value can be achieved simply through the BCS formula in Eq.(\ref{eq:1qp_E}) for the quasiparticle energy $E_k$. However, it is important to note that  if $K$ value is not specified, the resulting energy ordering of quasiparticle states might differ from that obtained with Eq.(\ref{eq:quasiparticle_energy_PNP}) for the energies of quasiparticle states with  PNP.

     \item  The restoration of rotational symmetry is crucial for accurately describing nuclear magnetic moments and transition strengths. By employing angular-momentum projection, it has been demonstrated that the electromagnetic properties of configurations exhibiting significant prolate deformations closely align with the predictions of the simple particle-rotor model (PRM). However, configurations with oblate or weakly prolate deformations tend to deviate from the predictions of the PRM.

    \item The predominant feature observed in the low-lying parity-doublet states of \nuclide[20]{Ne} is retained in those of \nuclide[21]{Ne}. Specifically, the $E2$ transition strengths between negative-parity states are consistently larger than those between positive-parity states. The excitation energies of negative-parity states are overestimated which is likely improved with the mixing of quasiparticle excitation configurations.
     
 \end{itemize}

Given the success of the current implementation of the MR-CDFT for low-lying states of  \nuclide[25]{Mg} and \nuclide[21]{Ne}, it is interesting to carry out a systematical study of the low-lying states of both even-even and odd-mass nuclei in the isotopes and isotones of interest, examining the evolution of nuclear shell structure and shapes towards driplines.  In particular, our method provides a tool of choice to study nuclear Schiff moments for the odd-mass nuclei with experiment interest, examining the impact of  static and dynamical octupole correlation effects. An accurate value of the Schiff moment is of importance to constrain the low-energy coupling constants of time-reversal violating nucleon-nucleon interactions based on latest measurement on the permanent electric dipole moment in corresponding atoms. In the some of the above applications, we may need to extend the current model space by including higher-order deformations, cranking states and many quasiparticle excitation configurations in the configuration-mixing calculation. Moreover, it is also interesting to extend the MR-CDFT with both the weight function and the mean-field configurations optimized simultaneously during the variation procedure as it has been proposed recently with a Skyrme EDF~\cite{Matsumoto:2023}.

\section*{Acknowledgments} 
 
 We would like to thank P. Ring for continuous support and encouragement throughout this work, and Q. B. Chen, H. Hergert, C. F. Jiao, and L. J. Wang  for fruitful discussions at different stages.  This work is partly supported by the National Natural Science Foundation of China (Grant Nos.  12375119, 12141501, and 12005804), Guangdong Basic and Applied Basic Research Foundation (2023A1515010936) and the Fundamental Research Funds for the Central Universities, Sun Yat-sen University.

\appendix

 \section{Evaluation of kernels}
 \label{app:kernels}

In this section, we present the detailed formulas for the norm and Hamiltonian kernels for the case that the mean-field configurations are restricted to have axial symmetry and can be labeled with   quadrupole-octupole deformation parameters, i.e., $\mathbf{q}=(\beta_{2}, \beta_{3}$). In this case, there is no $K$ mixing in the wave functions of the low-lying states. The kernels of the operators $\hat O=1, \hat H$ can be written as
 \beqn
\label{eq:H_kernel}
&&\mathscr{O}_{c_ac_b}^{NZJ\pi} \nonumber\\
&=&\cfrac{(2J+1)}{2}\int_0^{2\pi} 
\cfrac{d\varphi_N}{2\pi} e^{-iN\varphi_N}  \int_0^{2\pi} \cfrac{d\varphi_Z}{2\pi}e^{-iZ\varphi_Z}\nonumber\\
&&\times \int_0^\pi d\theta \sin \theta d_{KK}^{J*}(\theta) 
\bra{\Phi^{(\rm OA)}_{\kappa_a}(\mathbf{q}_a)}  \hat O \hat{\mathcal R}(\theta, \varphi_N, \varphi_Z) \hat P^\pi \ket{\Phi^{(\rm OA)}_{\kappa_b}(\mathbf{q}_b)}, \nonumber\\
\eeqn 
where the overlap is defined as
\beqn
&&\bra{\Phi^{(\rm OA)}_{\kappa_a}(\mathbf{q}_a)}  \hat O \hat{\mathcal R}(\theta, \varphi_N, \varphi_Z) \hat P^\pi \ket{\Phi^{(\rm OA)}_{\kappa_b}(\mathbf{q}_b)}\nonumber\\
&=&\frac{1}{2}\bra{\Phi^{(\rm OA)}_{\kappa_a}(\mathbf{q}_a)}  \hat O (1+\pi{\mathscr P}) e^{i\theta\hat J_y} 
e^{-i\varphi_N\hat N} e^{-i\varphi_Z\hat Z}\ket{\Phi^{(\rm OA)}_{\kappa_b}(\mathbf{q}_b)}\nonumber\\
\eeqn

For the norm kernel, the first term in the overlap is simply given by
\beqn
&& \frac{1}{2}\bra{\Phi^{(\rm OA)}_{\kappa_a}(\mathbf{q}_a)}    \hat{\mathcal R}(\theta, \varphi_N, \varphi_Z) \ket{\Phi^{(\rm OA)}_{\kappa_b}(\mathbf{q}_b)}\nonumber\\
&=&\frac{1}{2}\frac{\bra{\Phi^{(\rm OA)}_{\kappa_a}(\mathbf{q}_a)}    \hat{\mathcal R}(g) \ket{\Phi^{(\rm OA)}_{\kappa_b}(\mathbf{q}_b)}}
{\bra{\Phi_{(\kappa_a)}(\mathbf{q_a})}    \hat{\mathcal R}(g) \ket{\Phi_{(\kappa_b)}(\mathbf{q}_b)}}\cdot n(\kappa_a\mathbf{q}_a, \kappa_b\mathbf{q}_b; g),\nonumber\\
\eeqn 
where the norm overlap $n(\kappa_a\mathbf{q}_a, \kappa_b\mathbf{q}_b; g)$ of two quasiparticle vacua with the collective label $g=(\theta, \varphi_N, \varphi_Z)$ is determined by~\cite{Robledo:2009yd}
\beqn 
\label{eq:norm_overlap_pf}
n(\kappa_a\mathbf{q}_a, \kappa_b\mathbf{q}_b; g)
&\equiv& \bra{\Phi_{(\kappa_a)}(\mathbf{q_a})}    \hat{\mathcal R}(g) \ket{\Phi_{(\kappa_b)}(\mathbf{q}_b)}\nonumber\\
&=&(-1)^{M(M+1)/2}{\rm Pf}
\begin{pmatrix}
Z^{(b)}_g & -\mathbb{I} \\
\mathbb{I} & -Z^{(a)\ast}
\end{pmatrix}_{2M\times 2M},\nonumber\\
\eeqn 
 where  $\mathbb{I}$ is a unity matrix $\mathbb{I}_{ij}=\delta_{ij}$, and the skew-matrices are defined as
 \beq 
 Z^{(b)}_g=V(\mathbf{q}_b, g) U^{-1}(\mathbf{q}_b, g),\quad 
 Z^{(a)} = V(\mathbf{q}_a)U^{-1}(\mathbf{q}_a).
 \eeq 
 The normalized norm overlap between two mean-field configurations for odd-mass nuclei can be simplified as
 \beqn 
 && \frac{\bra{\Phi^{(\rm OA)}_{\kappa_a}(\mathbf{q}_a)}    \hat{\mathcal R}(g) \ket{\Phi^{(\rm OA)}_{\kappa_b}(\mathbf{q}_b)}}
{\bra{\Phi_{(\kappa_a)}(\mathbf{q})}    \hat{\mathcal R}(g) \ket{\Phi_{(\kappa_b)}(\mathbf{q}_b)}}\nonumber\\
&=&  \frac{\bra{\Phi_{(\kappa_a)}(\mathbf{q}_a)}  \alpha_{k_a}  \hat{\mathcal R}(g) \beta^\dagger_{k_b}\ket{\Phi_{(\kappa_b)}(\mathbf{q}_b)}}
{\bra{\Phi_{(\kappa_a)}(\mathbf{q})}    \hat{\mathcal R}(g) \ket{\Phi_{(\kappa_b)}(\mathbf{q}_b)}}\nonumber\\
&=&  \bra{\Phi_{(\kappa_a)}(\mathbf{q}_a)}  \alpha_{k_a}   \tilde\beta^\dagger_{k_b}\ket{\Phi_{(\kappa_b)}(\mathbf{q}_b; g)}, 
 \eeqn 
 where the rotated quasiparticle operator is introduced as
 \beq 
 \tilde\beta^\dagger_{k} \equiv 
 \hat{\mathcal R}(g)\beta^\dagger_{k} \hat{\mathcal R}^\dagger(g),
 \eeq  
 and the rotated quasiparticle vacuum 
 \beq 
 \label{eq:rotated_vacuum}
 \ket{\Phi_{(\kappa_b)}(\mathbf{q}_b; g)}
 \equiv  \frac{  \hat{\mathcal R}(g)  \ket{\Phi_{(\kappa_b)}(\mathbf{q}_b)}}
{\bra{\Phi_{(\kappa_a)}(\mathbf{q}_a)}    \hat{\mathcal R}(g) \ket{\Phi_{(\kappa_b)}(\mathbf{q}_b)}}.
 \eeq 
 The quasiparticle operators for the two states $\ket{\Phi_{(\kappa_a)}(\mathbf{q}_a)}$ and $\ket{\Phi_{(\kappa_b)}(\mathbf{q}_b; g)}$ are connected by, 
\begin{eqnarray}
\label{eq:Bogoliubov_transformation_rotation}
\left(
\begin{array}{cc}
\tilde\beta\\
\tilde\beta^{\dag}\\
\end{array}
\right)=\left(
\begin{array}{cc}
\mathbb{U}(g)&\mathbb{V}^{\ast}(g)\\
\mathbb{V}(g)&\mathbb{U}^\ast(g)\\
\end{array}
\right)\left(
\begin{array}{cc}
\alpha\\
\alpha^\dag\\
\end{array}
\right)
\end{eqnarray}
where  
\begin{eqnarray}
\mathbb{U}^\dag=U^\dag(\mathbf{q}_a) U(\mathbf{q}_b, g)+V^\dag(\mathbf{q}_a) V(\mathbf{q}_b, g),
\label{bbU}\\
\mathbb{V}^\dag =U^\dag(\mathbf{q}_a) V^*(\mathbf{q}_b, g)+V^\dag(\mathbf{q}_a) U^*(\mathbf{q}_b, g),
\end{eqnarray}
with the $U(\mathbf{q}, g), V(\mathbf{q}, g)$ for the rotated quasiparticle state $\ket{\Phi_{(\kappa_b)}(\mathbf{q}_b; g)}$ being given by
\begin{eqnarray}
    U(\mathbf{q}, g)=  \mathcal{R}(g)U(\mathbf{q}),\quad 
    V(\mathbf{q}, g)=  \mathcal{R}^\ast(g)V(\mathbf{q}).
\end{eqnarray}
On one hand, one has the following relation, 
 \beqn 
 \label{eq:overlap_2qp_1}
  &&\bra{\Phi_{(\kappa_a)}(\mathbf{q}_a)}  \hat{\mathcal R}(g) \beta_{k}   \beta^\dagger_{l}\ket{\Phi_{(\kappa_b)}(\mathbf{q}_b)}\nonumber\\
  &=&\bra{\Phi_{(\kappa_a)}(\mathbf{q}_a)}  \tilde\beta_{k}    \hat{\mathcal R}(g)\beta^\dagger_{l}\ket{\Phi_{(\kappa_b)}(\mathbf{q}_b)}\nonumber\\
    &=&\sum_{k'}\mathbb{U}_{kk'}(g)\bra{\Phi_{(\kappa_a)}(\mathbf{q}_a)}  \alpha_{k'}    \hat{\mathcal R}(g)\beta^\dagger_{l}\ket{\Phi_{(\kappa_b)}(\mathbf{q}_b)}.
    \eeqn 
 On the other hand, with the anticommutation relation $\{\beta_{k},   \beta^\dagger_{l}\}=\delta_{kl}$, the l.h.s. of the above expression is  simplified as 
 \beqn 
 \label{eq:overlap_2qp_2}
 &&\bra{\Phi_{(\kappa_a)}(\mathbf{q}_a)}  \hat{\mathcal R}(g) \beta_{k}   \beta^\dagger_{l}\ket{\Phi_{(\kappa_b)}(\mathbf{q}_b)}\nonumber\\
 &=&  \bra{\Phi_{(\kappa_a)}(\mathbf{q}_a)}  \hat{\mathcal R}(g) \delta_{kl}\ket{\Phi_{(\kappa_b)}(\mathbf{q}_b)},
 \eeqn 
 where the following relation is used 
\beq
 \beta_k \ket{\Phi_{(\kappa_b)}(\mathbf{q}_b)}=0.
\eeq
 Combining (\ref{eq:overlap_2qp_1}) and (\ref{eq:overlap_2qp_2}), one finds~\cite{Yao:2009traxi}
\beqn 
\label{eq:overlap_1qp}
  \bra{\Phi_{(\kappa_a)}(\mathbf{q}_a)}  \alpha_{k}    \hat{\mathcal R}(g)\beta^\dagger_{l}\ket{\Phi_{(\kappa_b)}(\mathbf{q}_b)}
  &=& [\mathbb{U}^{-1}]_{kl}.
 \eeqn

The second term in the norm kernel contains the space-inversion operator ${\mathscr P}$ which brings one additional phase $(-1)^\ell$ for the $(U, V)$~\cite{Zhou:2023_IJMPE},
\beq
U^{\mathscr{P}}_{mk} = (-1)^{\ell_m} U_{mk},\quad 
V^{\mathscr{P}}_{mk} = (-1)^{\ell_m} V_{mk},
\eeq 
where $m$ and $k$ are the indices for the spherical H.O. basis and quasiparticle state.

In the Hamiltonian kernel, the overlap becomes
\beqn
&&\bra{\Phi^{(\rm OA)}_{\kappa_a}(\mathbf{q}_a)}  \hat H \hat{\mathcal R}(\theta, \varphi_N, \varphi_Z) \hat P^\pi \ket{\Phi^{(\rm OA)}_{\kappa_b}(\mathbf{q}_b)}\nonumber\\
&=&\frac{1}{2}\bra{\Phi^{(\rm OA)}_{\kappa_a}(\mathbf{q}_a)}  \hat H (1+\pi{\mathscr P}) e^{i\theta\hat J_y} 
e^{-i\varphi_N\hat N} e^{-i\varphi_Z\hat Z}\ket{\Phi^{(\rm OA)}_{\kappa_b}(\mathbf{q}_b)},\nonumber\\
\eeqn 
where the first term is determined by
\beqn 
\label{eq:Hamiltonian_overlap_rotation}
&&\frac{1}{2}\bra{\Phi^{(\rm OA)}_{\kappa_a}(\mathbf{q}_a)}  \hat H e^{i\theta\hat J_y} 
e^{-i\varphi_N\hat N} e^{-i\varphi_Z\hat Z}\ket{\Phi^{(\rm OA)}_{\kappa_b}(\mathbf{q}_b)} \nonumber\\
&=&\frac{1}{2}
\bra{\Phi^{(\rm OA)}_{\kappa_a}(\mathbf{q}_a) }\hat H \hat{{\mathcal R}}(g)\ket{\Phi^{(\rm OA)}_{\kappa_b}(\mathbf{q}_b;g)} \cdot n(\kappa_a\mathbf{q}_a, \kappa_b\mathbf{q}_b; g).\nonumber\\
\eeqn 
Here, $n(\kappa_a\mathbf{q}_a, \kappa_b\mathbf{q}_b; g)$ has been given in (\ref{eq:norm_overlap_pf}). Besides, we have defined the rotated one-quasiparticle state 
 \beq 
 \label{eq:rotated_vacuum_1qp}
 \ket{\Phi^{\rm (OA)}_{\kappa_b}(\mathbf{q}; g)}
 \equiv  \frac{  \hat{\mathcal R}(g)  \ket{\Phi^{(\rm OA)}_{\kappa_b}(\mathbf{q})}}
{\bra{\Phi^{(\rm OA)}_{\kappa_a}(\mathbf{q}_a)}\hat{\mathcal R}(g)   \ket{\Phi^{(\rm OA)}_{\kappa_b}(\mathbf{q}_b)}}.
 \eeq

\section{Calculation of the Hamiltonian overlaps}
\label{app:overlaps}

With the help of generalized Wick theorem (GWT)~\cite{Balian:1969}, the Hamiltonian overlap in (\ref{eq:Hamiltonian_overlap_rotation}) for a given Hamiltonian composed of one-body and two-body interactions 
\beq 
\hat H = \sum_{ij} t_{ij} a^\dagger_i a_j + \frac{1}{4}\sum_{ijkl}
V_{ijkl}a^\dagger_i a^\dagger_j a_l a_k, 
\eeq 
is given by
\beqn 
\label{appendix_eq:hamiltonian_overlap}
&&\bra{\Phi^{(\rm OA)}_{\kappa_a}(\mathbf{q}_a) }\hat H \hat{{\mathcal R}}(g)\ket{\Phi^{(\rm OA)}_{\kappa_b}(\mathbf{q}_b;g)} \nonumber\\
&=&  \sum_{ij} t_{ij} \rho^{(ab)}_{ji}(g)+\frac{1}{2}\sum_{ijkl}
V_{ijkl} \rho^{(ab)}_{ki}(g)\rho^{(ab)}_{lj}(g) \nonumber\\
&&+\frac{1}{4}\sum_{ijkl}
V_{ijkl}{\mathcal K}^{(ab)\ast}_{ij}(g){\mathcal K}^{(ab)}_{kl}(g),
\eeqn 
where the one-body mixed density and pairing tensor are defined as
\bsub
\beqn 
\label{eq:one_body_density_g}
\rho^{(ab)}_{ji}(g)
&=& \bra{\Phi^{(\rm OA)}_{\kappa_a}(\mathbf{q}_a) } a^\dagger_i a_j \ket{\Phi^{(\rm OA)}_{\kappa_b}(\mathbf{q}_b;g)},\\
{\mathcal K}^{(ab)}_{ji}(g)
&=& \bra{\Phi^{(\rm OA)}_{\kappa_a}(\mathbf{q}_a) } a_i a_j  \ket{\Phi^{(\rm OA)}_{\kappa_b}(\mathbf{q}_b;g)},\\
{\mathcal K}^{(ab)\ast}_{ij}(g)
&=& \bra{\Phi^{(\rm OA)}_{\kappa_a}(\mathbf{q}_a) } a^\dagger_i a^\dagger_j  \ket{\Phi^{(\rm OA)}_{\kappa_b}(\mathbf{q}_b;g)}.
\eeqn 
\esub

According to the mix-density prescription~\cite{Robledo:2019JPG}, in the EDF-based calculation, the Hamiltonian overlap is replaced with an EDF which becomes a functional of mixed densities, pairing tensors and currents,
\beqn 
\bra{\Phi^{(\rm OA)}_{\kappa_a}(\mathbf{q}_a) }\hat H \hat{{\mathcal R}}(g)\ket{\Phi^{(\rm OA)}_{\kappa_b}(\mathbf{q}_b;g)} 
=E[\rho(g), \nabla^2\rho(g), \cdots].
\eeqn 

The matrix element of the mixed density can be evaluated with the GWT, 
\beqn
\label{eq:matrix_mxied_density}
\rho^{(ab)}_{ji}(g)
&=&\frac{\bra{\Phi_{(\kappa_a)}(\mathbf{q}_a)}  \alpha_{k_a}  a^\dagger_i a_j \hat{\mathcal R}(g) \beta^\dagger_{k_b}\ket{\Phi_{(\kappa_b)}(\mathbf{q}_b)}}
{\bra{\Phi_{(\kappa_a)}(\mathbf{q})}    \hat{\mathcal R}(g) \ket{\Phi_{(\kappa_b)}(\mathbf{q}_b)}}\nonumber\\
&&\times \frac{\bra{\Phi_{(\kappa_a)}(\mathbf{q})}    \hat{\mathcal R}(g) \ket{\Phi_{(\kappa_b)}(\mathbf{q}_b)}}
{\bra{\Phi_{(\kappa_a)}(\mathbf{q})} \alpha_{\kappa_a}   \hat{\mathcal R}(g) \beta^\dagger_{\kappa_b}\ket{\Phi_{(\kappa_b)}(\mathbf{q}_b)}}\nonumber\\
&=& \frac{1}{[\mathbb{U}^{-1}]_{k_a, k_b}} \bra{\Phi_{(\kappa_a)}(\mathbf{q}_a)}  \alpha_{k_a}  a^\dagger_i a_j   
\tilde\beta^\dagger_{k_b}\ket{\Phi_{(\kappa_b)}(\mathbf{q}_b, g)}\nonumber\\
&=&  \frac{1}{[\mathbb{U}^{-1}]_{k_a, k_b}} 
{\rm Pf}\Bigg(\mathcal{S}^\rho_{[\kappa_a, i, j, \kappa_b]}\Bigg),
\eeqn  
where the expression for the overlap of one-quasiparticle wave functions (\ref{eq:overlap_1qp}) is used. The elements of the $4\times 4$ skew matrix $\mathcal{S}^\rho_{[\kappa_a, i, j, \kappa_b]}$ are the contractions of two different operators with respect to quasiparticle vacuum states, determined by 
\begin{eqnarray}
[\mathcal{S}^\rho_{12}]_{\kappa_a i}&\equiv & \bra{\Phi_{(\kappa_a)}(\mathbf{q}_a)} \alpha_{\kappa_a} a_i^\dag \ket{\Phi_{(\kappa_b)}(\mathbf{q}_b, g)} \nonumber\\
&=& [\mathbb{U}^{-1}U^\dagger(\mathbf{q}_b,g)]_{\kappa_a i},\\ 
~[\mathcal{S}^\rho_{13}]_{\kappa_a j} &\equiv & \bra{\Phi_{(\kappa_a)}(\mathbf{q}_a)} \alpha_{\kappa_a} a_j^\dag \ket{\Phi_{(\kappa_b)}(\mathbf{q}_b, g)} \nonumber\\
&=& [\mathbb{U}^{-1}V^\dagger(\mathbf{q}_b,g)]_{\kappa_a j},\\
~[\mathcal{S}^\rho_{14}]_{\kappa_a,\kappa_b} &\equiv & \bra{\Phi_{(\kappa_a)}(\mathbf{q}_a)} \alpha_{\kappa_a} \tilde \beta_{\kappa_b}^\dag \ket{\Phi_{(\kappa_b)}(\mathbf{q}_b, g)} \nonumber\\
&=&[\mathbb{U}^{-1}]_{\kappa_a,\kappa_b}\\
~[\mathcal{S}^\rho_{23}]_{ij} &\equiv & \bra{\Phi_{(\kappa_a)}(\mathbf{q}_a)}  a^\dagger_i a_j  \ket{\Phi_{(\kappa_b)}(\mathbf{q}_b, g)} \nonumber\\
&=&[V(\mathbf{q}_a)\mathbb{U}^{-1}V^\dagger(\mathbf{q}_b,g)]_{ij}\\
~[\mathcal{S}^\rho_{24}]_{i\kappa_b} &\equiv & \bra{\Phi_{(\kappa_a)}(\mathbf{q}_a)}  a^\dagger_i   \tilde \beta_{\kappa_b}^\dag  \ket{\Phi_{(\kappa_b)}(\mathbf{q}_b, g)} \nonumber\\
&=&[V(\mathbf{q}_a)\mathbb{U}^{-1}]_{i\kappa_b}\\
~[\mathcal{S}^\rho_{34}]_{j\kappa_b} &\equiv & \bra{\Phi_{(\kappa_a)}(\mathbf{q}_a)}  a_j   \tilde \beta_{\kappa_b}^\dag  \ket{\Phi_{(\kappa_b)}(\mathbf{q}_b, g)} \nonumber\\
&=& [U(\mathbf{q}_a)\mathbb{U}^{-1}]_{j\kappa_b},
\end{eqnarray}
where the quasiparticle operators $(\alpha,\alpha^\dag)$, the rotated quasiparticle operators $(\tilde\alpha,\tilde\alpha^\dag)$, and the particle operators $(a,a^\dag)$ are connected by the Bogoliubov transformation. 

The matrix elements for the mixed pairing tensor are given by,
 \beqn
 \label{eq:matrix_mxied_tensor}
{\mathcal K}^{(ab)}_{ji}(g)
&=&\frac{\bra{\Phi_{(\kappa_a)}(\mathbf{q}_a)}  \alpha_{k_a}  a_i a_j \hat{\mathcal R}(g) \beta^\dagger_{k_b}\ket{\Phi_{(\kappa_b)}(\mathbf{q}_b)}}
{\bra{\Phi_{(\kappa_a)}(\mathbf{q})}    \hat{\mathcal R}(g) \ket{\Phi_{(\kappa_b)}(\mathbf{q}_b)}}\nonumber\\
&&\times \frac{\bra{\Phi_{(\kappa_a)}(\mathbf{q})}    \hat{\mathcal R}(g) \ket{\Phi_{(\kappa_b)}(\mathbf{q}_b)}}
{\bra{\Phi_{(\kappa_a)}(\mathbf{q})} \alpha_{\kappa_a}   \hat{\mathcal R}(g) \beta^\dagger_{\kappa_b}\ket{\Phi_{(\kappa_b)}(\mathbf{q}_b)}}\nonumber\\
&=& \frac{1}{[\mathbb{U}^{-1}]_{k_a, k_b}} \bra{\Phi_{(\kappa_a)}(\mathbf{q}_a)}  \alpha_{k_a}  a_i a_j   
\tilde\beta^\dagger_{k_b}\ket{\Phi_{(\kappa_b)}(\mathbf{q}_b, g)}\nonumber\\
&=&  \frac{1}{[\mathbb{U}^{-1}]_{k_a, k_b}} 
{\rm Pf}\Bigg(\mathcal{S}^{\mathcal K}_{[\kappa_a, i, j, \kappa_b]}\Bigg),
\eeqn 
where
\begin{eqnarray}
[\mathcal{S}^{\mathcal K}_{12}]_{\kappa_a i}
&=& [\mathbb{U}^{-1}V^\dagger(\mathbf{q}_b,g)]_{\kappa_a i},\\
~[\mathcal{S}^{\mathcal K}_{13}]_{\kappa_a j}&=&[\mathcal{S}^{\mathcal K}_{12}]_{\kappa_a j},\\
~[\mathcal{S}^{\mathcal K}_{14}]_{\kappa_a \kappa_b} &=&[\mathcal{S}^\rho_{14}]_{\kappa_a \kappa_b}, \\
~[\mathcal{S}^{\mathcal K}_{23}]_{ij} &=&  [U(\mathbf{q}_a)\mathbb{U}^{-1}V^\dagger(\mathbf{q}_b,g)]_{ij}, \\
~[\mathcal{S}^{\mathcal K}_{24}]_{i\kappa_b}  &=& [\mathcal{S}^\rho_{34}]_{i\kappa_b},\\
~[\mathcal{S}^{\mathcal K}_{34}]_{j\kappa_b} &=& [\mathcal{S}^{\mathcal K}_{24}]_{j\kappa_b}.
\end{eqnarray}
 Similarly, one finds 
 \beqn
 \label{eq:matrix_mxied_density_conjugate}
{\mathcal K}^{(ab)\ast}_{ij}(g)
&=& \frac{1}{[\mathbb{U}^{-1}]_{k_a, k_b}} \bra{\Phi_{(\kappa_a)}(\mathbf{q}_a)}  \alpha_{k_a}  a^\dagger_i a^\dagger_j   
\tilde\beta^\dagger_{k_b}\ket{\Phi_{(\kappa_b)}(\mathbf{q}_b, g)}\nonumber\\
&=&  \frac{1}{[\mathbb{U}^{-1}]_{k_a, k_b}} 
{\rm Pf}\Bigg(\mathcal{S}^{{\mathcal K}\ast}_{[\kappa_a, i, j, \kappa_b]}\Bigg),
\eeqn 
where
\begin{eqnarray}
~[\mathcal{S}^{{\mathcal K}\ast}_{12}]_{\kappa_a i}
&=& [\mathcal{S}^{\rho}_{12}]_{\kappa_a i},\\
~[\mathcal{S}^{{\mathcal K}\ast}_{13}]_{\kappa_a j}&=&[\mathcal{S}^{{\mathcal K}\ast}_{12}]_{\kappa_a j},\\
~[\mathcal{S}^{\mathcal K\ast}_{14}]_{\kappa_a \kappa_b} &=&[\mathcal{S}^\rho_{14}]_{\kappa_a \kappa_b}\nonumber\\
~[\mathcal{S}^{{\mathcal K}\ast}_{23}]_{ij} &=&  [V(\mathbf{q}_a)\mathbb{U}^{-1}U^\dagger(\mathbf{q}_b,g)]_{ij}, \\
~[\mathcal{S}^{{\mathcal K}\ast}_{24}]_{i\kappa_b}  &=& [\mathcal{S}^\rho_{24}]_{i\kappa_b},\\
~[\mathcal{S}^{{\mathcal K}\ast}_{34}]_{j\kappa_b} &=& [\mathcal{S}^{{\mathcal K}\ast}_{24}]_{j\kappa_b}.
\end{eqnarray}

For the most simple case that there is no rotation, i.e., ${\mathcal R}(g)=1$, and the bra and ket states are the same $\ket{\Phi(\mathbf{q}_a)}=\ket{\Phi(\mathbf{q}_b)}=\ket{\Phi(\mathbf{q})}$, one has $\mathbb{U}^{-1}=\mathbb{I}$, and obtains the density matrix for the one-quasiparticle state $\ket{\Phi^{\rm (OA)}_\kappa(\mathbf{q})}$,
\beqn  
\rho_{ji}
&=&{\rm Pf}\Bigg(\mathcal{S}^\rho_{[\kappa, i, j, \kappa]}\Bigg)\nonumber\\
&=& [\mathcal{S}^\rho_{12}]_{\kappa i}[\mathcal{S}^\rho_{34}]_{j\kappa}
-[\mathcal{S}^\rho_{13}]_{\kappa_a j}
[\mathcal{S}^\rho_{24}]_{i\kappa_b}
+ [\mathcal{S}^\rho_{14}]_{\kappa \kappa}[\mathcal{S}^\rho_{23}]_{ij}\nonumber\\
&=&U^\dagger_{\kappa i}  U_{j\kappa} 
-V^\dagger_{\kappa j}  V_{i\kappa} 
+[V^\ast V^T]_{ji},
\eeqn 
and the pairing tensors
\beqn  
{\mathcal K}_{ji}
&=&{\rm Pf}\Bigg(\mathcal{S}^{\mathcal K}_{[\kappa, i, j, \kappa]}\Bigg)\nonumber\\
&=& [\mathcal{S}^{\mathcal K}_{12}]_{\kappa i}[\mathcal{S}^{\mathcal K}_{34}]_{j\kappa}
-[\mathcal{S}^{\mathcal K}_{13}]_{\kappa j}
[\mathcal{S}^{\mathcal K}_{24}]_{i\kappa}
+ [\mathcal{S}^{\mathcal K}_{14}]_{\kappa \kappa}[\mathcal{S}^{\mathcal K}_{23}]_{ij}\nonumber\\
&=&U_{j\kappa}  V^\dagger_{\kappa i}  
-V^\ast_{j\kappa }  U^T_{\kappa i} 
+[ V^\ast U^T ]_{ji},
\eeqn 
and
\beqn  
{\mathcal K}^\ast_{ij}
&=&{\rm Pf}\Bigg(\mathcal{S}^{{\mathcal K}\ast}_{[\kappa, i, j, \kappa]}\Bigg)\nonumber\\
&=& [\mathcal{S}^{{\mathcal K}\ast}_{12}]_{\kappa i}[\mathcal{S}^{{\mathcal K}\ast}_{34}]_{j\kappa}
-[\mathcal{S}^{{\mathcal K}\ast}_{13}]_{\kappa j}
[\mathcal{S}^{{\mathcal K}\ast}_{24}]_{i\kappa}
+ [\mathcal{S}^{{\mathcal K}\ast}_{14}]_{\kappa \kappa}[\mathcal{S}^{{\mathcal K}\ast}_{23}]_{ij}\nonumber\\
&=&U^\ast_{i\kappa }  V^T_{\kappa j} 
-V_{i\kappa }  U^\dagger_{\kappa j}
+[ VU^\dagger]_{ij},
\eeqn 

As mentioned in the main text, the mixed-density prescription is employed in the MR-CDFT. In other words, the Hamiltonian overlap (\ref{appendix_eq:hamiltonian_overlap}) is replaced by the   energy density functional (\ref{eq:ECDFT}) provided that the densities and pairing tensors are replaced by the mixed ones  as discussed above.

\section{Examination of the problems of singularity  and finite steps in the MR-CDFT calculations}
\label{app:singularity_PNP}
  
 The restoration of broken symmetries with modern nuclear EDFs usually meets the problems of spurious divergences~\cite{Anguiano:2001,Tajima:1992,Dobaczewski:2007}  and finite steps~\cite{Bender:2009,Duguet:2009}. The solution to this problem in a proper way is still an open question. These issues can be avoided if the same interaction is employed for both the particle-hole and particle-particle channels. This is applied for Hamiltonian-based frameworks~\cite{Jiao:2017,Yao:2020,Bally:2021_sd}, but usually not for EDF frameworks.  In this section, we are examining the appearance and possible impact of the singularity problem and finite steps on the results of the CDFT calculation with PNP, where the EAC scheme is employed for \nuclide[25]{Mg}. In this scheme, the norm overlap of one quasiparticle  wave functions for the odd-mass nucleus is given by
\begin{eqnarray} 
  \label{eq:norm_overlap_gauge_angle}
 n(\kappa\mathbf{q},\kappa\mathbf{q}, \varphi) 
 &\equiv &\bra{\Phi^{\rm (OA)}_\kappa(\mathbf{q})} e^{i\varphi \hat N} \ket{\Phi^{\rm (OA)}_{\kappa}(\mathbf{q})} \nonumber\\
 &=& \prod_{i,j>0}\bra{0} (u_i+v_ia_{\bar i}a_i)\alpha_k e^{i\varphi \hat N} \alpha^\dagger_k(u_j+v_ja^\dagger_{j}a^\dagger_{\bar j})\ket{0}\nonumber\\
 &=& \frac{e^{i\varphi}}{(u_k^2+v_k^2e^{2i\varphi})}
  \prod_{j>0}( u^2_j +v^2_j e^{2i\varphi}),
\end{eqnarray}  
where the following relations were used 
\beq 
\alpha_k = u_k a^\dagger_k - v_k a_{\bar k},\quad 
\alpha^\dagger_k = u_k a_k - v_k a^\dagger_{\bar k},
\eeq 
and
 \beqn
 e^{i\varphi \hat N}a_ke^{-i\varphi \hat N}
 =e^{-i\varphi}a_k,\quad 
 e^{i\varphi \hat N}a^\dagger_ke^{-i\varphi \hat N}
 =e^{i\varphi}a^\dagger_k.
 \eeqn

 \begin{figure}[]
 \centering
 \includegraphics[width=0.8\columnwidth]{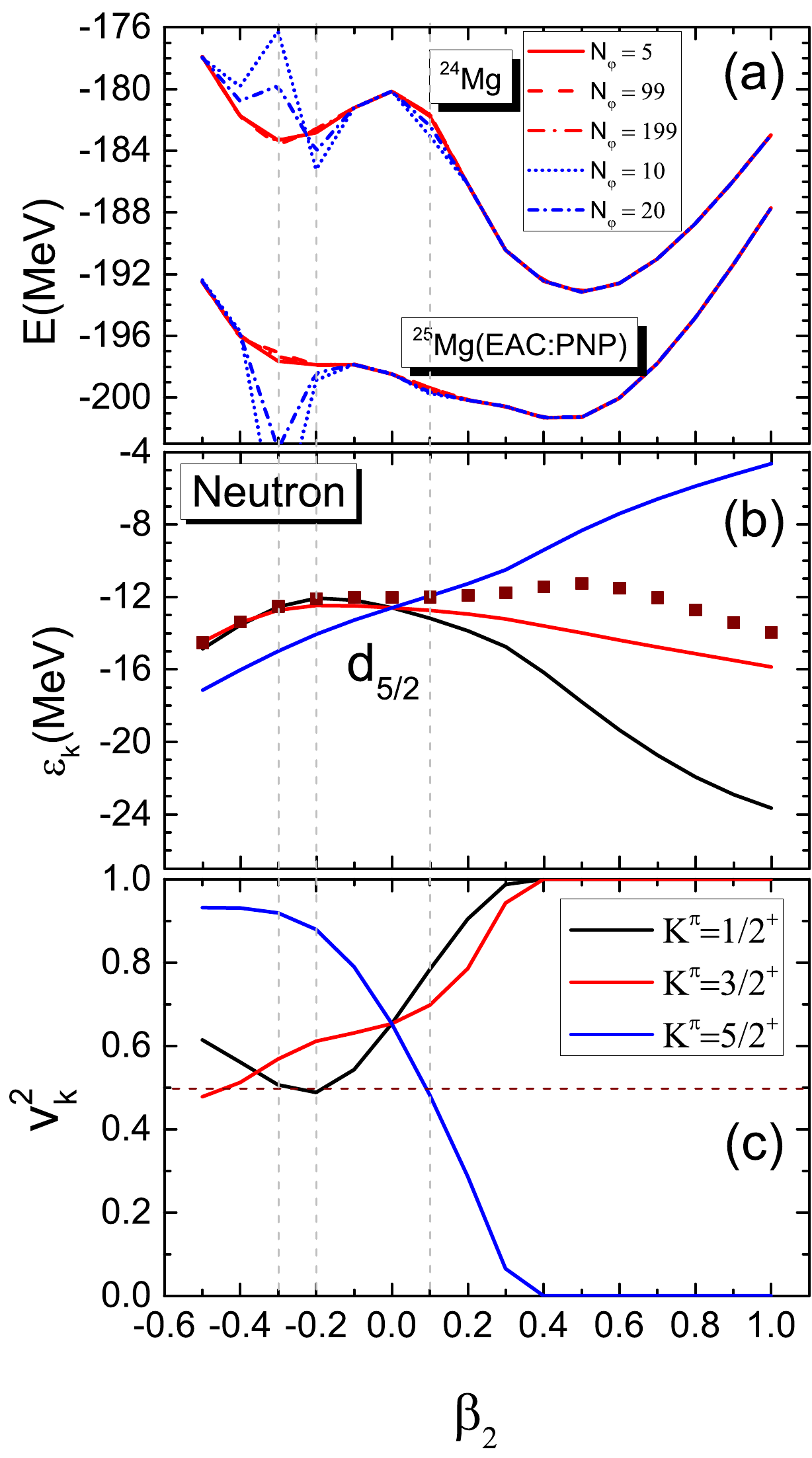} 
 \caption{\label{fig:PES_spe_v2} (Color online)  (a) Particle-number projected energy curves for \nuclide[24]{Mg} and \nuclide[25]{Mg} with an odd number $N_\varphi=5, 99, 199$, or an even number $N_\varphi=10, 20$ respectively. (b) The Nilsson diagram for the single-particle energies of neutron orbitals $\nu d_{5/2}$ in \nuclide[24]{Mg}. (c) The occupation probability of the neutron single-particle states with $K^\pi=1/2^+, 3/2^+$, and $5/2^+$, respectively. }
 \end{figure}
 
 \begin{figure}[]
 \centering
 \includegraphics[width=0.8\columnwidth]{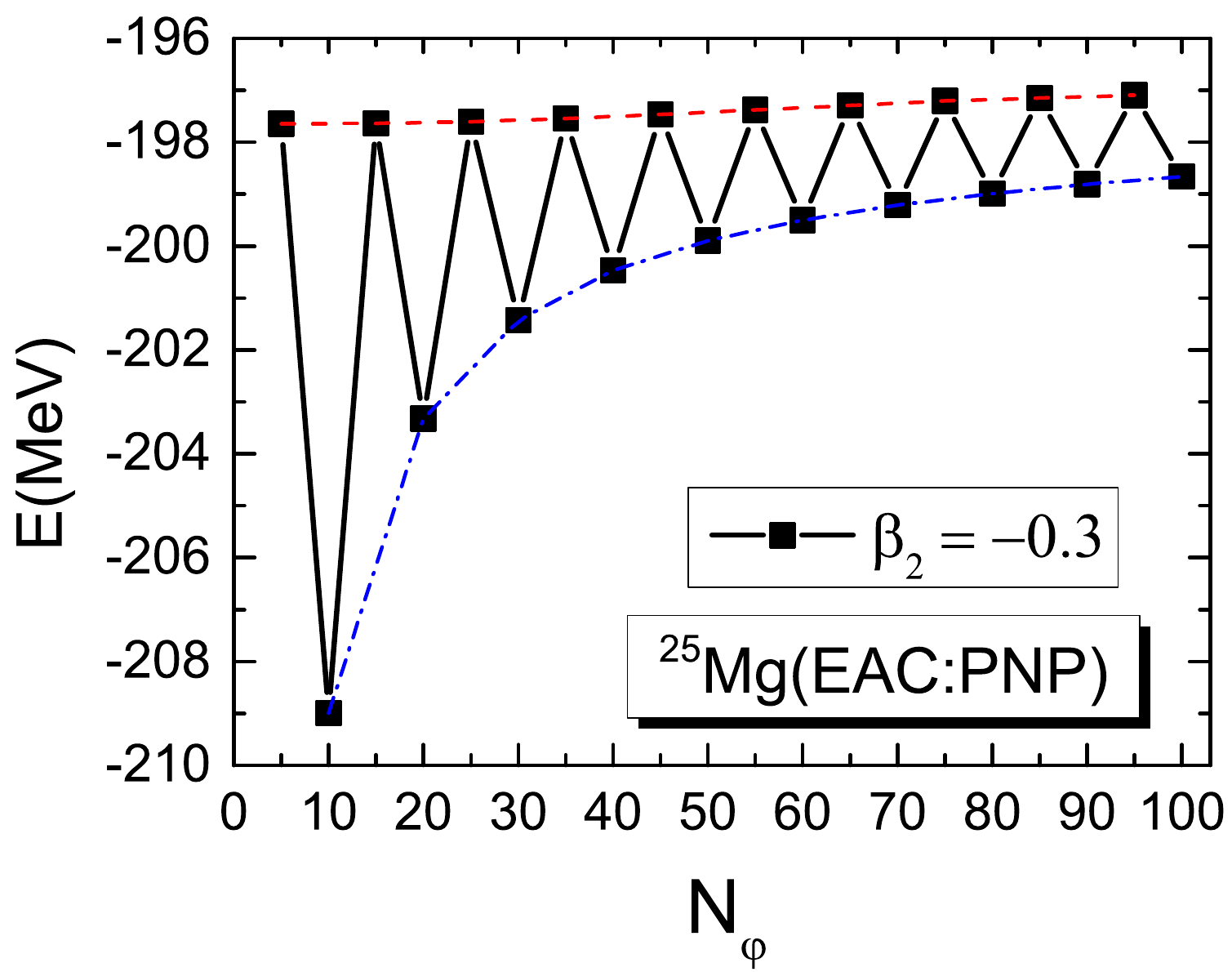} 
 \caption{\label{fig:projected_energy_beta03} (Color online) The energy of the particle-number projected state for \nuclide[25]{Mg} with $\beta_2=-0.3$ as a function of the $N_\varphi$.  }
 \end{figure}

 \begin{figure}[]
 \centering
 \includegraphics[width=\columnwidth]{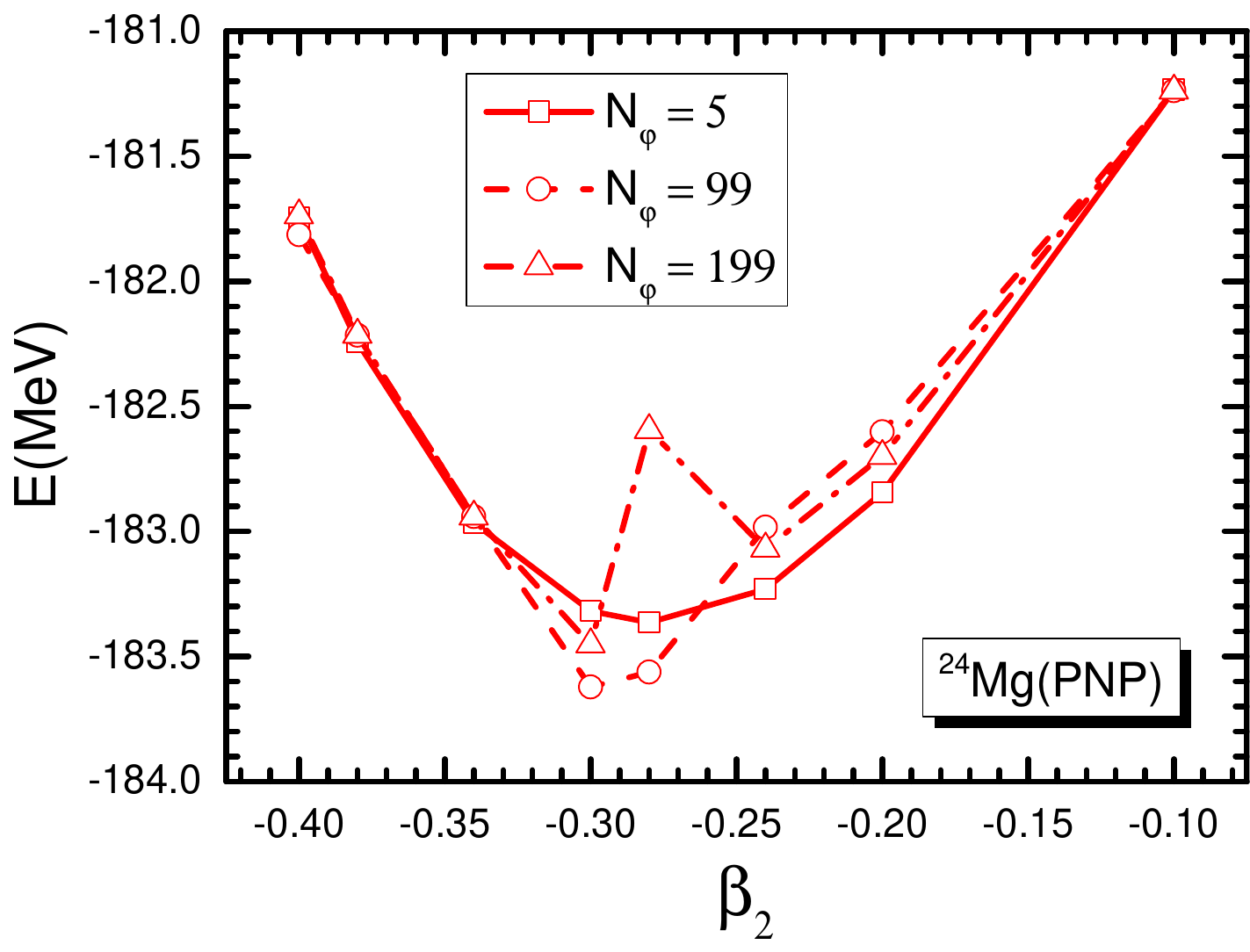} 
 \caption{\label{fig:finite_steps} (Color online) Same as Fig.~\ref{fig:PES_spe_v2}(a), but with denser mesh points around $\beta_2=-0.28$.  }
 \end{figure}

 The energy of the state with projection onto the particle numbers $N_0$ (neutrons and protons) is given by,
 \beqn
 \label{eq:E_PNP_append}
 E^{N_0}
 &=&\frac{\bra{\Phi^{\rm (OA)}_\kappa(\mathbf{q})} \hat H\hat P^{N_0} \ket{\Phi^{\rm (OA)}_{\kappa}(\mathbf{q})} }{\bra{\Phi^{\rm (OA)}_\kappa(\mathbf{q})} \hat P^{N_0} \ket{\Phi^{\rm (OA)}_{\kappa}(\mathbf{q})}}\nonumber\\
 &=&\frac{\int_0^{2\pi} e^{-i\varphi N_0}d\varphi\bra{\Phi^{\rm (OA)}_\kappa(\mathbf{q})} \hat H e^{i\varphi \hat N} \ket{\Phi^{\rm (OA)}_{\kappa}(\mathbf{q})} }
 {\int_0^{2\pi} e^{-i\varphi N_0}d\varphi\bra{\Phi^{\rm (OA)}_\kappa(\mathbf{q})} e^{i\varphi \hat N} \ket{\Phi^{\rm (OA)}_{\kappa}(\mathbf{q})}}\nonumber\\
  &=&\frac{\int_0^{2\pi} e^{-i\varphi N_0}d\varphi E[\rho(\varphi),{\mathcal K}(\varphi),{\mathcal K}^\ast(\varphi)] n(\kappa\mathbf{q},\kappa\mathbf{q}, \varphi) }
 {\int_0^{2\pi} e^{-i\varphi N_0} n(\kappa\mathbf{q},\kappa\mathbf{q}, \varphi) d\varphi}.\nonumber\\
 \eeqn
 Here, according to Eq.(\ref{eq:matrix_mxied_density}), and 
 \beq 
 U(\mathbf{q},\varphi) = e^{i\varphi} U(\mathbf{q}),\quad 
 V(\mathbf{q},\varphi) = e^{-i\varphi} V(\mathbf{q}),
 \eeq 
 the one-body mixed density (\ref{eq:one_body_density_g}) in the canonical basis is simplified as
\beqn 
\rho^{(\kappa\kappa)}_{lm}\left(\varphi\right)
&=&\frac{\bra{\Phi^{\rm (OA)}_\kappa(\mathbf{q})}a_m^{\dagger} a_l e^{i\varphi \hat{N}} \ket{\Phi^{\rm (OA)}_{\kappa}(\mathbf{q})} }
{\bra{\Phi^{\rm (OA)}_\kappa(\mathbf{q})}  e^{i\varphi \hat{N}} \ket{\Phi^{\rm (OA)}_{\kappa}(\mathbf{q})} } \nonumber\\
&=& \delta_{lm}e^{2i\varphi}\Bigg(   \frac{v^2_l +u^2_l\delta_{lk}-v^2_l\delta_{\bar l k}}{u_l^2+v_l^2 e^{2i\varphi}}\Bigg),
\eeqn
 and the pairing tensors from (\ref{eq:matrix_mxied_density}), (\ref{eq:matrix_mxied_density_conjugate}),
 \beqn 
\kappa^{(\kappa\kappa)}_{l \bar m}\left(\varphi\right)
&=&\frac{\bra{\Phi^{\rm (OA)}_\kappa(\mathbf{q})}a_{\bar m}  a_l e^{i\varphi \hat{N}} \ket{\Phi^{\rm (OA)}_{\kappa}(\mathbf{q})} }
{\bra{\Phi^{\rm (OA)}_\kappa(\mathbf{q})}  e^{i\varphi \hat{N}} \ket{\Phi^{\rm (OA)}_{\kappa}(\mathbf{q})} } \nonumber\\
&=& \delta_{lm}u_lv_le^{2i\varphi}\Bigg( \frac{ 1+ \delta_{lk} - \delta_{\bar lk}}{u_l^2+v_l^2 e^{2i\varphi}}\Bigg),
\eeqn 
 \beqn 
\kappa^{(\kappa\kappa)\ast}_{\bar m l}\left(\varphi\right)
&=&\frac{\bra{\Phi^{\rm (OA)}_\kappa(\mathbf{q})}a^\dagger_{\bar m}  a_l e^{i\varphi \hat{N}} \ket{\Phi^{\rm (OA)}_{\kappa}(\mathbf{q})} }
{\bra{\Phi^{\rm (OA)}_\kappa(\mathbf{q})}  e^{i\varphi \hat{N}} \ket{\Phi^{\rm (OA)}_{\kappa}(\mathbf{q})} } \nonumber\\
&=& \delta_{lm}u_lv_l \Bigg( \frac{ 1+ \delta_{lk} - \delta_{\bar lk}}{u_l^2+v_l^2 e^{2 i\varphi}}\Bigg).
\eeqn
Compared to the expressions for even-even nuclei~\cite{Anguiano:2001}, one has additional two terms (the second and third terms in the numerators) for odd-mass nuclei. In both cases, the mixed density is generally a complex number. It may cause an issue in the MR-EDF calculations with a non-integer power of densities~\cite{Bender:2009}.  This issue is not serious in the MR-CDFT as the EDF in (\ref{eq:ECDFT}) contains only integer powers of densities and currents~\cite{Yao:2009traxi,Yao:2010}.

The integration over the gauge angle $\varphi$ in Eq.(\ref{eq:E_PNP_append}) is  usually carried out using the trapezoidal rule based on the Fomenko expansion method~\cite{Fomenko:1970JPA} in which the PNP operator (\ref{eq:PNP_gauge}) becomes  
\begin{equation}
\label{eq:Fomenko}
\hat{P}^{N_\tau}=\frac{1}{N_\varphi} \sum_{m=1}^{N_\varphi} e^{i\left(\hat{N}_\tau-N_\tau\right) \varphi_m}, \quad \varphi_m=\frac{m}{N_\varphi}\pi,
\end{equation}
 where the gauge angle $\varphi_\tau\in[0,\pi]$  is discretized with  $N_\varphi$  points.  In the case with $u_l = v_l$, and $\varphi=\pi/2$, the factor $1/(u_l^2+v_l^2 e^{2 i\varphi})$ in the matrix elements  diverges. This phenomenon happens in the calculation with an even number of $N_\varphi$, as shown in Fig.~\ref{fig:PES_spe_v2}. The energy curves of both $^{24}$Mg and $^{25}$Mg present the divergent behavior around $\beta_2=-0.3$,  where one of the spherical $\nu d_{5/2}$ orbitals has occupation probability $v^2_k$ close to $0.5$.  To avoid the numerical ambiguity from the division by a small number, an odd number of mesh points in the integration over the gauge angles for PNP is usually recommended.  It is seen that the energy curves by the $N_\varphi=5, 99$, and $199$ become normal in the sense the divergence does not show up. To see it more clearly, we plot the energy of the particle-number projected state for \nuclide[25]{Mg} with $\beta_2=-0.3$ as a function of the $N_\varphi$ in   Fig.~\ref{fig:projected_energy_beta03}. Of particular interest is the observation of the staggering of the energy with the choices of odd and even numbers of the $N_\varphi$. One can see that the energy by using an even value of $N_\varphi$ is systematically lower than that by an odd value of $N_\varphi$, and the discrepancy is decreasing with the increase of $N_\varphi$. 
 
 Besides, we also examine the problem of finite steps caused by the self-interaction and self-pairing \cite{Bender:2009} in our MR-CDFT. Figure~\ref{fig:finite_steps} displays the particle-number projected energy curves for \nuclide[24]{Mg}. One can see that the energy curve by $N_\varphi=199$ is not smooth around $\beta_2=-0.28$.  In contrast, the discontinuity in the projected energy curve is not shown evidently in the calculations with a small value of $N_\varphi$ and spare discretization points in the deformation space, cf. Fig.~\ref{fig:PES_spe_v2}. The impact of these problems on the  GCM calculation for the properties of nuclear low-lying states is negligible, as the low-lying states are predominated by the prolate configurations.

 \bibliographystyle{apsrev4-1} 

\begin{thebibliography}{97}%
\makeatletter
\providecommand \@ifxundefined [1]{%
 \@ifx{#1\undefined}
}%
\providecommand \@ifnum [1]{%
 \ifnum #1\expandafter \@firstoftwo
 \else \expandafter \@secondoftwo
 \fi
}%
\providecommand \@ifx [1]{%
 \ifx #1\expandafter \@firstoftwo
 \else \expandafter \@secondoftwo
 \fi
}%
\providecommand \natexlab [1]{#1}%
\providecommand \enquote  [1]{``#1''}%
\providecommand \bibnamefont  [1]{#1}%
\providecommand \bibfnamefont [1]{#1}%
\providecommand \citenamefont [1]{#1}%
\providecommand \href@noop [0]{\@secondoftwo}%
\providecommand \href [0]{\begingroup \@sanitize@url \@href}%
\providecommand \@href[1]{\@@startlink{#1}\@@href}%
\providecommand \@@href[1]{\endgroup#1\@@endlink}%
\providecommand \@sanitize@url [0]{\catcode `\\12\catcode `\$12\catcode
  `\&12\catcode `\#12\catcode `\^12\catcode `\_12\catcode `\%12\relax}%
\providecommand \@@startlink[1]{}%
\providecommand \@@endlink[0]{}%
\providecommand \url  [0]{\begingroup\@sanitize@url \@url }%
\providecommand \@url [1]{\endgroup\@href {#1}{\urlprefix }}%
\providecommand \urlprefix  [0]{URL }%
\providecommand \Eprint [0]{\href }%
\providecommand \doibase [0]{http://dx.doi.org/}%
\providecommand \selectlanguage [0]{\@gobble}%
\providecommand \bibinfo  [0]{\@secondoftwo}%
\providecommand \bibfield  [0]{\@secondoftwo}%
\providecommand \translation [1]{[#1]}%
\providecommand \BibitemOpen [0]{}%
\providecommand \bibitemStop [0]{}%
\providecommand \bibitemNoStop [0]{.\EOS\space}%
\providecommand \EOS [0]{\spacefactor3000\relax}%
\providecommand \BibitemShut  [1]{\csname bibitem#1\endcsname}%
\let\auto@bib@innerbib\@empty
\bibitem [{\citenamefont {Bohr}\ and\ \citenamefont
  {Mottelson}(1998)}]{Bohr:1998}%
  \BibitemOpen
  \bibfield  {author} {\bibinfo {author} {\bibfnamefont {A.}~\bibnamefont
  {Bohr}}\ and\ \bibinfo {author} {\bibfnamefont {B.~R.}\ \bibnamefont
  {Mottelson}},\ }\href@noop {} {\emph {\bibinfo {title} {Nuclear structure.
  Volume II. Nuclear deformations}}}\ (\bibinfo  {publisher} {Word
  Scientific},\ \bibinfo {address} {Singapore},\ \bibinfo {year}
  {1998})\BibitemShut {NoStop}%
\bibitem [{\citenamefont {Ring}\ and\ \citenamefont
  {Schuck}(1980)}]{Ring:1980}%
  \BibitemOpen
  \bibfield  {author} {\bibinfo {author} {\bibfnamefont {P.}~\bibnamefont
  {Ring}}\ and\ \bibinfo {author} {\bibfnamefont {P.}~\bibnamefont {Schuck}},\
  }\href@noop {} {\emph {\bibinfo {title} {The nuclear many-body problem}}}\
  (\bibinfo  {publisher} {Springer-Verlag},\ \bibinfo {address} {New York},\
  \bibinfo {year} {1980})\BibitemShut {NoStop}%
\bibitem [{\citenamefont {Klos}\ \emph {et~al.}(2013)\citenamefont {Klos},
  \citenamefont {Men\'endez}, \citenamefont {Gazit},\ and\ \citenamefont
  {Schwenk}}]{Klos:2013}%
  \BibitemOpen
  \bibfield  {author} {\bibinfo {author} {\bibfnamefont {P.}~\bibnamefont
  {Klos}}, \bibinfo {author} {\bibfnamefont {J.}~\bibnamefont {Men\'endez}},
  \bibinfo {author} {\bibfnamefont {D.}~\bibnamefont {Gazit}}, \ and\ \bibinfo
  {author} {\bibfnamefont {A.}~\bibnamefont {Schwenk}},\ }\href {\doibase
  10.1103/PhysRevD.88.083516} {\bibfield  {journal} {\bibinfo  {journal} {Phys.
  Rev. D}\ }\textbf {\bibinfo {volume} {88}},\ \bibinfo {pages} {083516}
  (\bibinfo {year} {2013})}\BibitemShut {NoStop}%
\bibitem [{\citenamefont {Engel}\ \emph {et~al.}(2013)\citenamefont {Engel},
  \citenamefont {Ramsey-Musolf},\ and\ \citenamefont {van
  Kolck}}]{Engel:2013PPNP}%
  \BibitemOpen
  \bibfield  {author} {\bibinfo {author} {\bibfnamefont {J.}~\bibnamefont
  {Engel}}, \bibinfo {author} {\bibfnamefont {M.~J.}\ \bibnamefont
  {Ramsey-Musolf}}, \ and\ \bibinfo {author} {\bibfnamefont {U.}~\bibnamefont
  {van Kolck}},\ }\href {\doibase 10.1016/j.ppnp.2013.03.003} {\bibfield
  {journal} {\bibinfo  {journal} {Prog. Part. Nucl. Phys.}\ }\textbf {\bibinfo
  {volume} {71}},\ \bibinfo {pages} {21} (\bibinfo {year} {2013})},\ \Eprint
  {http://arxiv.org/abs/1303.2371} {arXiv:1303.2371 [nucl-th]} \BibitemShut
  {NoStop}%
\bibitem [{\citenamefont {Malov}\ \emph {et~al.}(2023)\citenamefont {Malov},
  \citenamefont {Bezbakh}, \citenamefont {Adamian}, \citenamefont {Antonenko},\
  and\ \citenamefont {Jolos}}]{Malov:2023}%
  \BibitemOpen
  \bibfield  {author} {\bibinfo {author} {\bibfnamefont {L.~A.}\ \bibnamefont
  {Malov}}, \bibinfo {author} {\bibfnamefont {A.~N.}\ \bibnamefont {Bezbakh}},
  \bibinfo {author} {\bibfnamefont {G.~G.}\ \bibnamefont {Adamian}}, \bibinfo
  {author} {\bibfnamefont {N.~V.}\ \bibnamefont {Antonenko}}, \ and\ \bibinfo
  {author} {\bibfnamefont {R.~V.}\ \bibnamefont {Jolos}},\ }\href {\doibase
  10.1103/PhysRevC.108.044302} {\bibfield  {journal} {\bibinfo  {journal}
  {Phys. Rev. C}\ }\textbf {\bibinfo {volume} {108}},\ \bibinfo {pages}
  {044302} (\bibinfo {year} {2023})}\BibitemShut {NoStop}%
\bibitem [{\citenamefont {Inglis}(1954)}]{Inglis:1954}%
  \BibitemOpen
  \bibfield  {author} {\bibinfo {author} {\bibfnamefont {D.~R.}\ \bibnamefont
  {Inglis}},\ }\href {\doibase 10.1103/PhysRev.96.1059} {\bibfield  {journal}
  {\bibinfo  {journal} {Phys. Rev.}\ }\textbf {\bibinfo {volume} {96}},\
  \bibinfo {pages} {1059} (\bibinfo {year} {1954})}\BibitemShut {NoStop}%
\bibitem [{\citenamefont {Inglis}(1956)}]{Inglis:1956}%
  \BibitemOpen
  \bibfield  {author} {\bibinfo {author} {\bibfnamefont {D.~R.}\ \bibnamefont
  {Inglis}},\ }\href {\doibase 10.1103/PhysRev.103.1786} {\bibfield  {journal}
  {\bibinfo  {journal} {Phys. Rev.}\ }\textbf {\bibinfo {volume} {103}},\
  \bibinfo {pages} {1786} (\bibinfo {year} {1956})}\BibitemShut {NoStop}%
\bibitem [{\citenamefont {Bonche}\ \emph {et~al.}(1987)\citenamefont {Bonche},
  \citenamefont {Flocard},\ and\ \citenamefont {Heenen}}]{Bonche:1987}%
  \BibitemOpen
  \bibfield  {author} {\bibinfo {author} {\bibfnamefont {P.}~\bibnamefont
  {Bonche}}, \bibinfo {author} {\bibfnamefont {H.}~\bibnamefont {Flocard}}, \
  and\ \bibinfo {author} {\bibfnamefont {P.~H.}\ \bibnamefont {Heenen}},\
  }\href {\doibase 10.1016/0375-9474(87)90331-9} {\bibfield  {journal}
  {\bibinfo  {journal} {Nucl. Phys. A}\ }\textbf {\bibinfo {volume} {467}},\
  \bibinfo {pages} {115} (\bibinfo {year} {1987})}\BibitemShut {NoStop}%
\bibitem [{\citenamefont {Koepf}\ and\ \citenamefont
  {Ring}(1989)}]{Koepf:1989}%
  \BibitemOpen
  \bibfield  {author} {\bibinfo {author} {\bibfnamefont {W.}~\bibnamefont
  {Koepf}}\ and\ \bibinfo {author} {\bibfnamefont {P.}~\bibnamefont {Ring}},\
  }\href {\doibase 10.1016/0375-9474(89)90532-0} {\bibfield  {journal}
  {\bibinfo  {journal} {Nucl. Phys. A}\ }\textbf {\bibinfo {volume} {493}},\
  \bibinfo {pages} {61} (\bibinfo {year} {1989})}\BibitemShut {NoStop}%
\bibitem [{\citenamefont {Konig}\ and\ \citenamefont
  {Ring}(1993)}]{Konig:1993}%
  \BibitemOpen
  \bibfield  {author} {\bibinfo {author} {\bibfnamefont {J.}~\bibnamefont
  {Konig}}\ and\ \bibinfo {author} {\bibfnamefont {P.}~\bibnamefont {Ring}},\
  }\href {\doibase 10.1103/PhysRevLett.71.3079} {\bibfield  {journal} {\bibinfo
   {journal} {Phys. Rev. Lett.}\ }\textbf {\bibinfo {volume} {71}},\ \bibinfo
  {pages} {3079} (\bibinfo {year} {1993})},\ \Eprint
  {http://arxiv.org/abs/nucl-th/9310020} {arXiv:nucl-th/9310020} \BibitemShut
  {NoStop}%
\bibitem [{\citenamefont {Egido}\ and\ \citenamefont
  {Robledo}(1993)}]{Egido:1993}%
  \BibitemOpen
  \bibfield  {author} {\bibinfo {author} {\bibfnamefont {J.~L.}\ \bibnamefont
  {Egido}}\ and\ \bibinfo {author} {\bibfnamefont {L.~M.}\ \bibnamefont
  {Robledo}},\ }\href {\doibase 10.1103/PhysRevLett.70.2876} {\bibfield
  {journal} {\bibinfo  {journal} {Phys. Rev. Lett.}\ }\textbf {\bibinfo
  {volume} {70}},\ \bibinfo {pages} {2876} (\bibinfo {year}
  {1993})}\BibitemShut {NoStop}%
\bibitem [{\citenamefont {Vretenar}\ \emph {et~al.}(2005)\citenamefont
  {Vretenar}, \citenamefont {Afanasjev}, \citenamefont {Lalazissis},\ and\
  \citenamefont {Ring}}]{Vretenar:2005}%
  \BibitemOpen
  \bibfield  {author} {\bibinfo {author} {\bibfnamefont {D.}~\bibnamefont
  {Vretenar}}, \bibinfo {author} {\bibfnamefont {A.~V.}\ \bibnamefont
  {Afanasjev}}, \bibinfo {author} {\bibfnamefont {G.~A.}\ \bibnamefont
  {Lalazissis}}, \ and\ \bibinfo {author} {\bibfnamefont {P.}~\bibnamefont
  {Ring}},\ }\href {\doibase 10.1016/j.physrep.2004.10.001} {\bibfield
  {journal} {\bibinfo  {journal} {Phys. Rept.}\ }\textbf {\bibinfo {volume}
  {409}},\ \bibinfo {pages} {101} (\bibinfo {year} {2005})}\BibitemShut
  {NoStop}%
\bibitem [{\citenamefont {Almberger}\ \emph {et~al.}(1980)\citenamefont
  {Almberger}, \citenamefont {Hamamoto},\ and\ \citenamefont
  {Leander}}]{Almberger:1980}%
  \BibitemOpen
  \bibfield  {author} {\bibinfo {author} {\bibfnamefont {J.}~\bibnamefont
  {Almberger}}, \bibinfo {author} {\bibfnamefont {I.}~\bibnamefont {Hamamoto}},
  \ and\ \bibinfo {author} {\bibfnamefont {G.}~\bibnamefont {Leander}},\ }\href
  {\doibase https://doi.org/10.1016/0375-9474(80)90228-6} {\bibfield  {journal}
  {\bibinfo  {journal} {Nucl. Phys. A}\ }\textbf {\bibinfo {volume} {333}},\
  \bibinfo {pages} {184} (\bibinfo {year} {1980})}\BibitemShut {NoStop}%
\bibitem [{\citenamefont {Bonneau}\ \emph {et~al.}(2007)\citenamefont
  {Bonneau}, \citenamefont {Quentin},\ and\ \citenamefont
  {Moller}}]{Bonneau:2007}%
  \BibitemOpen
  \bibfield  {author} {\bibinfo {author} {\bibfnamefont {L.}~\bibnamefont
  {Bonneau}}, \bibinfo {author} {\bibfnamefont {P.}~\bibnamefont {Quentin}}, \
  and\ \bibinfo {author} {\bibfnamefont {P.}~\bibnamefont {Moller}},\ }\href
  {\doibase 10.1103/PhysRevC.76.024320} {\bibfield  {journal} {\bibinfo
  {journal} {Phys. Rev. C}\ }\textbf {\bibinfo {volume} {76}},\ \bibinfo
  {pages} {024320} (\bibinfo {year} {2007})},\ \Eprint
  {http://arxiv.org/abs/nucl-th/0703092} {arXiv:nucl-th/0703092} \BibitemShut
  {NoStop}%
\bibitem [{\citenamefont {Perez-Martin}\ and\ \citenamefont
  {Robledo}(2008)}]{Martin:2008EFA}%
  \BibitemOpen
  \bibfield  {author} {\bibinfo {author} {\bibfnamefont {S.}~\bibnamefont
  {Perez-Martin}}\ and\ \bibinfo {author} {\bibfnamefont {L.~M.}\ \bibnamefont
  {Robledo}},\ }\href {\doibase 10.1103/PhysRevC.78.014304} {\bibfield
  {journal} {\bibinfo  {journal} {Phys. Rev. C}\ }\textbf {\bibinfo {volume}
  {78}},\ \bibinfo {pages} {014304} (\bibinfo {year} {2008})},\ \Eprint
  {http://arxiv.org/abs/0805.4318} {arXiv:0805.4318 [nucl-th]} \BibitemShut
  {NoStop}%
\bibitem [{\citenamefont {Schunck}\ \emph {et~al.}(2010)\citenamefont
  {Schunck}, \citenamefont {Dobaczewski}, \citenamefont {McDonnell},
  \citenamefont {More}, \citenamefont {Nazarewicz}, \citenamefont {Sarich},\
  and\ \citenamefont {Stoitsov}}]{Schunck:2010EFA}%
  \BibitemOpen
  \bibfield  {author} {\bibinfo {author} {\bibfnamefont {N.}~\bibnamefont
  {Schunck}}, \bibinfo {author} {\bibfnamefont {J.}~\bibnamefont
  {Dobaczewski}}, \bibinfo {author} {\bibfnamefont {J.}~\bibnamefont
  {McDonnell}}, \bibinfo {author} {\bibfnamefont {J.}~\bibnamefont {More}},
  \bibinfo {author} {\bibfnamefont {W.}~\bibnamefont {Nazarewicz}}, \bibinfo
  {author} {\bibfnamefont {J.}~\bibnamefont {Sarich}}, \ and\ \bibinfo {author}
  {\bibfnamefont {M.~V.}\ \bibnamefont {Stoitsov}},\ }\href {\doibase
  10.1103/PhysRevC.81.024316} {\bibfield  {journal} {\bibinfo  {journal} {Phys.
  Rev. C}\ }\textbf {\bibinfo {volume} {81}},\ \bibinfo {pages} {024316}
  (\bibinfo {year} {2010})},\ \Eprint {http://arxiv.org/abs/0910.2164}
  {arXiv:0910.2164 [nucl-th]} \BibitemShut {NoStop}%
\bibitem [{\citenamefont {Pan}\ \emph {et~al.}(2022)\citenamefont {Pan},
  \citenamefont {Cheoun}, \citenamefont {Choi}, \citenamefont {Dong},
  \citenamefont {Du}, \citenamefont {Fan}, \citenamefont {Gao}, \citenamefont
  {Geng}, \citenamefont {Ha}, \citenamefont {He}, \citenamefont {Huang},
  \citenamefont {Huang}, \citenamefont {Kim}, \citenamefont {Kim},
  \citenamefont {Lee}, \citenamefont {Lee}, \citenamefont {Li}, \citenamefont
  {Liu}, \citenamefont {Ma}, \citenamefont {Meng}, \citenamefont {Mun},
  \citenamefont {Niu}, \citenamefont {Papakonstantinou}, \citenamefont {Shang},
  \citenamefont {Shen}, \citenamefont {Shen}, \citenamefont {Sun},
  \citenamefont {Sun}, \citenamefont {Wu}, \citenamefont {Wu}, \citenamefont
  {Xia}, \citenamefont {Yan}, \citenamefont {Yiu}, \citenamefont {Zhang},
  \citenamefont {Zhang}, \citenamefont {Zhang}, \citenamefont {Zhang},
  \citenamefont {Zhao}, \citenamefont {Zheng},\ and\ \citenamefont
  {Zhou}}]{Pan:2022}%
  \BibitemOpen
  \bibfield  {author} {\bibinfo {author} {\bibfnamefont {C.}~\bibnamefont
  {Pan}}, \bibinfo {author} {\bibfnamefont {M.-K.}\ \bibnamefont {Cheoun}},
  \bibinfo {author} {\bibfnamefont {Y.-B.}\ \bibnamefont {Choi}}, \bibinfo
  {author} {\bibfnamefont {J.}~\bibnamefont {Dong}}, \bibinfo {author}
  {\bibfnamefont {X.}~\bibnamefont {Du}}, \bibinfo {author} {\bibfnamefont
  {X.-H.}\ \bibnamefont {Fan}}, \bibinfo {author} {\bibfnamefont
  {W.}~\bibnamefont {Gao}}, \bibinfo {author} {\bibfnamefont {L.}~\bibnamefont
  {Geng}}, \bibinfo {author} {\bibfnamefont {E.}~\bibnamefont {Ha}}, \bibinfo
  {author} {\bibfnamefont {X.-T.}\ \bibnamefont {He}}, \bibinfo {author}
  {\bibfnamefont {J.}~\bibnamefont {Huang}}, \bibinfo {author} {\bibfnamefont
  {K.}~\bibnamefont {Huang}}, \bibinfo {author} {\bibfnamefont
  {S.}~\bibnamefont {Kim}}, \bibinfo {author} {\bibfnamefont {Y.}~\bibnamefont
  {Kim}}, \bibinfo {author} {\bibfnamefont {C.-H.}\ \bibnamefont {Lee}},
  \bibinfo {author} {\bibfnamefont {J.}~\bibnamefont {Lee}}, \bibinfo {author}
  {\bibfnamefont {Z.}~\bibnamefont {Li}}, \bibinfo {author} {\bibfnamefont
  {Z.-R.}\ \bibnamefont {Liu}}, \bibinfo {author} {\bibfnamefont
  {Y.}~\bibnamefont {Ma}}, \bibinfo {author} {\bibfnamefont {J.}~\bibnamefont
  {Meng}}, \bibinfo {author} {\bibfnamefont {M.-H.}\ \bibnamefont {Mun}},
  \bibinfo {author} {\bibfnamefont {Z.}~\bibnamefont {Niu}}, \bibinfo {author}
  {\bibfnamefont {P.}~\bibnamefont {Papakonstantinou}}, \bibinfo {author}
  {\bibfnamefont {X.}~\bibnamefont {Shang}}, \bibinfo {author} {\bibfnamefont
  {C.}~\bibnamefont {Shen}}, \bibinfo {author} {\bibfnamefont {G.}~\bibnamefont
  {Shen}}, \bibinfo {author} {\bibfnamefont {W.}~\bibnamefont {Sun}}, \bibinfo
  {author} {\bibfnamefont {X.-X.}\ \bibnamefont {Sun}}, \bibinfo {author}
  {\bibfnamefont {J.}~\bibnamefont {Wu}}, \bibinfo {author} {\bibfnamefont
  {X.}~\bibnamefont {Wu}}, \bibinfo {author} {\bibfnamefont {X.}~\bibnamefont
  {Xia}}, \bibinfo {author} {\bibfnamefont {Y.}~\bibnamefont {Yan}}, \bibinfo
  {author} {\bibfnamefont {T.~C.}\ \bibnamefont {Yiu}}, \bibinfo {author}
  {\bibfnamefont {K.}~\bibnamefont {Zhang}}, \bibinfo {author} {\bibfnamefont
  {S.}~\bibnamefont {Zhang}}, \bibinfo {author} {\bibfnamefont
  {W.}~\bibnamefont {Zhang}}, \bibinfo {author} {\bibfnamefont
  {X.}~\bibnamefont {Zhang}}, \bibinfo {author} {\bibfnamefont
  {Q.}~\bibnamefont {Zhao}}, \bibinfo {author} {\bibfnamefont {R.}~\bibnamefont
  {Zheng}}, \ and\ \bibinfo {author} {\bibfnamefont {S.-G.}\ \bibnamefont
  {Zhou}} (\bibinfo {collaboration} {DRHBc Mass Table Collaboration}),\ }\href
  {\doibase 10.1103/PhysRevC.106.014316} {\bibfield  {journal} {\bibinfo
  {journal} {Phys. Rev. C}\ }\textbf {\bibinfo {volume} {106}},\ \bibinfo
  {pages} {014316} (\bibinfo {year} {2022})}\BibitemShut {NoStop}%
\bibitem [{\citenamefont {Giuliani}\ and\ \citenamefont
  {Robledo}(2023)}]{Giuliani:2023}%
  \BibitemOpen
  \bibfield  {author} {\bibinfo {author} {\bibfnamefont {S.~A.}\ \bibnamefont
  {Giuliani}}\ and\ \bibinfo {author} {\bibfnamefont {L.~M.}\ \bibnamefont
  {Robledo}},\ }\href@noop {} {\  (\bibinfo {year} {2023})},\ \Eprint
  {http://arxiv.org/abs/2311.03882} {arXiv:2311.03882 [nucl-th]} \BibitemShut
  {NoStop}%
\bibitem [{\citenamefont {Rutz}\ \emph {et~al.}(1998)\citenamefont {Rutz},
  \citenamefont {Bender}, \citenamefont {Reinhard}, \citenamefont {Maruhn},\
  and\ \citenamefont {Greiner}}]{Rutz:1998}%
  \BibitemOpen
  \bibfield  {author} {\bibinfo {author} {\bibfnamefont {K.}~\bibnamefont
  {Rutz}}, \bibinfo {author} {\bibfnamefont {M.}~\bibnamefont {Bender}},
  \bibinfo {author} {\bibfnamefont {P.~G.}\ \bibnamefont {Reinhard}}, \bibinfo
  {author} {\bibfnamefont {J.~A.}\ \bibnamefont {Maruhn}}, \ and\ \bibinfo
  {author} {\bibfnamefont {W.}~\bibnamefont {Greiner}},\ }\href {\doibase
  10.1016/S0375-9474(98)00153-5} {\bibfield  {journal} {\bibinfo  {journal}
  {Nucl. Phys. A}\ }\textbf {\bibinfo {volume} {634}},\ \bibinfo {pages} {67}
  (\bibinfo {year} {1998})}\BibitemShut {NoStop}%
\bibitem [{\citenamefont {Hofmann}\ and\ \citenamefont
  {Ring}(1988)}]{Hofmann:1988}%
  \BibitemOpen
  \bibfield  {author} {\bibinfo {author} {\bibfnamefont {U.}~\bibnamefont
  {Hofmann}}\ and\ \bibinfo {author} {\bibfnamefont {P.}~\bibnamefont {Ring}},\
  }\href {\doibase 10.1016/0370-2693(88)91367-6} {\bibfield  {journal}
  {\bibinfo  {journal} {Phys. Lett. B}\ }\textbf {\bibinfo {volume} {214}},\
  \bibinfo {pages} {307} (\bibinfo {year} {1988})}\BibitemShut {NoStop}%
\bibitem [{\citenamefont {Yao}\ \emph {et~al.}(2006)\citenamefont {Yao},
  \citenamefont {Chen},\ and\ \citenamefont {Meng}}]{Yao:2006}%
  \BibitemOpen
  \bibfield  {author} {\bibinfo {author} {\bibfnamefont {J.~M.}\ \bibnamefont
  {Yao}}, \bibinfo {author} {\bibfnamefont {H.}~\bibnamefont {Chen}}, \ and\
  \bibinfo {author} {\bibfnamefont {J.}~\bibnamefont {Meng}},\ }\href {\doibase
  10.1103/PhysRevC.74.024307} {\bibfield  {journal} {\bibinfo  {journal} {Phys.
  Rev. C}\ }\textbf {\bibinfo {volume} {74}},\ \bibinfo {pages} {024307}
  (\bibinfo {year} {2006})},\ \Eprint {http://arxiv.org/abs/nucl-th/0606040}
  {arXiv:nucl-th/0606040} \BibitemShut {NoStop}%
\bibitem [{\citenamefont {Bender}\ \emph
  {et~al.}(2000{\natexlab{a}})\citenamefont {Bender}, \citenamefont {Rutz},
  \citenamefont {Reinhard},\ and\ \citenamefont {Maruhn}}]{Bender:2000fqv}%
  \BibitemOpen
  \bibfield  {author} {\bibinfo {author} {\bibfnamefont {M.}~\bibnamefont
  {Bender}}, \bibinfo {author} {\bibfnamefont {K.}~\bibnamefont {Rutz}},
  \bibinfo {author} {\bibfnamefont {P.~G.}\ \bibnamefont {Reinhard}}, \ and\
  \bibinfo {author} {\bibfnamefont {J.~A.}\ \bibnamefont {Maruhn}},\ }\href
  {\doibase 10.1007/s10050-000-4504-z} {\bibfield  {journal} {\bibinfo
  {journal} {Eur. Phys. J. A}\ }\textbf {\bibinfo {volume} {8}},\ \bibinfo
  {pages} {59} (\bibinfo {year} {2000}{\natexlab{a}})},\ \Eprint
  {http://arxiv.org/abs/nucl-th/0005028} {arXiv:nucl-th/0005028} \BibitemShut
  {NoStop}%
\bibitem [{\citenamefont {Rodriguez-Guzman}\ \emph {et~al.}(2010)\citenamefont
  {Rodriguez-Guzman}, \citenamefont {Sarriguren}, \citenamefont {Robledo},\
  and\ \citenamefont {Perez-Martin}}]{Guzman:2010EFA}%
  \BibitemOpen
  \bibfield  {author} {\bibinfo {author} {\bibfnamefont {R.}~\bibnamefont
  {Rodriguez-Guzman}}, \bibinfo {author} {\bibfnamefont {P.}~\bibnamefont
  {Sarriguren}}, \bibinfo {author} {\bibfnamefont {L.~M.}\ \bibnamefont
  {Robledo}}, \ and\ \bibinfo {author} {\bibfnamefont {S.}~\bibnamefont
  {Perez-Martin}},\ }\href {\doibase 10.1016/j.physletb.2010.06.035} {\bibfield
   {journal} {\bibinfo  {journal} {Phys. Lett. B}\ }\textbf {\bibinfo {volume}
  {691}},\ \bibinfo {pages} {202} (\bibinfo {year} {2010})},\ \Eprint
  {http://arxiv.org/abs/1006.1975} {arXiv:1006.1975 [nucl-th]} \BibitemShut
  {NoStop}%
\bibitem [{\citenamefont {Li}\ \emph {et~al.}(2012)\citenamefont {Li},
  \citenamefont {Meng}, \citenamefont {Ring}, \citenamefont {Zhao},\ and\
  \citenamefont {Zhou}}]{Li:2012EFA}%
  \BibitemOpen
  \bibfield  {author} {\bibinfo {author} {\bibfnamefont {L.}~\bibnamefont
  {Li}}, \bibinfo {author} {\bibfnamefont {J.}~\bibnamefont {Meng}}, \bibinfo
  {author} {\bibfnamefont {P.}~\bibnamefont {Ring}}, \bibinfo {author}
  {\bibfnamefont {E.-G.}\ \bibnamefont {Zhao}}, \ and\ \bibinfo {author}
  {\bibfnamefont {S.-G.}\ \bibnamefont {Zhou}},\ }\href {\doibase
  10.1088/0256-307X/29/4/042101} {\bibfield  {journal} {\bibinfo  {journal}
  {Chin. Phys. Lett.}\ }\textbf {\bibinfo {volume} {29}},\ \bibinfo {pages}
  {042101} (\bibinfo {year} {2012})},\ \Eprint {http://arxiv.org/abs/1203.1363}
  {arXiv:1203.1363 [nucl-th]} \BibitemShut {NoStop}%
\bibitem [{\citenamefont {Duguet}\ \emph {et~al.}(2001)\citenamefont {Duguet},
  \citenamefont {Bonche}, \citenamefont {Heenen},\ and\ \citenamefont
  {Meyer}}]{Duguet:2001}%
  \BibitemOpen
  \bibfield  {author} {\bibinfo {author} {\bibfnamefont {T.}~\bibnamefont
  {Duguet}}, \bibinfo {author} {\bibfnamefont {P.}~\bibnamefont {Bonche}},
  \bibinfo {author} {\bibfnamefont {P.-H.}\ \bibnamefont {Heenen}}, \ and\
  \bibinfo {author} {\bibfnamefont {J.}~\bibnamefont {Meyer}},\ }\href
  {\doibase 10.1103/PhysRevC.65.014310} {\bibfield  {journal} {\bibinfo
  {journal} {Phys. Rev. C}\ }\textbf {\bibinfo {volume} {65}},\ \bibinfo
  {pages} {014310} (\bibinfo {year} {2001})}\BibitemShut {NoStop}%
\bibitem [{\citenamefont {Meyer}\ \emph {et~al.}(1979)\citenamefont {Meyer},
  \citenamefont {Dani\'ere}, \citenamefont {Letessier},\ and\ \citenamefont
  {Quentin}}]{Meyer:1979NPA}%
  \BibitemOpen
  \bibfield  {author} {\bibinfo {author} {\bibfnamefont {M.}~\bibnamefont
  {Meyer}}, \bibinfo {author} {\bibfnamefont {J.}~\bibnamefont {Dani\'ere}},
  \bibinfo {author} {\bibfnamefont {J.}~\bibnamefont {Letessier}}, \ and\
  \bibinfo {author} {\bibfnamefont {P.}~\bibnamefont {Quentin}},\ }\href
  {\doibase 10.1016/0375-9474(79)90674-2} {\bibfield  {journal} {\bibinfo
  {journal} {Nucl. Phys. A}\ }\textbf {\bibinfo {volume} {316}},\ \bibinfo
  {pages} {93} (\bibinfo {year} {1979})}\BibitemShut {NoStop}%
\bibitem [{\citenamefont {Libert}\ \emph {et~al.}(1982)\citenamefont {Libert},
  \citenamefont {Meyer},\ and\ \citenamefont {Quentin}}]{Libert:1982PRC}%
  \BibitemOpen
  \bibfield  {author} {\bibinfo {author} {\bibfnamefont {J.}~\bibnamefont
  {Libert}}, \bibinfo {author} {\bibfnamefont {M.}~\bibnamefont {Meyer}}, \
  and\ \bibinfo {author} {\bibfnamefont {P.}~\bibnamefont {Quentin}},\ }\href
  {\doibase 10.1103/PhysRevC.25.586} {\bibfield  {journal} {\bibinfo  {journal}
  {Phys. Rev. C}\ }\textbf {\bibinfo {volume} {25}},\ \bibinfo {pages} {586}
  (\bibinfo {year} {1982})}\BibitemShut {NoStop}%
\bibitem [{\citenamefont {Quan}\ \emph {et~al.}(2017)\citenamefont {Quan},
  \citenamefont {Liu}, \citenamefont {Li},\ and\ \citenamefont
  {Smith}}]{Quan:2017CQC}%
  \BibitemOpen
  \bibfield  {author} {\bibinfo {author} {\bibfnamefont {S.}~\bibnamefont
  {Quan}}, \bibinfo {author} {\bibfnamefont {W.~P.}\ \bibnamefont {Liu}},
  \bibinfo {author} {\bibfnamefont {Z.~P.}\ \bibnamefont {Li}}, \ and\ \bibinfo
  {author} {\bibfnamefont {M.~S.}\ \bibnamefont {Smith}},\ }\href {\doibase
  10.1103/PhysRevC.96.054309} {\bibfield  {journal} {\bibinfo  {journal} {Phys.
  Rev. C}\ }\textbf {\bibinfo {volume} {96}},\ \bibinfo {pages} {054309}
  (\bibinfo {year} {2017})},\ \Eprint {http://arxiv.org/abs/1710.08172}
  {arXiv:1710.08172 [nucl-th]} \BibitemShut {NoStop}%
\bibitem [{\citenamefont {Sun}\ \emph {et~al.}(2019)\citenamefont {Sun},
  \citenamefont {Quan}, \citenamefont {Li}, \citenamefont {Zhao}, \citenamefont
  {Nik\v{s}i\'c},\ and\ \citenamefont {Vretenar}}]{Sun:2019CQC}%
  \BibitemOpen
  \bibfield  {author} {\bibinfo {author} {\bibfnamefont {W.}~\bibnamefont
  {Sun}}, \bibinfo {author} {\bibfnamefont {S.}~\bibnamefont {Quan}}, \bibinfo
  {author} {\bibfnamefont {Z.~P.}\ \bibnamefont {Li}}, \bibinfo {author}
  {\bibfnamefont {J.}~\bibnamefont {Zhao}}, \bibinfo {author} {\bibfnamefont
  {T.}~\bibnamefont {Nik\v{s}i\'c}}, \ and\ \bibinfo {author} {\bibfnamefont
  {D.}~\bibnamefont {Vretenar}},\ }\href {\doibase 10.1103/PhysRevC.100.044319}
  {\bibfield  {journal} {\bibinfo  {journal} {Phys. Rev. C}\ }\textbf {\bibinfo
  {volume} {100}},\ \bibinfo {pages} {044319} (\bibinfo {year} {2019})},\
  \Eprint {http://arxiv.org/abs/1911.02422} {arXiv:1911.02422 [nucl-th]}
  \BibitemShut {NoStop}%
\bibitem [{\citenamefont {Nomura}\ \emph {et~al.}(2016)\citenamefont {Nomura},
  \citenamefont {Nik\v{s}i\'c},\ and\ \citenamefont
  {Vretenar}}]{Nomura:2016IBFM}%
  \BibitemOpen
  \bibfield  {author} {\bibinfo {author} {\bibfnamefont {K.}~\bibnamefont
  {Nomura}}, \bibinfo {author} {\bibfnamefont {T.}~\bibnamefont
  {Nik\v{s}i\'c}}, \ and\ \bibinfo {author} {\bibfnamefont {D.}~\bibnamefont
  {Vretenar}},\ }\href {\doibase 10.1103/PhysRevC.93.054305} {\bibfield
  {journal} {\bibinfo  {journal} {Phys. Rev. C}\ }\textbf {\bibinfo {volume}
  {93}},\ \bibinfo {pages} {054305} (\bibinfo {year} {2016})},\ \Eprint
  {http://arxiv.org/abs/1605.00755} {arXiv:1605.00755 [nucl-th]} \BibitemShut
  {NoStop}%
\bibitem [{\citenamefont {Nomura}\ \emph {et~al.}(2017)\citenamefont {Nomura},
  \citenamefont {Rodr\'\i{}guez-Guzm\'an},\ and\ \citenamefont
  {Robledo}}]{Nomura:2017IBFM}%
  \BibitemOpen
  \bibfield  {author} {\bibinfo {author} {\bibfnamefont {K.}~\bibnamefont
  {Nomura}}, \bibinfo {author} {\bibfnamefont {R.}~\bibnamefont
  {Rodr\'\i{}guez-Guzm\'an}}, \ and\ \bibinfo {author} {\bibfnamefont {L.~M.}\
  \bibnamefont {Robledo}},\ }\href {\doibase 10.1103/PhysRevC.96.014314}
  {\bibfield  {journal} {\bibinfo  {journal} {Phys. Rev. C}\ }\textbf {\bibinfo
  {volume} {96}},\ \bibinfo {pages} {014314} (\bibinfo {year} {2017})},\
  \Eprint {http://arxiv.org/abs/1705.05539} {arXiv:1705.05539 [nucl-th]}
  \BibitemShut {NoStop}%
\bibitem [{\citenamefont {Bender}\ \emph {et~al.}(2003)\citenamefont {Bender},
  \citenamefont {Heenen},\ and\ \citenamefont {Reinhard}}]{Bender:2003SMF}%
  \BibitemOpen
  \bibfield  {author} {\bibinfo {author} {\bibfnamefont {M.}~\bibnamefont
  {Bender}}, \bibinfo {author} {\bibfnamefont {P.-H.}\ \bibnamefont {Heenen}},
  \ and\ \bibinfo {author} {\bibfnamefont {P.-G.}\ \bibnamefont {Reinhard}},\
  }\href {\doibase 10.1103/RevModPhys.75.121} {\bibfield  {journal} {\bibinfo
  {journal} {Rev. Mod. Phys.}\ }\textbf {\bibinfo {volume} {75}},\ \bibinfo
  {pages} {121} (\bibinfo {year} {2003})}\BibitemShut {NoStop}%
\bibitem [{\citenamefont {Niksic}\ \emph {et~al.}(2011)\citenamefont {Niksic},
  \citenamefont {Vretenar},\ and\ \citenamefont {Ring}}]{Niksic:2011_PPNP}%
  \BibitemOpen
  \bibfield  {author} {\bibinfo {author} {\bibfnamefont {T.}~\bibnamefont
  {Niksic}}, \bibinfo {author} {\bibfnamefont {D.}~\bibnamefont {Vretenar}}, \
  and\ \bibinfo {author} {\bibfnamefont {P.}~\bibnamefont {Ring}},\ }\href
  {\doibase 10.1016/j.ppnp.2011.01.055} {\bibfield  {journal} {\bibinfo
  {journal} {Prog. Part. Nucl. Phys.}\ }\textbf {\bibinfo {volume} {66}},\
  \bibinfo {pages} {519} (\bibinfo {year} {2011})},\ \Eprint
  {http://arxiv.org/abs/1102.4193} {arXiv:1102.4193 [nucl-th]} \BibitemShut
  {NoStop}%
\bibitem [{\citenamefont {Robledo}\ \emph {et~al.}(2019)\citenamefont
  {Robledo}, \citenamefont {Rodr\'\i{}guez},\ and\ \citenamefont
  {Rodr\'\i{}guez-Guzm\'an}}]{Robledo:2019JPG}%
  \BibitemOpen
  \bibfield  {author} {\bibinfo {author} {\bibfnamefont {L.~M.}\ \bibnamefont
  {Robledo}}, \bibinfo {author} {\bibfnamefont {T.~R.}\ \bibnamefont
  {Rodr\'\i{}guez}}, \ and\ \bibinfo {author} {\bibfnamefont {R.~R.}\
  \bibnamefont {Rodr\'\i{}guez-Guzm\'an}},\ }\href {\doibase
  10.1088/1361-6471/aadebd} {\bibfield  {journal} {\bibinfo  {journal} {J.
  Phys. G}\ }\textbf {\bibinfo {volume} {46}},\ \bibinfo {pages} {013001}
  (\bibinfo {year} {2019})},\ \Eprint {http://arxiv.org/abs/1807.02518}
  {arXiv:1807.02518 [nucl-th]} \BibitemShut {NoStop}%
\bibitem [{\citenamefont {Sheikh}\ \emph {et~al.}(2021)\citenamefont {Sheikh},
  \citenamefont {Dobaczewski}, \citenamefont {Ring}, \citenamefont {Robledo},\
  and\ \citenamefont {Yannouleas}}]{Sheikh:2021JPG}%
  \BibitemOpen
  \bibfield  {author} {\bibinfo {author} {\bibfnamefont {J.~A.}\ \bibnamefont
  {Sheikh}}, \bibinfo {author} {\bibfnamefont {J.}~\bibnamefont {Dobaczewski}},
  \bibinfo {author} {\bibfnamefont {P.}~\bibnamefont {Ring}}, \bibinfo {author}
  {\bibfnamefont {L.~M.}\ \bibnamefont {Robledo}}, \ and\ \bibinfo {author}
  {\bibfnamefont {C.}~\bibnamefont {Yannouleas}},\ }\href {\doibase
  10.1088/1361-6471/ac288a} {\bibfield  {journal} {\bibinfo  {journal} {J.
  Phys. G}\ }\textbf {\bibinfo {volume} {48}},\ \bibinfo {pages} {123001}
  (\bibinfo {year} {2021})},\ \Eprint {http://arxiv.org/abs/1901.06992}
  {arXiv:1901.06992 [nucl-th]} \BibitemShut {NoStop}%
\bibitem [{\citenamefont {Yao}(2022)}]{Yao:2022_HB}%
  \BibitemOpen
  \bibfield  {author} {\bibinfo {author} {\bibfnamefont {J.~M.}\ \bibnamefont
  {Yao}},\ }\enquote {\bibinfo {title} {{Symmetry Restoration Methods}},}\ in\
  \href {\doibase 10.1007/978-981-15-8818-1_18-1} {\emph {\bibinfo {booktitle}
  {{Handbook of Nuclear Physics}}}},\ \bibinfo {editor} {edited by\ \bibinfo
  {editor} {\bibfnamefont {I.}~\bibnamefont {Tanihata}}, \bibinfo {editor}
  {\bibfnamefont {H.}~\bibnamefont {Toki}}, \ and\ \bibinfo {editor}
  {\bibfnamefont {T.}~\bibnamefont {Kajino}}}\ (\bibinfo {year} {2022})\ pp.\
  \bibinfo {pages} {1--36},\ \Eprint {http://arxiv.org/abs/2204.12126}
  {arXiv:2204.12126 [nucl-th]} \BibitemShut {NoStop}%
\bibitem [{\citenamefont {Bassichis}\ \emph {et~al.}(1965)\citenamefont
  {Bassichis}, \citenamefont {Giraud},\ and\ \citenamefont
  {Ripka}}]{Bassichis:1965}%
  \BibitemOpen
  \bibfield  {author} {\bibinfo {author} {\bibfnamefont {W.~H.}\ \bibnamefont
  {Bassichis}}, \bibinfo {author} {\bibfnamefont {B.}~\bibnamefont {Giraud}}, \
  and\ \bibinfo {author} {\bibfnamefont {G.}~\bibnamefont {Ripka}},\ }\href
  {\doibase 10.1103/PhysRevLett.15.980} {\bibfield  {journal} {\bibinfo
  {journal} {Phys. Rev. Lett.}\ }\textbf {\bibinfo {volume} {15}},\ \bibinfo
  {pages} {980} (\bibinfo {year} {1965})}\BibitemShut {NoStop}%
\bibitem [{\citenamefont {Gunye}\ and\ \citenamefont
  {Warke}(1967)}]{Gunye:1967}%
  \BibitemOpen
  \bibfield  {author} {\bibinfo {author} {\bibfnamefont {M.~R.}\ \bibnamefont
  {Gunye}}\ and\ \bibinfo {author} {\bibfnamefont {C.~S.}\ \bibnamefont
  {Warke}},\ }\href {\doibase 10.1103/PhysRev.156.1087} {\bibfield  {journal}
  {\bibinfo  {journal} {Phys. Rev.}\ }\textbf {\bibinfo {volume} {156}},\
  \bibinfo {pages} {1087} (\bibinfo {year} {1967})}\BibitemShut {NoStop}%
\bibitem [{\citenamefont {Rath}\ \emph {et~al.}(1993)\citenamefont {Rath},
  \citenamefont {Praharaj},\ and\ \citenamefont {Khadkikar}}]{Rath:1993}%
  \BibitemOpen
  \bibfield  {author} {\bibinfo {author} {\bibfnamefont {A.~K.}\ \bibnamefont
  {Rath}}, \bibinfo {author} {\bibfnamefont {C.~R.}\ \bibnamefont {Praharaj}},
  \ and\ \bibinfo {author} {\bibfnamefont {S.~B.}\ \bibnamefont {Khadkikar}},\
  }\href {\doibase 10.1103/PhysRevC.47.1990} {\bibfield  {journal} {\bibinfo
  {journal} {Phys. Rev. C}\ }\textbf {\bibinfo {volume} {47}},\ \bibinfo
  {pages} {1990} (\bibinfo {year} {1993})}\BibitemShut {NoStop}%
\bibitem [{\citenamefont {Hara}\ and\ \citenamefont
  {Iwasaki}(1984)}]{Hara:1984}%
  \BibitemOpen
  \bibfield  {author} {\bibinfo {author} {\bibfnamefont {K.}~\bibnamefont
  {Hara}}\ and\ \bibinfo {author} {\bibfnamefont {S.}~\bibnamefont {Iwasaki}},\
  }\href {\doibase 10.1016/0375-9474(84)90199-4} {\bibfield  {journal}
  {\bibinfo  {journal} {Nucl. Phys. A}\ }\textbf {\bibinfo {volume} {430}},\
  \bibinfo {pages} {175} (\bibinfo {year} {1984})}\BibitemShut {NoStop}%
\bibitem [{\citenamefont {Hara}\ and\ \citenamefont {Sun}(1995)}]{Hara:1995}%
  \BibitemOpen
  \bibfield  {author} {\bibinfo {author} {\bibfnamefont {K.}~\bibnamefont
  {Hara}}\ and\ \bibinfo {author} {\bibfnamefont {Y.}~\bibnamefont {Sun}},\
  }\href {\doibase 10.1142/S0218301395000250} {\bibfield  {journal} {\bibinfo
  {journal} {Int. J. Mod. Phys. E}\ }\textbf {\bibinfo {volume} {4}},\ \bibinfo
  {pages} {637} (\bibinfo {year} {1995})}\BibitemShut {NoStop}%
\bibitem [{\citenamefont {Hammar\'en}\ \emph {et~al.}(1985)\citenamefont
  {Hammar\'en}, \citenamefont {Schmid}, \citenamefont {Gr\"ummer},
  \citenamefont {Faessler},\ and\ \citenamefont {Fladt}}]{Hammaren:1985}%
  \BibitemOpen
  \bibfield  {author} {\bibinfo {author} {\bibfnamefont {E.}~\bibnamefont
  {Hammar\'en}}, \bibinfo {author} {\bibfnamefont {K.~W.}\ \bibnamefont
  {Schmid}}, \bibinfo {author} {\bibfnamefont {F.}~\bibnamefont {Gr\"ummer}},
  \bibinfo {author} {\bibfnamefont {A.}~\bibnamefont {Faessler}}, \ and\
  \bibinfo {author} {\bibfnamefont {B.}~\bibnamefont {Fladt}},\ }\href
  {\doibase 10.1016/0375-9474(85)90225-8} {\bibfield  {journal} {\bibinfo
  {journal} {Nucl. Phys. A}\ }\textbf {\bibinfo {volume} {437}},\ \bibinfo
  {pages} {1} (\bibinfo {year} {1985})}\BibitemShut {NoStop}%
\bibitem [{\citenamefont {Kimura}(2007)}]{Kimura:2007}%
  \BibitemOpen
  \bibfield  {author} {\bibinfo {author} {\bibfnamefont {M.}~\bibnamefont
  {Kimura}},\ }\href {\doibase 10.1103/PhysRevC.75.041302} {\bibfield
  {journal} {\bibinfo  {journal} {Phys. Rev. C}\ }\textbf {\bibinfo {volume}
  {75}},\ \bibinfo {pages} {041302} (\bibinfo {year} {2007})},\ \Eprint
  {http://arxiv.org/abs/nucl-th/0702012} {arXiv:nucl-th/0702012} \BibitemShut
  {NoStop}%
\bibitem [{\citenamefont {Kimura}(2011)}]{Kimura:2011}%
  \BibitemOpen
  \bibfield  {author} {\bibinfo {author} {\bibfnamefont {M.}~\bibnamefont
  {Kimura}},\ }\href@noop {} {\  (\bibinfo {year} {2011})},\ \Eprint
  {http://arxiv.org/abs/1105.3281} {arXiv:1105.3281 [nucl-th]} \BibitemShut
  {NoStop}%
\bibitem [{\citenamefont {Minomo}\ \emph {et~al.}(2012)\citenamefont {Minomo},
  \citenamefont {Sumi}, \citenamefont {Kimura}, \citenamefont {Ogata},
  \citenamefont {Shimizu},\ and\ \citenamefont {Yahiro}}]{Minomo:2012}%
  \BibitemOpen
  \bibfield  {author} {\bibinfo {author} {\bibfnamefont {K.}~\bibnamefont
  {Minomo}}, \bibinfo {author} {\bibfnamefont {T.}~\bibnamefont {Sumi}},
  \bibinfo {author} {\bibfnamefont {M.}~\bibnamefont {Kimura}}, \bibinfo
  {author} {\bibfnamefont {K.}~\bibnamefont {Ogata}}, \bibinfo {author}
  {\bibfnamefont {Y.~R.}\ \bibnamefont {Shimizu}}, \ and\ \bibinfo {author}
  {\bibfnamefont {M.}~\bibnamefont {Yahiro}},\ }\href {\doibase
  10.1103/PhysRevLett.108.052503} {\bibfield  {journal} {\bibinfo  {journal}
  {Phys. Rev. Lett.}\ }\textbf {\bibinfo {volume} {108}},\ \bibinfo {pages}
  {052503} (\bibinfo {year} {2012})},\ \Eprint {http://arxiv.org/abs/1110.3867}
  {arXiv:1110.3867 [nucl-th]} \BibitemShut {NoStop}%
\bibitem [{\citenamefont {Bally}\ \emph {et~al.}(2014)\citenamefont {Bally},
  \citenamefont {Avez}, \citenamefont {Bender},\ and\ \citenamefont
  {Heenen}}]{Bally:2014prl}%
  \BibitemOpen
  \bibfield  {author} {\bibinfo {author} {\bibfnamefont {B.}~\bibnamefont
  {Bally}}, \bibinfo {author} {\bibfnamefont {B.}~\bibnamefont {Avez}},
  \bibinfo {author} {\bibfnamefont {M.}~\bibnamefont {Bender}}, \ and\ \bibinfo
  {author} {\bibfnamefont {P.~H.}\ \bibnamefont {Heenen}},\ }\href {\doibase
  10.1103/PhysRevLett.113.162501} {\bibfield  {journal} {\bibinfo  {journal}
  {Phys. Rev. Lett.}\ }\textbf {\bibinfo {volume} {113}},\ \bibinfo {pages}
  {162501} (\bibinfo {year} {2014})},\ \Eprint {http://arxiv.org/abs/1406.5984}
  {arXiv:1406.5984 [nucl-th]} \BibitemShut {NoStop}%
\bibitem [{\citenamefont {Bally}\ \emph {et~al.}(2022)\citenamefont {Bally},
  \citenamefont {Giacalone},\ and\ \citenamefont {Bender}}]{Bally:2022_EPJA}%
  \BibitemOpen
  \bibfield  {author} {\bibinfo {author} {\bibfnamefont {B.}~\bibnamefont
  {Bally}}, \bibinfo {author} {\bibfnamefont {G.}~\bibnamefont {Giacalone}}, \
  and\ \bibinfo {author} {\bibfnamefont {M.}~\bibnamefont {Bender}},\ }\href
  {\doibase 10.1140/epja/s10050-022-00833-4} {\bibfield  {journal} {\bibinfo
  {journal} {Eur. Phys. J. A}\ }\textbf {\bibinfo {volume} {58}},\ \bibinfo
  {pages} {187} (\bibinfo {year} {2022})},\ \Eprint
  {http://arxiv.org/abs/2207.13576} {arXiv:2207.13576 [nucl-th]} \BibitemShut
  {NoStop}%
\bibitem [{\citenamefont {Bally}\ \emph {et~al.}(2023)\citenamefont {Bally},
  \citenamefont {Giacalone},\ and\ \citenamefont {Bender}}]{Bally:2023_EPJA}%
  \BibitemOpen
  \bibfield  {author} {\bibinfo {author} {\bibfnamefont {B.}~\bibnamefont
  {Bally}}, \bibinfo {author} {\bibfnamefont {G.}~\bibnamefont {Giacalone}}, \
  and\ \bibinfo {author} {\bibfnamefont {M.}~\bibnamefont {Bender}},\ }\href
  {\doibase 10.1140/epja/s10050-023-00955-3} {\bibfield  {journal} {\bibinfo
  {journal} {Eur. Phys. J. A}\ }\textbf {\bibinfo {volume} {59}},\ \bibinfo
  {pages} {58} (\bibinfo {year} {2023})},\ \Eprint
  {http://arxiv.org/abs/2301.02420} {arXiv:2301.02420 [nucl-th]} \BibitemShut
  {NoStop}%
\bibitem [{\citenamefont {Borrajo}\ and\ \citenamefont
  {Egido}(2016)}]{Borrajo:2016vfz}%
  \BibitemOpen
  \bibfield  {author} {\bibinfo {author} {\bibfnamefont {M.}~\bibnamefont
  {Borrajo}}\ and\ \bibinfo {author} {\bibfnamefont {J.~L.}\ \bibnamefont
  {Egido}},\ }\href {\doibase 10.1140/epja/i2016-16277-8} {\bibfield  {journal}
  {\bibinfo  {journal} {Eur. Phys. J. A}\ }\textbf {\bibinfo {volume} {52}},\
  \bibinfo {pages} {277} (\bibinfo {year} {2016})},\ \bibinfo {note} {[Erratum:
  Eur.Phys.J.A 53, 38 (2017)]},\ \Eprint {http://arxiv.org/abs/1609.02472}
  {arXiv:1609.02472 [nucl-th]} \BibitemShut {NoStop}%
\bibitem [{\citenamefont {Borrajo}\ and\ \citenamefont
  {Egido}(2017)}]{Borrajo:2017}%
  \BibitemOpen
  \bibfield  {author} {\bibinfo {author} {\bibfnamefont {M.}~\bibnamefont
  {Borrajo}}\ and\ \bibinfo {author} {\bibfnamefont {J.~L.}\ \bibnamefont
  {Egido}},\ }\href {\doibase 10.1016/j.physletb.2016.11.037} {\bibfield
  {journal} {\bibinfo  {journal} {Phys. Lett. B}\ }\textbf {\bibinfo {volume}
  {764}},\ \bibinfo {pages} {328} (\bibinfo {year} {2017})},\ \Eprint
  {http://arxiv.org/abs/1611.06982} {arXiv:1611.06982 [nucl-th]} \BibitemShut
  {NoStop}%
\bibitem [{\citenamefont {Borrajo}\ and\ \citenamefont
  {Egido}(2018)}]{Borrajo:2018sqj}%
  \BibitemOpen
  \bibfield  {author} {\bibinfo {author} {\bibfnamefont {M.}~\bibnamefont
  {Borrajo}}\ and\ \bibinfo {author} {\bibfnamefont {J.~L.}\ \bibnamefont
  {Egido}},\ }\href {\doibase 10.1103/PhysRevC.98.044317} {\bibfield  {journal}
  {\bibinfo  {journal} {Phys. Rev. C}\ }\textbf {\bibinfo {volume} {98}},\
  \bibinfo {pages} {044317} (\bibinfo {year} {2018})},\ \Eprint
  {http://arxiv.org/abs/1809.04287} {arXiv:1809.04287 [nucl-th]} \BibitemShut
  {NoStop}%
\bibitem [{\citenamefont {Meng}\ \emph {et~al.}(2017)\citenamefont {Meng},
  \citenamefont {Song},\ and\ \citenamefont {Yao}}]{Meng:2017}%
  \BibitemOpen
  \bibfield  {author} {\bibinfo {author} {\bibfnamefont {J.}~\bibnamefont
  {Meng}}, \bibinfo {author} {\bibfnamefont {L.~S.}\ \bibnamefont {Song}}, \
  and\ \bibinfo {author} {\bibfnamefont {J.~M.}\ \bibnamefont {Yao}},\ }\href
  {\doibase 10.1142/S0218301317400201} {\bibfield  {journal} {\bibinfo
  {journal} {International Journal of Modern Physics E}\ }\textbf {\bibinfo
  {volume} {26}},\ \bibinfo {pages} {1740020} (\bibinfo {year}
  {2017})}\BibitemShut {NoStop}%
\bibitem [{\citenamefont {Ring}(1996)}]{Ring:1996PPNP}%
  \BibitemOpen
  \bibfield  {author} {\bibinfo {author} {\bibfnamefont {P.}~\bibnamefont
  {Ring}},\ }\href {\doibase 10.1016/0146-6410(96)00054-3} {\bibfield
  {journal} {\bibinfo  {journal} {Prog. Part. Nucl. Phys.}\ }\textbf {\bibinfo
  {volume} {37}},\ \bibinfo {pages} {193} (\bibinfo {year} {1996})}\BibitemShut
  {NoStop}%
\bibitem [{\citenamefont {Meng}\ \emph {et~al.}(2006)\citenamefont {Meng},
  \citenamefont {Toki}, \citenamefont {Zhou}, \citenamefont {Zhang},
  \citenamefont {Long},\ and\ \citenamefont {Geng}}]{Meng:2006}%
  \BibitemOpen
  \bibfield  {author} {\bibinfo {author} {\bibfnamefont {J.}~\bibnamefont
  {Meng}}, \bibinfo {author} {\bibfnamefont {H.}~\bibnamefont {Toki}}, \bibinfo
  {author} {\bibfnamefont {S.~G.}\ \bibnamefont {Zhou}}, \bibinfo {author}
  {\bibfnamefont {S.~Q.}\ \bibnamefont {Zhang}}, \bibinfo {author}
  {\bibfnamefont {W.~H.}\ \bibnamefont {Long}}, \ and\ \bibinfo {author}
  {\bibfnamefont {L.~S.}\ \bibnamefont {Geng}},\ }\href {\doibase
  10.1016/j.ppnp.2005.06.001} {\bibfield  {journal} {\bibinfo  {journal} {Prog.
  Part. Nucl. Phys.}\ }\textbf {\bibinfo {volume} {57}},\ \bibinfo {pages}
  {470} (\bibinfo {year} {2006})},\ \Eprint
  {http://arxiv.org/abs/nucl-th/0508020} {arXiv:nucl-th/0508020} \BibitemShut
  {NoStop}%
\bibitem [{\citenamefont {Niksic}\ \emph {et~al.}(2006)\citenamefont {Niksic},
  \citenamefont {Vretenar},\ and\ \citenamefont {Ring}}]{Niksic:2006PRC}%
  \BibitemOpen
  \bibfield  {author} {\bibinfo {author} {\bibfnamefont {T.}~\bibnamefont
  {Niksic}}, \bibinfo {author} {\bibfnamefont {D.}~\bibnamefont {Vretenar}}, \
  and\ \bibinfo {author} {\bibfnamefont {P.}~\bibnamefont {Ring}},\ }\href
  {\doibase 10.1103/PhysRevC.73.034308} {\bibfield  {journal} {\bibinfo
  {journal} {Phys. Rev. C}\ }\textbf {\bibinfo {volume} {73}},\ \bibinfo
  {pages} {034308} (\bibinfo {year} {2006})},\ \Eprint
  {http://arxiv.org/abs/nucl-th/0601005} {arXiv:nucl-th/0601005} \BibitemShut
  {NoStop}%
\bibitem [{\citenamefont {Yao}\ \emph {et~al.}(2010{\natexlab{a}})\citenamefont
  {Yao}, \citenamefont {Meng}, \citenamefont {Ring},\ and\ \citenamefont
  {Vretenar}}]{Yao:2009traxi}%
  \BibitemOpen
  \bibfield  {author} {\bibinfo {author} {\bibfnamefont {J.~M.}\ \bibnamefont
  {Yao}}, \bibinfo {author} {\bibfnamefont {J.}~\bibnamefont {Meng}}, \bibinfo
  {author} {\bibfnamefont {P.}~\bibnamefont {Ring}}, \ and\ \bibinfo {author}
  {\bibfnamefont {D.}~\bibnamefont {Vretenar}},\ }\href {\doibase
  10.1103/PhysRevC.81.044311} {\bibfield  {journal} {\bibinfo  {journal} {Phys.
  Rev. C}\ }\textbf {\bibinfo {volume} {81}},\ \bibinfo {pages} {044311}
  (\bibinfo {year} {2010}{\natexlab{a}})},\ \Eprint
  {http://arxiv.org/abs/0912.2650} {arXiv:0912.2650 [nucl-th]} \BibitemShut
  {NoStop}%
\bibitem [{\citenamefont {Zhao}\ \emph {et~al.}(2016)\citenamefont {Zhao},
  \citenamefont {Ring},\ and\ \citenamefont {Meng}}]{Zhao:2016}%
  \BibitemOpen
  \bibfield  {author} {\bibinfo {author} {\bibfnamefont {P.~W.}\ \bibnamefont
  {Zhao}}, \bibinfo {author} {\bibfnamefont {P.}~\bibnamefont {Ring}}, \ and\
  \bibinfo {author} {\bibfnamefont {J.}~\bibnamefont {Meng}},\ }\href {\doibase
  10.1103/PhysRevC.94.041301} {\bibfield  {journal} {\bibinfo  {journal} {Phys.
  Rev. C}\ }\textbf {\bibinfo {volume} {94}},\ \bibinfo {pages} {041301}
  (\bibinfo {year} {2016})},\ \Eprint {http://arxiv.org/abs/1607.04241}
  {arXiv:1607.04241 [nucl-th]} \BibitemShut {NoStop}%
\bibitem [{\citenamefont {Sun}\ and\ \citenamefont {Zhou}(2021)}]{SunXX:2021}%
  \BibitemOpen
  \bibfield  {author} {\bibinfo {author} {\bibfnamefont {X.-X.}\ \bibnamefont
  {Sun}}\ and\ \bibinfo {author} {\bibfnamefont {S.-G.}\ \bibnamefont {Zhou}},\
  }\href {\doibase 10.1103/PhysRevC.104.064319} {\bibfield  {journal} {\bibinfo
   {journal} {Phys. Rev. C}\ }\textbf {\bibinfo {volume} {104}},\ \bibinfo
  {pages} {064319} (\bibinfo {year} {2021})}\BibitemShut {NoStop}%
\bibitem [{\citenamefont {Yao}\ \emph {et~al.}(2011)\citenamefont {Yao},
  \citenamefont {Mei}, \citenamefont {Chen}, \citenamefont {Meng},
  \citenamefont {Ring},\ and\ \citenamefont {Vretenar}}]{Yao:2011_Mg}%
  \BibitemOpen
  \bibfield  {author} {\bibinfo {author} {\bibfnamefont {J.~M.}\ \bibnamefont
  {Yao}}, \bibinfo {author} {\bibfnamefont {H.}~\bibnamefont {Mei}}, \bibinfo
  {author} {\bibfnamefont {H.}~\bibnamefont {Chen}}, \bibinfo {author}
  {\bibfnamefont {J.}~\bibnamefont {Meng}}, \bibinfo {author} {\bibfnamefont
  {P.}~\bibnamefont {Ring}}, \ and\ \bibinfo {author} {\bibfnamefont
  {D.}~\bibnamefont {Vretenar}},\ }\href {\doibase 10.1103/PhysRevC.83.014308}
  {\bibfield  {journal} {\bibinfo  {journal} {Phys. Rev. C}\ }\textbf {\bibinfo
  {volume} {83}},\ \bibinfo {pages} {014308} (\bibinfo {year} {2011})},\
  \Eprint {http://arxiv.org/abs/1006.1400} {arXiv:1006.1400 [nucl-th]}
  \BibitemShut {NoStop}%
\bibitem [{\citenamefont {Yao}\ \emph {et~al.}(2014)\citenamefont {Yao},
  \citenamefont {Hagino}, \citenamefont {Li}, \citenamefont {Meng},\ and\
  \citenamefont {Ring}}]{Yao:2014_76Kr}%
  \BibitemOpen
  \bibfield  {author} {\bibinfo {author} {\bibfnamefont {J.~M.}\ \bibnamefont
  {Yao}}, \bibinfo {author} {\bibfnamefont {K.}~\bibnamefont {Hagino}},
  \bibinfo {author} {\bibfnamefont {Z.~P.}\ \bibnamefont {Li}}, \bibinfo
  {author} {\bibfnamefont {J.}~\bibnamefont {Meng}}, \ and\ \bibinfo {author}
  {\bibfnamefont {P.}~\bibnamefont {Ring}},\ }\href {\doibase
  10.1103/PhysRevC.89.054306} {\bibfield  {journal} {\bibinfo  {journal} {Phys.
  Rev. C}\ }\textbf {\bibinfo {volume} {89}},\ \bibinfo {pages} {054306}
  (\bibinfo {year} {2014})},\ \Eprint {http://arxiv.org/abs/1403.4812}
  {arXiv:1403.4812 [nucl-th]} \BibitemShut {NoStop}%
\bibitem [{\citenamefont {Wang}\ \emph {et~al.}(2019)\citenamefont {Wang},
  \citenamefont {Chen}, \citenamefont {Zhao}, \citenamefont {Zhang},\ and\
  \citenamefont {Meng}}]{Wang:2019_TPSM}%
  \BibitemOpen
  \bibfield  {author} {\bibinfo {author} {\bibfnamefont {Y.~K.}\ \bibnamefont
  {Wang}}, \bibinfo {author} {\bibfnamefont {F.~Q.}\ \bibnamefont {Chen}},
  \bibinfo {author} {\bibfnamefont {P.~W.}\ \bibnamefont {Zhao}}, \bibinfo
  {author} {\bibfnamefont {S.~Q.}\ \bibnamefont {Zhang}}, \ and\ \bibinfo
  {author} {\bibfnamefont {J.}~\bibnamefont {Meng}},\ }\href {\doibase
  10.1103/PhysRevC.99.054303} {\bibfield  {journal} {\bibinfo  {journal} {Phys.
  Rev. C}\ }\textbf {\bibinfo {volume} {99}},\ \bibinfo {pages} {054303}
  (\bibinfo {year} {2019})},\ \Eprint {http://arxiv.org/abs/1903.03497}
  {arXiv:1903.03497 [nucl-th]} \BibitemShut {NoStop}%
\bibitem [{\citenamefont {Wang}\ \emph {et~al.}(2021)\citenamefont {Wang},
  \citenamefont {Zhao},\ and\ \citenamefont {Meng}}]{Wang:2021triaxial}%
  \BibitemOpen
  \bibfield  {author} {\bibinfo {author} {\bibfnamefont {Y.~K.}\ \bibnamefont
  {Wang}}, \bibinfo {author} {\bibfnamefont {P.~W.}\ \bibnamefont {Zhao}}, \
  and\ \bibinfo {author} {\bibfnamefont {J.}~\bibnamefont {Meng}},\ }\href
  {\doibase 10.1103/PhysRevC.104.014320} {\bibfield  {journal} {\bibinfo
  {journal} {Phys. Rev. C}\ }\textbf {\bibinfo {volume} {104}},\ \bibinfo
  {pages} {014320} (\bibinfo {year} {2021})},\ \Eprint
  {http://arxiv.org/abs/2105.02649} {arXiv:2105.02649 [nucl-th]} \BibitemShut
  {NoStop}%
\bibitem [{\citenamefont {Yao}\ \emph {et~al.}(2015{\natexlab{a}})\citenamefont
  {Yao}, \citenamefont {Zhou},\ and\ \citenamefont {Li}}]{Yao:2015_Octupole}%
  \BibitemOpen
  \bibfield  {author} {\bibinfo {author} {\bibfnamefont {J.~M.}\ \bibnamefont
  {Yao}}, \bibinfo {author} {\bibfnamefont {E.~F.}\ \bibnamefont {Zhou}}, \
  and\ \bibinfo {author} {\bibfnamefont {Z.~P.}\ \bibnamefont {Li}},\ }\href
  {\doibase 10.1103/PhysRevC.92.041304} {\bibfield  {journal} {\bibinfo
  {journal} {Phys. Rev. C}\ }\textbf {\bibinfo {volume} {92}},\ \bibinfo
  {pages} {041304} (\bibinfo {year} {2015}{\natexlab{a}})},\ \Eprint
  {http://arxiv.org/abs/1507.03298} {arXiv:1507.03298 [nucl-th]} \BibitemShut
  {NoStop}%
\bibitem [{\citenamefont {Zhou}\ \emph {et~al.}(2016)\citenamefont {Zhou},
  \citenamefont {Yao}, \citenamefont {Li}, \citenamefont {Meng},\ and\
  \citenamefont {Ring}}]{Zhou:2016_Ne20}%
  \BibitemOpen
  \bibfield  {author} {\bibinfo {author} {\bibfnamefont {E.~F.}\ \bibnamefont
  {Zhou}}, \bibinfo {author} {\bibfnamefont {J.~M.}\ \bibnamefont {Yao}},
  \bibinfo {author} {\bibfnamefont {Z.~P.}\ \bibnamefont {Li}}, \bibinfo
  {author} {\bibfnamefont {J.}~\bibnamefont {Meng}}, \ and\ \bibinfo {author}
  {\bibfnamefont {P.}~\bibnamefont {Ring}},\ }\href {\doibase
  10.1016/j.physletb.2015.12.028} {\bibfield  {journal} {\bibinfo  {journal}
  {Phys. Lett. B}\ }\textbf {\bibinfo {volume} {753}},\ \bibinfo {pages} {227}
  (\bibinfo {year} {2016})},\ \Eprint {http://arxiv.org/abs/1510.05232}
  {arXiv:1510.05232 [nucl-th]} \BibitemShut {NoStop}%
\bibitem [{\citenamefont {Rong}\ \emph {et~al.}(2023)\citenamefont {Rong},
  \citenamefont {Wu}, \citenamefont {Lu},\ and\ \citenamefont
  {Yao}}]{Rong:2023}%
  \BibitemOpen
  \bibfield  {author} {\bibinfo {author} {\bibfnamefont {Y.-T.}\ \bibnamefont
  {Rong}}, \bibinfo {author} {\bibfnamefont {X.-Y.}\ \bibnamefont {Wu}},
  \bibinfo {author} {\bibfnamefont {B.-N.}\ \bibnamefont {Lu}}, \ and\ \bibinfo
  {author} {\bibfnamefont {J.-M.}\ \bibnamefont {Yao}},\ }\href {\doibase
  10.1016/j.physletb.2023.137896} {\bibfield  {journal} {\bibinfo  {journal}
  {Phys. Lett. B}\ }\textbf {\bibinfo {volume} {840}},\ \bibinfo {pages}
  {137896} (\bibinfo {year} {2023})},\ \Eprint
  {http://arxiv.org/abs/2201.02114} {arXiv:2201.02114 [nucl-th]} \BibitemShut
  {NoStop}%
\bibitem [{\citenamefont {Mei}\ \emph {et~al.}(2016)\citenamefont {Mei},
  \citenamefont {Hagino},\ and\ \citenamefont {Yao}}]{Mei:2016PRC}%
  \BibitemOpen
  \bibfield  {author} {\bibinfo {author} {\bibfnamefont {H.}~\bibnamefont
  {Mei}}, \bibinfo {author} {\bibfnamefont {K.}~\bibnamefont {Hagino}}, \ and\
  \bibinfo {author} {\bibfnamefont {J.~M.}\ \bibnamefont {Yao}},\ }\href
  {\doibase 10.1103/PhysRevC.93.011301} {\bibfield  {journal} {\bibinfo
  {journal} {Phys. Rev. C}\ }\textbf {\bibinfo {volume} {93}},\ \bibinfo
  {pages} {011301} (\bibinfo {year} {2016})}\BibitemShut {NoStop}%
\bibitem [{\citenamefont {Xia}\ \emph {et~al.}(2023)\citenamefont {Xia},
  \citenamefont {Wu}, \citenamefont {Mei},\ and\ \citenamefont
  {Yao}}]{Xia:2023}%
  \BibitemOpen
  \bibfield  {author} {\bibinfo {author} {\bibfnamefont {H.}~\bibnamefont
  {Xia}}, \bibinfo {author} {\bibfnamefont {X.}~\bibnamefont {Wu}}, \bibinfo
  {author} {\bibfnamefont {H.}~\bibnamefont {Mei}}, \ and\ \bibinfo {author}
  {\bibfnamefont {J.}~\bibnamefont {Yao}},\ }\href {\doibase
  10.1007/s11433-022-2045-x} {\bibfield  {journal} {\bibinfo  {journal} {Sci.
  China Phys. Mech. Astron.}\ }\textbf {\bibinfo {volume} {66}},\ \bibinfo
  {pages} {252011} (\bibinfo {year} {2023})},\ \Eprint
  {http://arxiv.org/abs/2211.09356} {arXiv:2211.09356 [nucl-th]} \BibitemShut
  {NoStop}%
\bibitem [{\citenamefont {Song}\ \emph {et~al.}(2014)\citenamefont {Song},
  \citenamefont {Yao}, \citenamefont {Ring},\ and\ \citenamefont
  {Meng}}]{Song:2014}%
  \BibitemOpen
  \bibfield  {author} {\bibinfo {author} {\bibfnamefont {L.~S.}\ \bibnamefont
  {Song}}, \bibinfo {author} {\bibfnamefont {J.~M.}\ \bibnamefont {Yao}},
  \bibinfo {author} {\bibfnamefont {P.}~\bibnamefont {Ring}}, \ and\ \bibinfo
  {author} {\bibfnamefont {J.}~\bibnamefont {Meng}},\ }\href {\doibase
  https://doi.org/10.1103/PhysRevC.90.054309} {\bibfield  {journal} {\bibinfo
  {journal} {Phys. Rev. C}\ }\textbf {\bibinfo {volume} {90}},\ \bibinfo
  {pages} {054309} (\bibinfo {year} {2014})}\BibitemShut {NoStop}%
\bibitem [{\citenamefont {Yao}\ \emph {et~al.}(2015{\natexlab{b}})\citenamefont
  {Yao}, \citenamefont {Song}, \citenamefont {Hagino}, \citenamefont {Ring},\
  and\ \citenamefont {Meng}}]{Yao:2015_NLDBD}%
  \BibitemOpen
  \bibfield  {author} {\bibinfo {author} {\bibfnamefont {J.~M.}\ \bibnamefont
  {Yao}}, \bibinfo {author} {\bibfnamefont {L.~S.}\ \bibnamefont {Song}},
  \bibinfo {author} {\bibfnamefont {K.}~\bibnamefont {Hagino}}, \bibinfo
  {author} {\bibfnamefont {P.}~\bibnamefont {Ring}}, \ and\ \bibinfo {author}
  {\bibfnamefont {J.}~\bibnamefont {Meng}},\ }\href {\doibase
  10.1103/PhysRevC.91.024316} {\bibfield  {journal} {\bibinfo  {journal} {Phys.
  Rev. C}\ }\textbf {\bibinfo {volume} {91}},\ \bibinfo {pages} {024316}
  (\bibinfo {year} {2015}{\natexlab{b}})}\BibitemShut {NoStop}%
\bibitem [{\citenamefont {Song}\ \emph {et~al.}(2017)\citenamefont {Song},
  \citenamefont {Yao}, \citenamefont {Ring},\ and\ \citenamefont
  {Meng}}]{Song:2017}%
  \BibitemOpen
  \bibfield  {author} {\bibinfo {author} {\bibfnamefont {L.~S.}\ \bibnamefont
  {Song}}, \bibinfo {author} {\bibfnamefont {J.~M.}\ \bibnamefont {Yao}},
  \bibinfo {author} {\bibfnamefont {P.}~\bibnamefont {Ring}}, \ and\ \bibinfo
  {author} {\bibfnamefont {J.}~\bibnamefont {Meng}},\ }\href {\doibase
  10.1103/PhysRevC.95.024305} {\bibfield  {journal} {\bibinfo  {journal} {Phys.
  Rev. C}\ }\textbf {\bibinfo {volume} {95}},\ \bibinfo {pages} {024305}
  (\bibinfo {year} {2017})}\BibitemShut {NoStop}%
\bibitem [{\citenamefont {Ding}\ \emph {et~al.}(2023)\citenamefont {Ding},
  \citenamefont {Zhang}, \citenamefont {Yao}, \citenamefont {Ring},\ and\
  \citenamefont {Meng}}]{Ding:2023}%
  \BibitemOpen
  \bibfield  {author} {\bibinfo {author} {\bibfnamefont {C.~R.}\ \bibnamefont
  {Ding}}, \bibinfo {author} {\bibfnamefont {X.}~\bibnamefont {Zhang}},
  \bibinfo {author} {\bibfnamefont {J.~M.}\ \bibnamefont {Yao}}, \bibinfo
  {author} {\bibfnamefont {P.}~\bibnamefont {Ring}}, \ and\ \bibinfo {author}
  {\bibfnamefont {J.}~\bibnamefont {Meng}},\ }\href {\doibase
  10.1103/PhysRevC.108.054304} {\bibfield  {journal} {\bibinfo  {journal}
  {Phys. Rev. C}\ }\textbf {\bibinfo {volume} {108}},\ \bibinfo {pages}
  {054304} (\bibinfo {year} {2023})},\ \Eprint
  {http://arxiv.org/abs/2305.00742} {arXiv:2305.00742 [nucl-th]} \BibitemShut
  {NoStop}%
\bibitem [{\citenamefont {Yao}\ \emph {et~al.}(2022)\citenamefont {Yao},
  \citenamefont {Meng}, \citenamefont {Niu},\ and\ \citenamefont
  {Ring}}]{Yao:2022PPNP}%
  \BibitemOpen
  \bibfield  {author} {\bibinfo {author} {\bibfnamefont {J.~M.}\ \bibnamefont
  {Yao}}, \bibinfo {author} {\bibfnamefont {J.}~\bibnamefont {Meng}}, \bibinfo
  {author} {\bibfnamefont {Y.~F.}\ \bibnamefont {Niu}}, \ and\ \bibinfo
  {author} {\bibfnamefont {P.}~\bibnamefont {Ring}},\ }\href {\doibase
  10.1016/j.ppnp.2022.103965} {\bibfield  {journal} {\bibinfo  {journal} {Prog.
  Part. Nucl. Phys.}\ }\textbf {\bibinfo {volume} {126}},\ \bibinfo {pages}
  {103965} (\bibinfo {year} {2022})},\ \Eprint
  {http://arxiv.org/abs/2111.15543} {arXiv:2111.15543 [nucl-th]} \BibitemShut
  {NoStop}%
\bibitem [{\citenamefont {Zhou}\ and\ \citenamefont
  {Yao}(2023)}]{Zhou:2023_IJMPE}%
  \BibitemOpen
  \bibfield  {author} {\bibinfo {author} {\bibfnamefont {E.~F.}\ \bibnamefont
  {Zhou}}\ and\ \bibinfo {author} {\bibfnamefont {J.~M.}\ \bibnamefont {Yao}},\
  }\href@noop {} {\  (\bibinfo {year} {2023})},\ \Eprint
  {http://arxiv.org/abs/2309.09488} {arXiv:2309.09488 [nucl-th]} \BibitemShut
  {NoStop}%
\bibitem [{\citenamefont {Hill}\ and\ \citenamefont
  {Wheeler}(1953)}]{Hill:1953}%
  \BibitemOpen
  \bibfield  {author} {\bibinfo {author} {\bibfnamefont {D.~L.}\ \bibnamefont
  {Hill}}\ and\ \bibinfo {author} {\bibfnamefont {J.~A.}\ \bibnamefont
  {Wheeler}},\ }\href {\doibase 10.1103/PhysRev.89.1102} {\bibfield  {journal}
  {\bibinfo  {journal} {Phys. Rev.}\ }\textbf {\bibinfo {volume} {89}},\
  \bibinfo {pages} {1102} (\bibinfo {year} {1953})}\BibitemShut {NoStop}%
\bibitem [{\citenamefont {Bonche}\ \emph {et~al.}(1990)\citenamefont {Bonche},
  \citenamefont {Dobaczewski}, \citenamefont {Flocard}, \citenamefont
  {Heenen},\ and\ \citenamefont {Meyer}}]{Bonche:1990ghh}%
  \BibitemOpen
  \bibfield  {author} {\bibinfo {author} {\bibfnamefont {P.}~\bibnamefont
  {Bonche}}, \bibinfo {author} {\bibfnamefont {J.}~\bibnamefont {Dobaczewski}},
  \bibinfo {author} {\bibfnamefont {H.}~\bibnamefont {Flocard}}, \bibinfo
  {author} {\bibfnamefont {P.~H.}\ \bibnamefont {Heenen}}, \ and\ \bibinfo
  {author} {\bibfnamefont {J.}~\bibnamefont {Meyer}},\ }\href {\doibase
  10.1016/0375-9474(90)90062-Q} {\bibfield  {journal} {\bibinfo  {journal}
  {Nucl. Phys. A}\ }\textbf {\bibinfo {volume} {510}},\ \bibinfo {pages} {466}
  (\bibinfo {year} {1990})}\BibitemShut {NoStop}%
\bibitem [{\citenamefont {Yao}\ \emph {et~al.}(2010{\natexlab{b}})\citenamefont
  {Yao}, \citenamefont {Meng}, \citenamefont {Ring},\ and\ \citenamefont
  {Vretenar}}]{Yao:2010}%
  \BibitemOpen
  \bibfield  {author} {\bibinfo {author} {\bibfnamefont {J.~M.}\ \bibnamefont
  {Yao}}, \bibinfo {author} {\bibfnamefont {J.}~\bibnamefont {Meng}}, \bibinfo
  {author} {\bibfnamefont {P.}~\bibnamefont {Ring}}, \ and\ \bibinfo {author}
  {\bibfnamefont {D.}~\bibnamefont {Vretenar}},\ }\href {\doibase
  10.1103/PhysRevC.81.044311} {\bibfield  {journal} {\bibinfo  {journal} {Phys.
  Rev. C}\ }\textbf {\bibinfo {volume} {81}},\ \bibinfo {pages} {044311}
  (\bibinfo {year} {2010}{\natexlab{b}})},\ \Eprint
  {http://arxiv.org/abs/0912.2650} {arXiv:0912.2650 [nucl-th]} \BibitemShut
  {NoStop}%
\bibitem [{\citenamefont {Burvenich}\ \emph {et~al.}(2002)\citenamefont
  {Burvenich}, \citenamefont {Madland}, \citenamefont {Maruhn},\ and\
  \citenamefont {Reinhard}}]{Burvenich:2001rh}%
  \BibitemOpen
  \bibfield  {author} {\bibinfo {author} {\bibfnamefont {T.}~\bibnamefont
  {Burvenich}}, \bibinfo {author} {\bibfnamefont {D.~G.}\ \bibnamefont
  {Madland}}, \bibinfo {author} {\bibfnamefont {J.~A.}\ \bibnamefont {Maruhn}},
  \ and\ \bibinfo {author} {\bibfnamefont {P.~G.}\ \bibnamefont {Reinhard}},\
  }\href {\doibase 10.1103/PhysRevC.65.044308} {\bibfield  {journal} {\bibinfo
  {journal} {Phys. Rev. C}\ }\textbf {\bibinfo {volume} {65}},\ \bibinfo
  {pages} {044308} (\bibinfo {year} {2002})},\ \Eprint
  {http://arxiv.org/abs/nucl-th/0111012} {arXiv:nucl-th/0111012} \BibitemShut
  {NoStop}%
\bibitem [{\citenamefont {Zhao}\ \emph {et~al.}(2010)\citenamefont {Zhao},
  \citenamefont {Li}, \citenamefont {Yao},\ and\ \citenamefont
  {Meng}}]{Zhao:2010}%
  \BibitemOpen
  \bibfield  {author} {\bibinfo {author} {\bibfnamefont {P.~W.}\ \bibnamefont
  {Zhao}}, \bibinfo {author} {\bibfnamefont {Z.~P.}\ \bibnamefont {Li}},
  \bibinfo {author} {\bibfnamefont {J.~M.}\ \bibnamefont {Yao}}, \ and\
  \bibinfo {author} {\bibfnamefont {J.}~\bibnamefont {Meng}},\ }\href {\doibase
  10.1103/PhysRevC.82.054319} {\bibfield  {journal} {\bibinfo  {journal} {Phys.
  Rev. C}\ }\textbf {\bibinfo {volume} {82}},\ \bibinfo {pages} {054319}
  (\bibinfo {year} {2010})},\ \Eprint {http://arxiv.org/abs/1002.1789}
  {arXiv:1002.1789 [nucl-th]} \BibitemShut {NoStop}%
\bibitem [{\citenamefont {Bender}\ \emph
  {et~al.}(2000{\natexlab{b}})\citenamefont {Bender}, \citenamefont {Rutz},
  \citenamefont {Reinhard},\ and\ \citenamefont {Maruhn}}]{Bender:2000com}%
  \BibitemOpen
  \bibfield  {author} {\bibinfo {author} {\bibfnamefont {M.}~\bibnamefont
  {Bender}}, \bibinfo {author} {\bibfnamefont {K.}~\bibnamefont {Rutz}},
  \bibinfo {author} {\bibfnamefont {P.~G.}\ \bibnamefont {Reinhard}}, \ and\
  \bibinfo {author} {\bibfnamefont {J.~A.}\ \bibnamefont {Maruhn}},\ }\href
  {\doibase 10.1007/s100500050419} {\bibfield  {journal} {\bibinfo  {journal}
  {Eur. Phys. J. A}\ }\textbf {\bibinfo {volume} {7}},\ \bibinfo {pages} {467}
  (\bibinfo {year} {2000}{\natexlab{b}})},\ \Eprint
  {http://arxiv.org/abs/nucl-th/9910025} {arXiv:nucl-th/9910025} \BibitemShut
  {NoStop}%
\bibitem [{\citenamefont {Chasman}\ and\ \citenamefont
  {Wahlborn}(1967)}]{Chasman:1967}%
  \BibitemOpen
  \bibfield  {author} {\bibinfo {author} {\bibfnamefont {R.~R.}\ \bibnamefont
  {Chasman}}\ and\ \bibinfo {author} {\bibfnamefont {S.}~\bibnamefont
  {Wahlborn}},\ }\href {\doibase 10.1016/0375-9474(67)90242-4} {\bibfield
  {journal} {\bibinfo  {journal} {Nucl. Phys. A}\ }\textbf {\bibinfo {volume}
  {90}},\ \bibinfo {pages} {401} (\bibinfo {year} {1967})}\BibitemShut
  {NoStop}%
\bibitem [{\citenamefont {Serot}(1981)}]{Serot:1981}%
  \BibitemOpen
  \bibfield  {author} {\bibinfo {author} {\bibfnamefont {B.~D.}\ \bibnamefont
  {Serot}},\ }\href {\doibase 10.1016/0370-2693(81)90826-1} {\bibfield
  {journal} {\bibinfo  {journal} {Phys. Lett. B}\ }\textbf {\bibinfo {volume}
  {107}},\ \bibinfo {pages} {263} (\bibinfo {year} {1981})}\BibitemShut
  {NoStop}%
\bibitem [{\citenamefont {{National Nuclear Data Center}}(2020)}]{NNDC}%
  \BibitemOpen
  \bibfield  {author} {\bibinfo {author} {\bibnamefont {{National Nuclear Data
  Center}}},\ }\href {https://www.nndc.bnl.gov/nudat2} {\enquote {\bibinfo
  {title} {{NuDat 2 Database}},}\ } (\bibinfo {year} {2020}),\ \bibinfo {note}
  {\url{https://www.nndc.bnl.gov/nudat2}}\BibitemShut {NoStop}%
\bibitem [{\citenamefont {Firestone}(2009)}]{Firestone:2009}%
  \BibitemOpen
  \bibfield  {author} {\bibinfo {author} {\bibfnamefont {R.}~\bibnamefont
  {Firestone}},\ }\href {\doibase https://doi.org/10.1016/j.nds.2009.06.001}
  {\bibfield  {journal} {\bibinfo  {journal} {Nuclear Data Sheets}\ }\textbf
  {\bibinfo {volume} {110}},\ \bibinfo {pages} {1691} (\bibinfo {year}
  {2009})}\BibitemShut {NoStop}%
\bibitem [{\citenamefont {Yao}\ \emph {et~al.}(2023)\citenamefont {Yao},
  \citenamefont {Wu},\ and\ \citenamefont {Mei}}]{Yao:2023_NPA}%
  \BibitemOpen
  \bibfield  {author} {\bibinfo {author} {\bibfnamefont {Y.~G.}\ \bibnamefont
  {Yao}}, \bibinfo {author} {\bibfnamefont {X.~Y.}\ \bibnamefont {Wu}}, \ and\
  \bibinfo {author} {\bibfnamefont {H.}~\bibnamefont {Mei}},\ }\href {\doibase
  https://doi.org/10.1016/j.nuclphysa.2023.122794} {\bibfield  {journal}
  {\bibinfo  {journal} {Nucl. Phys. A}\ } (\bibinfo {year} {2023}),\
  https://doi.org/10.1016/j.nuclphysa.2023.122794}\BibitemShut {NoStop}%
\bibitem [{\citenamefont {Firestone}(2015)}]{Firestone:2015}%
  \BibitemOpen
  \bibfield  {author} {\bibinfo {author} {\bibfnamefont {R.~B.}\ \bibnamefont
  {Firestone}},\ }\href {\doibase 10.1016/j.nds.2015.07.001} {\bibfield
  {journal} {\bibinfo  {journal} {Nucl. Data Sheets}\ }\textbf {\bibinfo
  {volume} {127}},\ \bibinfo {pages} {1} (\bibinfo {year} {2015})}\BibitemShut
  {NoStop}%
\bibitem [{\citenamefont {Matsumoto}\ \emph {et~al.}(2023)\citenamefont
  {Matsumoto}, \citenamefont {Tanimura},\ and\ \citenamefont
  {Hagino}}]{Matsumoto:2023}%
  \BibitemOpen
  \bibfield  {author} {\bibinfo {author} {\bibfnamefont {M.}~\bibnamefont
  {Matsumoto}}, \bibinfo {author} {\bibfnamefont {Y.}~\bibnamefont {Tanimura}},
  \ and\ \bibinfo {author} {\bibfnamefont {K.}~\bibnamefont {Hagino}},\ }\href
  {\doibase 10.1103/PhysRevC.108.L051302} {\bibfield  {journal} {\bibinfo
  {journal} {Phys. Rev. C}\ }\textbf {\bibinfo {volume} {108}},\ \bibinfo
  {pages} {L051302} (\bibinfo {year} {2023})}\BibitemShut {NoStop}%
\bibitem [{\citenamefont {Robledo}(2009)}]{Robledo:2009yd}%
  \BibitemOpen
  \bibfield  {author} {\bibinfo {author} {\bibfnamefont {L.~M.}\ \bibnamefont
  {Robledo}},\ }\href {\doibase 10.1103/PhysRevC.79.021302} {\bibfield
  {journal} {\bibinfo  {journal} {Phys. Rev. C}\ }\textbf {\bibinfo {volume}
  {79}},\ \bibinfo {pages} {021302} (\bibinfo {year} {2009})},\ \Eprint
  {http://arxiv.org/abs/0901.3213} {arXiv:0901.3213 [nucl-th]} \BibitemShut
  {NoStop}%
\bibitem [{\citenamefont {Balian}\ and\ \citenamefont
  {Brezin}(1969)}]{Balian:1969}%
  \BibitemOpen
  \bibfield  {author} {\bibinfo {author} {\bibfnamefont {R.}~\bibnamefont
  {Balian}}\ and\ \bibinfo {author} {\bibfnamefont {E.}~\bibnamefont
  {Brezin}},\ }\href {\doibase 10.1007/BF02710281} {\bibfield  {journal}
  {\bibinfo  {journal} {Nuovo Cim. B}\ }\textbf {\bibinfo {volume} {64}},\
  \bibinfo {pages} {37} (\bibinfo {year} {1969})}\BibitemShut {NoStop}%
\bibitem [{\citenamefont {Anguiano}\ \emph {et~al.}(2001)\citenamefont
  {Anguiano}, \citenamefont {Egido},\ and\ \citenamefont
  {Robledo}}]{Anguiano:2001}%
  \BibitemOpen
  \bibfield  {author} {\bibinfo {author} {\bibfnamefont {M.}~\bibnamefont
  {Anguiano}}, \bibinfo {author} {\bibfnamefont {J.~L.}\ \bibnamefont {Egido}},
  \ and\ \bibinfo {author} {\bibfnamefont {L.~M.}\ \bibnamefont {Robledo}},\
  }\href {\doibase 10.1016/S0375-9474(01)01219-2} {\bibfield  {journal}
  {\bibinfo  {journal} {Nucl. Phys. A}\ }\textbf {\bibinfo {volume} {696}},\
  \bibinfo {pages} {467} (\bibinfo {year} {2001})},\ \Eprint
  {http://arxiv.org/abs/nucl-th/0105003} {arXiv:nucl-th/0105003} \BibitemShut
  {NoStop}%
\bibitem [{\citenamefont {Tajima}\ \emph {et~al.}(1992)\citenamefont {Tajima},
  \citenamefont {Flocard}, \citenamefont {Bonche}, \citenamefont
  {Dobaczewski},\ and\ \citenamefont {Heenen}}]{Tajima:1992}%
  \BibitemOpen
  \bibfield  {author} {\bibinfo {author} {\bibfnamefont {N.}~\bibnamefont
  {Tajima}}, \bibinfo {author} {\bibfnamefont {H.}~\bibnamefont {Flocard}},
  \bibinfo {author} {\bibfnamefont {P.}~\bibnamefont {Bonche}}, \bibinfo
  {author} {\bibfnamefont {J.}~\bibnamefont {Dobaczewski}}, \ and\ \bibinfo
  {author} {\bibfnamefont {P.~H.}\ \bibnamefont {Heenen}},\ }\href {\doibase
  10.1016/0375-9474(92)90101-O} {\bibfield  {journal} {\bibinfo  {journal}
  {Nucl. Phys. A}\ }\textbf {\bibinfo {volume} {542}},\ \bibinfo {pages} {355}
  (\bibinfo {year} {1992})}\BibitemShut {NoStop}%
\bibitem [{\citenamefont {Dobaczewski}\ \emph {et~al.}(2007)\citenamefont
  {Dobaczewski}, \citenamefont {Stoitsov}, \citenamefont {Nazarewicz},\ and\
  \citenamefont {Reinhard}}]{Dobaczewski:2007}%
  \BibitemOpen
  \bibfield  {author} {\bibinfo {author} {\bibfnamefont {J.}~\bibnamefont
  {Dobaczewski}}, \bibinfo {author} {\bibfnamefont {M.~V.}\ \bibnamefont
  {Stoitsov}}, \bibinfo {author} {\bibfnamefont {W.}~\bibnamefont
  {Nazarewicz}}, \ and\ \bibinfo {author} {\bibfnamefont {P.-G.}\ \bibnamefont
  {Reinhard}},\ }\href {\doibase 10.1103/PhysRevC.76.054315} {\bibfield
  {journal} {\bibinfo  {journal} {Phys. Rev. C}\ }\textbf {\bibinfo {volume}
  {76}},\ \bibinfo {pages} {054315} (\bibinfo {year} {2007})}\BibitemShut
  {NoStop}%
\bibitem [{\citenamefont {Bender}\ \emph {et~al.}(2009)\citenamefont {Bender},
  \citenamefont {Duguet},\ and\ \citenamefont {Lacroix}}]{Bender:2009}%
  \BibitemOpen
  \bibfield  {author} {\bibinfo {author} {\bibfnamefont {M.}~\bibnamefont
  {Bender}}, \bibinfo {author} {\bibfnamefont {T.}~\bibnamefont {Duguet}}, \
  and\ \bibinfo {author} {\bibfnamefont {D.}~\bibnamefont {Lacroix}},\ }\href
  {\doibase 10.1103/PhysRevC.79.044319} {\bibfield  {journal} {\bibinfo
  {journal} {Phys. Rev. C}\ }\textbf {\bibinfo {volume} {79}},\ \bibinfo
  {pages} {044319} (\bibinfo {year} {2009})},\ \Eprint
  {http://arxiv.org/abs/0809.2045} {arXiv:0809.2045 [nucl-th]} \BibitemShut
  {NoStop}%
\bibitem [{\citenamefont {Duguet}\ \emph {et~al.}(2009)\citenamefont {Duguet},
  \citenamefont {Bender}, \citenamefont {Bennaceur}, \citenamefont {Lacroix},\
  and\ \citenamefont {Lesinski}}]{Duguet:2009}%
  \BibitemOpen
  \bibfield  {author} {\bibinfo {author} {\bibfnamefont {T.}~\bibnamefont
  {Duguet}}, \bibinfo {author} {\bibfnamefont {M.}~\bibnamefont {Bender}},
  \bibinfo {author} {\bibfnamefont {K.}~\bibnamefont {Bennaceur}}, \bibinfo
  {author} {\bibfnamefont {D.}~\bibnamefont {Lacroix}}, \ and\ \bibinfo
  {author} {\bibfnamefont {T.}~\bibnamefont {Lesinski}},\ }\href {\doibase
  10.1103/PhysRevC.79.044320} {\bibfield  {journal} {\bibinfo  {journal} {Phys.
  Rev. C}\ }\textbf {\bibinfo {volume} {79}},\ \bibinfo {pages} {044320}
  (\bibinfo {year} {2009})},\ \Eprint {http://arxiv.org/abs/0809.2049}
  {arXiv:0809.2049 [nucl-th]} \BibitemShut {NoStop}%
\bibitem [{\citenamefont {Jiao}\ \emph {et~al.}(2017)\citenamefont {Jiao},
  \citenamefont {Engel},\ and\ \citenamefont {Holt}}]{Jiao:2017}%
  \BibitemOpen
  \bibfield  {author} {\bibinfo {author} {\bibfnamefont {C.~F.}\ \bibnamefont
  {Jiao}}, \bibinfo {author} {\bibfnamefont {J.}~\bibnamefont {Engel}}, \ and\
  \bibinfo {author} {\bibfnamefont {J.~D.}\ \bibnamefont {Holt}},\ }\href
  {\doibase 10.1103/PhysRevC.96.054310} {\bibfield  {journal} {\bibinfo
  {journal} {Phys. Rev. C}\ }\textbf {\bibinfo {volume} {96}},\ \bibinfo
  {pages} {054310} (\bibinfo {year} {2017})}\BibitemShut {NoStop}%
\bibitem [{\citenamefont {Yao}\ \emph {et~al.}(2020)\citenamefont {Yao},
  \citenamefont {Bally}, \citenamefont {Engel}, \citenamefont {Wirth},
  \citenamefont {Rodr\'{\i}guez},\ and\ \citenamefont {Hergert}}]{Yao:2020}%
  \BibitemOpen
  \bibfield  {author} {\bibinfo {author} {\bibfnamefont {J.~M.}\ \bibnamefont
  {Yao}}, \bibinfo {author} {\bibfnamefont {B.}~\bibnamefont {Bally}}, \bibinfo
  {author} {\bibfnamefont {J.}~\bibnamefont {Engel}}, \bibinfo {author}
  {\bibfnamefont {R.}~\bibnamefont {Wirth}}, \bibinfo {author} {\bibfnamefont
  {T.~R.}\ \bibnamefont {Rodr\'{\i}guez}}, \ and\ \bibinfo {author}
  {\bibfnamefont {H.}~\bibnamefont {Hergert}},\ }\href {\doibase
  10.1103/PhysRevLett.124.232501} {\bibfield  {journal} {\bibinfo  {journal}
  {Phys. Rev. Lett.}\ }\textbf {\bibinfo {volume} {124}},\ \bibinfo {pages}
  {232501} (\bibinfo {year} {2020})}\BibitemShut {NoStop}%
\bibitem [{\citenamefont {S\'anchez-Fern\'andez}\ \emph
  {et~al.}(2021)\citenamefont {S\'anchez-Fern\'andez}, \citenamefont {Bally},\
  and\ \citenamefont {Rodr\'{\i}guez}}]{Bally:2021_sd}%
  \BibitemOpen
  \bibfield  {author} {\bibinfo {author} {\bibfnamefont {A.}~\bibnamefont
  {S\'anchez-Fern\'andez}}, \bibinfo {author} {\bibfnamefont {B.}~\bibnamefont
  {Bally}}, \ and\ \bibinfo {author} {\bibfnamefont {T.~R.}\ \bibnamefont
  {Rodr\'{\i}guez}},\ }\href {\doibase 10.1103/PhysRevC.104.054306} {\bibfield
  {journal} {\bibinfo  {journal} {Phys. Rev. C}\ }\textbf {\bibinfo {volume}
  {104}},\ \bibinfo {pages} {054306} (\bibinfo {year} {2021})}\BibitemShut
  {NoStop}%
\bibitem [{\citenamefont {Fomenko}(1970)}]{Fomenko:1970JPA}%
  \BibitemOpen
  \bibfield  {author} {\bibinfo {author} {\bibfnamefont {V.~N.}\ \bibnamefont
  {Fomenko}},\ }\href {\doibase 10.1088/0305-4470/3/1/002} {\bibfield
  {journal} {\bibinfo  {journal} {J. Phys. A: General Physics}\ }\textbf
  {\bibinfo {volume} {3}},\ \bibinfo {pages} {8} (\bibinfo {year}
  {1970})}\BibitemShut {NoStop}%
\end{thebibliography}
  
%

\end{document}